\documentclass[prb,amsmath,amssymb,superscriptaddress,twocolumn]{revtex4}
\usepackage{graphicx}

\newcommand{\br}{{\bf r}}

\newcommand{\bk}{{\bf k}}

\newcommand{\bj}{{\bf j}}
\newcommand{\bp}{{\bf p}}

\newcommand{\bq}{{\bf q}}

\newcommand{\bJ}{{\bf J}}

\newcommand{\eps}{\epsilon}

\newcommand{\cde}{\mathcal D}
\newcommand{\mesk}{\int\frac{d^D\bk}{(2\pi)^D}\int_{-\infty}^{\infty}\frac{d\omega}{2\pi}}
\newcommand{\mesp}{\int\frac{d^D\bp}{(2\pi)^D}\int_{-\infty}^{\infty}\frac{d\omega'}{2\pi}}
\DeclareMathAlphabet{\mathpzc}{OT1}{pzc}{m}{it} 
\begin{document}
\title{Conductivity of interacting massless Dirac particles in graphene: collisionless regime}
\author{Vladimir Juri\v ci\'c}
\affiliation{Instituut-Lorentz for Theoretical Physics, Universiteit
Leiden,
 P.O. Box 9506, 2300 RA Leiden, The Netherlands}
\affiliation{Max-Planck-Institute for Solid State Research,
Heisenbergstr. 1, D-70569 Stuttgart, Germany}
\author{Oskar Vafek}
\affiliation{National High Magnetic Field Laboratory and Department
of Physics,\\ Florida State University, Tallahasse, Florida 32306,
USA}
\author{Igor F. Herbut}
\affiliation{Department of Physics, Simon Fraser University,
Burnaby, British Columbia, Canada V5A 1S6}

\begin{abstract}
We provide detailed calculation of the a.c. conductivity in the case
of $1/r$ Coulomb interacting massless Dirac particles in graphene in
the collisionless limit when $\omega\gg T$. The analysis of the
electron self-energy, current vertex function and polarization
function, which enter into the calculation of physical quantities
including the a.c. conductivity, is carried out by checking the
Ward-Takahashi identities associated with the electrical charge
conservation and making sure that they are satisfied at each step.
We adopt a variant of the dimensional regularization of Veltman and
't Hooft by taking the spatial dimension $D=2-\eps$ for $\eps>0$. The
procedure adopted here yields a result for the conductivity
correction which, while explicitly preserving charge conservation
laws, is nevertheless different from the results reported previously
in literature.
\end{abstract} \maketitle

\section{Introduction}
The role of Coulomb electron-electron interactions in systems
described by massless two-dimensional Dirac fermions has been a
subject of interest for some
time.\cite{Gonzalez1994,dima,igor,VafekPRL2007,aleiner,son,
MishchenkoPRL,SheehySchmalianPRL,HerbutJuricicVafekPRL08,
MishchenkoEPL08,ijr,muller,igorPHYS,juricic,wang} Discovery of graphene,
a single atomic layer of sp$^2$ hybridized carbon, and more recently
of topological insulators, both of which support such massless Dirac
fermions, brought this issue into sharp focus. In particular, which
physically measurable quantities are modified from their
non-interacting values, and by how much, would allow deeper
understanding of the  physics governed by electron-electron
interactions in these systems.

When weak, the unscreened $1/r$ Coulomb interactions are expected to
modify the velocity of the Dirac fermions as $v_{F}\rightarrow
v_F+\frac{e^2}{4}\ln\Lambda/k$, where $k$ is the wavenumber measured
from the Dirac point. This modification of the electronic dispersion
is expected to lead to logarithmic suppression of the density of
states near the Dirac point, an effect in principle observable in
tunneling experiments. In addition, the low temperature electronic
contribution to the specific heat should be suppressed from $T^2$ to
$T^2/\log^2T$, as shown in Ref.\ \onlinecite{VafekPRL2007}, and the strength of this suppression is
related to the strength of the Coulomb interaction.

The role of Coulomb interaction in a.c. electrical transport was
investigated by Mishchenko in Ref.\ \onlinecite{MishchenkoPRL}, who
originally concluded that the a.c. conductivity $\sigma(\omega)$
vanishes as $\omega\rightarrow 0$ and the system is a (weak) {\it
insulator}. Were this the case, the interactions would have dramatic
effect on transport since the a.c. conductivity of the
non-interacting system is finite,\cite{Peres} i.e., $\sigma_0(\omega)=\pi e^2/2h$ for $\omega\gg T$.
This was later argued to be incorrect by Sheehy and
Schmalian\cite{SheehySchmalianPRL}, and independently by the present
authors\cite{HerbutJuricicVafekPRL08} using Renormalization Group (RG)
scaling analysis. While the former presented only a scaling
argument, without calculating the correction to transport, the
latter reported on an explicit calculation where
\begin{eqnarray}\label{conductivity-final}
\sigma(\omega)=\sigma_0\left(1+{\cal C}\frac{e^2}{v_F+\frac{e^2}{4}\ln\frac{\Lambda}{\omega}}\right)
\end{eqnarray}
with the coefficient found to be ${\cal C}=(25-6\pi)/12\simeq0.5125$. 
Note that, since $e^2$ does not renormalize,\cite{HerbutPRL2001,HerbutJuricicVafekPRL08} any change 
in the cutoff in the above expression for the
conductivity may be compensated by a redefinition of the Fermi velocity, $v_F$.
At small $\omega$ the
correction vanishes and the non-interacting value of $\sigma$ is
recovered. At small but finite frequencies, the correction scales as
$1/|\log\omega|$, with the {\it interaction independent} prefactor
determined by ${\cal C}$. The numerical value of this correction, which can
be understood as correction to scaling near the Gaussian fixed point
and which is expected to be universal, has since been a subject of
debate. In subsequent work, Mishchenko\cite{MishchenkoEPL08}
recovered the functional form (\ref{conductivity-final}), which gives
metallic conductivity at small $\omega$, but argued for a different
value of ${\cal C}=(19-6\pi)/12\simeq0.01254$ which happens to be much smaller than the one found
by us. Technically, the difference originated from different
regularization adopted in the two approaches. The standard
momentum space cutoff, motivated by the underlying discrete lattice
structure and reported in Ref.\ \onlinecite{HerbutJuricicVafekPRL08} was questioned
in Ref.\ \onlinecite{MishchenkoEPL08}, where the correction to conductivity
was calculated using a cutoff on the $1/r$ interaction,  and argued to
be the same regardless of whether it is calculated using Kubo
formula (current-current correlator) or continuity equation and
density-density correlator. The same calculational
procedure was later advocated by Sheehy and
Schmalian,\cite{Sheehy-SchmalianPRB09} who argued that unlike
hard-cutoff in momentum space, cutoff on the interaction leads to
expressions obeying Ward-Takahashi  identity. In addition, they
claimed the result obtained in such way is consistent with the
experimentally measured optical conductivity, where, surprisingly, no discernible
correction to the non-interacting value was reported.\cite{exp-transparency}

In quantum field theories, it seems
reasonable that if two ultraviolet (UV) regularization
schemes give different results for physical quantities, then the
regularization that is typically chosen is the one which respects
charge U(1) symmetry, as is the case for chiral anomaly in (3+1)-dimensional
massless quantum electrodynamics (QED), for instance.\cite{Peskin}
Here we argue that  the regularization of the electron-electron
interaction alone is incomplete and cannot serve as a consistent
regularization of the theory. We also show by explicit calculation that
the dimensional regularization used here preserves the Ward-Takahashi
identity, i.e., that it is consistent with $U(1)$ gauge symmetry of
the theory,
and that, moreover, has the additional advantage of serving as an
interaction-independent regularization scheme for the whole field
theory. The interaction correction to the conductivity within this
regularization scheme is calculated independently using the
current-current and the density-density correlators,  which both yield
the same number ${\cal C}=(11-3\pi)/6\simeq0.2625$ in Eq.\
(\ref{conductivity-final}), precisely as a consequence of explicitly preserved
$U(1)$ gauge symmetry. Furthermore, we show that while the hard-cutoff regularization
in principle violates the Ward-Takahashi identity, the original integral expression\cite{HerbutJuricicVafekPRL08}
 for the constant ${\cal C}$ is in fact UV convergent, and when computed with necessary care it
unambiguously leads to the same value as quoted above.

A comparison with experiment which has been performed at {\it high}
frequencies near the cutoff\cite{exp-transparency,Basov} (see also Ref.\ \onlinecite{Heinz})
may be misleading,  since the result for the leading
logarithmic correction to the conductivity in
Eq.(\ref{conductivity-final}) is valid only at frequencies of the
order of $1$meV, much {\it smaller} than the cutoff.
As the Coulomb coupling constant in graphene $e^2 /v_F $ is believed to be of order one,
we expect that in this region the
interaction corrections to different observables, relative to the values in the noninteracting theory, should be
significant. Why the interaction correction to the conductivity, in particular,
appears to be small even in the high-frequency regime is unclear at the moment.

Whereas the results in the collisionless limit ($\omega\gg T$)
discussed here at least in principle follow from a straightforward
application of the perturbative renormalization group, transport in
the collision-dominated regime ($\omega\ll T$) requires re-summation
of an infinite series of Feynman diagrams. This is easily seen in
the non-interacting limit where a finite temperature $T$ produces a
finite, linear in $T$, "Drude" $\delta-$function response in
conductivity, $\sim T\delta(\omega)$. Collisions due to the
electron-electron scattering lead to broadening of the
$\delta-$function and clearly the result must be non-analytic in
$e^2/v_F$ as the interaction $V(\br)\rightarrow 0$. Alternative
approach has been advanced in
Refs.\ \onlinecite{FritzPRB08,MullerPRB08} where the leading
correction is argued to be captured by the solution to the quantum
Boltzman equation with the collision integral calculated
perturbatively in the interaction strength. In the clean limit, the
conductivity in the collision dominated regime is found to {\em
increase} with decreasing $T$ and proportional to $\ln^2 (T/\Lambda)$.
Interestingly, experiments on suspended samples at the neutrality
point\cite{BolotinPRL08} find conductivity which {\em decreases} with decreasing
$T$. Finally, $T-$linear increase of the d.c.
conductivity observed in small devices\cite{EvaAndreiNature08}, has
been argued to arise from purely ballistic
transport\cite{MullerPRL09} where conductivity grows with the sample
size $L$ and temperature $T$ as $\sigma\sim TL/\hbar v_F$.

The paper is organized as follows: in section II we introduce the
Lagrangian and the response functions, and in section III we discuss
different regularization schemes for massless Dirac fermions. In
section IV, we review some well-known results regarding the polarization tensor and the conductivity.
In section V, we
explicitly construct polarization tensor for the
non-interacting theory, and in section VI
 we consider the same problem for the contact
interactions to first order in the interaction strength and to
$\mathcal{O}(N)$. We do not discuss the (RPA-like) contribution to the order $N^2$
which, while simple to calculate, does not
contribute to transport.
The main results of the paper are presented in section VII where we show
that the Coulomb correction to the polarization tensor is transversal, as well as that the Coulomb vertex function obeys the Ward-Takahashi
identity within the dimensional regularization. In this section, we also present calculations of the Coulomb correction to the a.c. conductivity using both the current-current correlator (Kubo formula) and the density-density correlator, within the dimensional regularization.
Section VIII is reserved for
further discussion of these results and comparison with previous
results reported in the literature. Various technical
details of the calculations are presented in the appendices.

\section{Hamiltonian, Lagrangian and the response functions}
We start with the Hamiltonian
\begin{eqnarray}
\hat{H}&=&\int d^D\br \psi^{\dagger}(\br)v_F\sigma_a p_a
\psi(\br)\nonumber\\
&+& \int d^D \br d^D\br'
\psi^{\dagger}(\br)\psi(\br)V(|\br-\br'|)\psi^{\dagger}(\br')\psi(\br'),
\end{eqnarray}
where we consider $N$ copies of two-component Fermi fields
$\psi(\br,\tau)$ (which therefore has $2N$ components), the momentum
operator $p_a=-i\hbar\partial_a$ and $\sigma_a$ are Pauli matrices.
Operators in the interaction term are assumed normal ordered.
Hereafter, the Latin letters $a, b$ are used only for the spatial
indices, while the Greek letters $\mu,\nu$ are reserved for the
spacetime ones, and summation over the repeated indices is assumed.
$V(\br)$ is the two-body interaction potential, which is left
unspecified at the moment. Later we will consider two different
cases: a short-range contact interaction $V(\br)=u\delta(\br)$ and
the 3D Coulomb potential $V(\br)=e^2/r$ 
with $e^2/v_F$ as the dimensionless Coulomb coupling constant.
To simplify the notation,
we will work in the natural units $\hbar=c=k_B=1$. When the speed of
light $c$ does not appear, we will also set $v_F=1$. In our final
results we will restore the physical units.

The corresponding imaginary time Lagrangian is
\begin{eqnarray}\label{lagrangian}
\mathcal{L}=\mathcal{L}_0+\mathcal{L}_{int}
\end{eqnarray}
where
\begin{eqnarray}
\mathcal{L}_0=\int d^D\br
\psi^{\dagger}(\tau,\br)\left[\frac{\partial}{\partial
\tau}+v_F\sigma\cdot \bp \right]\psi(\tau,\br)
\end{eqnarray}
and
\begin{eqnarray}
\mathcal{L}_{int}\!=\!\frac{1}{2}\int\!\! d^D\!\!\br d^D\br'
\rho(\tau,\br)V(|\br-\br'|)\rho(\tau,\br'),
\end{eqnarray}
where $\rho(\tau,{\br})\equiv
\psi^{\dagger}(\tau,\br)\psi(\tau,\br)$ is the density of fermions.
The quantum partition function can then be written as the imaginary
time Grassman path integral\cite{Negele-Orland}
\begin{eqnarray}
Z=\int \mathcal{D}\psi^{\dagger}{\mathcal D}\psi
\exp\left({-\int_0^{\beta}d\tau\mathcal{L}}\right)
\end{eqnarray}
where the inverse temperature factor $\beta=1/(k_BT)$. In the
sections which follow, the additional imaginary-time index on the
Fermi field $\psi(\tau,\br)$ inside a path integral automatically
means that they are considered to be coherent state Grassman fields.
We will take $T\rightarrow 0$ first and then perform the
calculations. Note that in light of the discussion in the
Introduction, taking $T\rightarrow 0$ first automatically sets the
collisionless limit.

By the standard spectral representation theorems we can first
calculate the correlation functions as imaginary time ordered
products, Fourier transform over time, and then analytically
continue to find the physical retarded (or advanced) response
functions.\cite{Negele-Orland} Specifically, for some bosonic
operator $\hat{\mathcal{O}}_a(t,\br)$ in the (real time) Heisenberg
representation, the retarded correlation function
\begin{eqnarray}\label{Sret}
S^{ret}_{ab}(t-t',\br,\br')=-i\theta(t-t')\langle
\left[\hat{\mathcal{O}}_a(t,\br),\hat{\mathcal{O}}_b(t',\br')\right]\rangle
\end{eqnarray}
can be related to the imaginary time ordered correlation function
\begin{eqnarray}\label{Sim}
S_{ab}(\tau-\tau',\br,\br')=-\langle T_{\tau}
\hat{\mathcal{O}}_{a}(\tau,\br)\hat{\mathcal{O}}_{b}(\tau',\br')
\rangle,
\end{eqnarray}
where
\begin{eqnarray}
\hat{\mathcal{O}}_{a}(\tau,\br)=e^{\beta\hat{H}}\hat{\mathcal{O}}_{a}(\br)e^{-\beta\hat{H}}.
\end{eqnarray}
In the Eqs.(\ref{Sret}-\ref{Sim}) the angular brackets denote
thermal averaging
\begin{eqnarray}
\langle \ldots\rangle=\frac{1}{Z}\mbox{Tr}\left(e^{-\beta
\hat{H}}\ldots\right).
\end{eqnarray}
Specifically, the frequency Fourier transforms
\begin{eqnarray}
\label{eq:reSFourier}
S^{ret}_{ab}(\omega;\br,\br')&=&\int_{-\infty}^{\infty} d t
e^{i\Omega
t}S^{ret}_{ab}(t,\br,\br'),\\
\label{eq:imSFourier} S_{ab}(i\Omega_n;\br,\br')&=&\int_0^{\beta} d
\tau e^{i\Omega_n \tau}S_{ab}(\tau,\br,\br'),
\end{eqnarray}
satisfy
\begin{eqnarray}
S^{ret}_{ab}(\Omega;\br,\br')=S_{ab}(i\Omega_n\rightarrow\Omega+i0^+;\br,\br'),
\end{eqnarray}
where the bosonic Matsubara frequency is $\Omega_n=2\pi n/\beta$ for
$n=0,\pm1,\pm2,\ldots$. We will use the above relations in what
follows when we focus on the electrical conductivity, in which case
the bosonic operator $\hat{\mathcal{O}}$ of interest will be either
charge density or charge current.

For completeness we note that for $V(\br)=0$ the two-particle
imaginary time Green's function is
\begin{eqnarray}
\langle
\psi(i\omega,\bk)\psi^{\dagger}(i\omega',\bk')\rangle=\beta\delta_{\omega,\omega'}(2\pi)^2\delta(\bk-\bk')G_{\bk}(i\omega)
\end{eqnarray}
where
\begin{eqnarray}\label{GreensFxn}
G_{\bk}(i\omega)&=& \frac{i\omega+\sigma\cdot\bk}{\omega^2+\bk^2},
\end{eqnarray}
which will be used extensively in the later sections. Strictly
speaking, in any solid state system which supports massless Dirac
particles, the above propagator is valid only for wavevectors
smaller than some cutoff $\Lambda$, which depends on the physical
situation. In the case of electrons on the honeycomb lattice, the
order of magnitude of the cutoff, $\AA^{-1}$, is determined by the
requirement that the true electronic dispersion does not deviate
appreciably from the conical (Dirac).

\section{Regularization schemes for massless Dirac fermions}

Since we are interested in the long distance (low frequency)
behavior of physical quantities, we can use the above low energy
field theory, given by the above Lagrangian with the corresponding
propagators, provided that divergent terms in the perturbation
theory are properly regularized. In the context of high energy
physics it is also well known that a quantum field theory of Dirac
fermions needs to be regularized,\cite{Peskin} and typically there
is no unique way of doing so. Additional requirements, usually based
on the symmetries of the theory, determine what type of
regularization should be employed.

In case of the theory of the Coulomb interacting Dirac fermions, we
will require that the $U(1)$ gauge symmetry must be preserved, or
equivalently, that the charge must be conserved. As we show below,
dimensional regularization introduced by 't Hooft and
Veltman\cite{thooft} is consistent with this requirement. Before
discussing this regularization scheme, let us briefly review the
hard cutoff and the Pauli-Villars regularization schemes in the
context of the fermionic field theory considered here.

\subsection{Hard cutoff}

The idea of the hard cutoff regularization is to impose a cutoff in
the upper limit of an otherwise divergent momentum integral.
Physically, this is due to the $\bk-$space restriction on the modes
which appear in the theory, a condition which appears naturally
within Wilson formulation of the RG.\cite{book} The singular part of
the integral then appears dependent on the cutoff scale. Although
very simple to implement, this regularization scheme is known to
violate $U(1)$ gauge symmetry of QED, for instance.\cite{Peskin}
Terms that violate the gauge symmetry appear as a power of the
cutoff scale and must be subtracted in order to insure that the
gauge symmetry is preserved. On the other hand, the typical
divergent terms appear as the logarithm of the cutoff scale. Of
course, the cutoff scale must not appear explicitly in any
observable quantity in order for the theory to be physically
meaningful. The disappearance of the cutoff scale $\Lambda$ indeed
occurs in the calculation of the interaction correction to the a.c.
conductivity within quantum field theory of the Coulomb interacting
Dirac fermions, as discussed below Eq.\ (\ref{conductivity-final}).
However, as we show in Appendix D, and
as was anticipated in Ref.\onlinecite{Sheehy-SchmalianPRB09}, the
hard-cutoff regularization violates the Ward-Takahashi identity. We
are thus led to conclude that this regularization scheme is in
principle not consistent with $U(1)$ gauge symmetry of the theory.
This conclusion notwithstanding, the particular coefficient  ${\cal
C}$ from the introduction may be written as an integral which is
unambiguous and perfectly convergent in the upper limit, provided
the momentum cutoff is taken to infinity after {\it all} the
integrals have been performed (see Appendix H).

\subsection{Pauli-Villars regularization}

Another way to regularize divergent self-energy and vertex diagrams
in QED is to introduce an additional artificial "heavy
photon".\cite{Peskin} In Euclidean spacetime this leads to the
following replacement of the photon propagator
$$
\frac{1}{\Omega^2+\bk^2}\rightarrow
\frac{1}{\Omega^2+\bk^2}-\frac{1}{\Omega^2+\bk^2+M^2}
$$
and the mass parameter $M$ is sent to $\infty$ at the end of the
calculation. Since the additional fictitious particle couples
minimally to the fermions, the regularization preserves
Ward-Takahashi identities which relate the self-energy to the
current vertex. However, as such, this regularization is unable to
render photon polarization diagrams finite. This can be avoided by
introducing additional Pauli-Villars fermions,\cite{BjorkenDrell} at the
expense of making the method complicated.\cite{Peskin}

In the context of the (2+1)D massless Dirac fermions interacting with
static (non-retarded) $1/r$ Coulomb interaction, the analog of the
Pauli-Villars regularization is
$$
\frac{1}{|\bk|}\rightarrow
\frac{1}{|\bk|}-\frac{1}{\sqrt{\bk^2+M^2}}.
$$
Physically, this corresponds to cutting-off the short-distance
divergence of the $1/r$ interaction, without affecting its long
range tail. This modified interaction preserves Ward-Takahashi
identities relating vertex and the
self-energy,\cite{Sheehy-SchmalianPRB09} but, just as in the case of
QED, it fails to regularize the polarization function without
introducing additional Pauli-Villars fermions. Therefore, as such it
cannot serve as a complete regularization of the theory.

\subsection{Dimensional regularization}

Originally introduced in the context of relativistic quantum field
theory, the basic idea of the dimensional regularization is to
regularize four-momentum integrals by lowering the number of
spacetime dimensions over which the integral is performed. This
procedure was introduced by 't Hooft and Veltman\cite{thooft} to
preserve the symmetries of gauge theories. It also bypasses the necessity to
introduce Pauli-Villars fermions and bosons.

Here we employ a variant of the dimensional regularization scheme in
that the frequency integrals are performed from $-\infty$ to
$+\infty$ while the momentum integrals are analytically continued
from $D=2$ to $D=2-\epsilon$ dimensions. Such separation of time
from space is used because in the case considered here the Lorentz
invariance is violated by the interaction terms. A momentum integral
is therefore calculated for an arbitrary number of dimensions $D$,
and expanded in the parameter $\eps$. Singular parts of the integral
then appear as the first-order poles in the Laurent expansion over
the parameter $\eps$, i.e., as terms of the form $1/\eps$, and the
finite part is the term of order $\eps^0$ in this expansion.

The following $D$-dimensional (Euclidean) integrals are frequently
encountered in this regularization scheme\cite{Peskin}
\begin{eqnarray}\label{dim-reg-int}
&&\int\frac{d^D\ell}{(2\pi)^D}\frac{1}{(\ell^2+\Delta)^n}=\frac{\Gamma\left(n-\frac{D}{2}\right)}{(4\pi)^{D/2}\Gamma(n)}
\frac{1}{\Delta^{n-\frac{D}{2}}},
\end{eqnarray}
and
\begin{eqnarray}\label{dim-reg-int1}
&&\int\frac{d^D\ell}{(2\pi)^D}\frac{\ell^2}{(\ell^2+\Delta)^n}=\frac{1}{(4\pi)^{D/2}}\frac{D}{2}\frac{\Gamma\left(n-\frac{D}{2}-1\right)}{\Gamma(n)}\nonumber\\
&\times&\frac{1}{\Delta^{n-\frac{D}{2}-1}},
\end{eqnarray}
where $\Gamma(x)$ is the Euler gamma function and $\Delta\geq0$.

Furthermore, Pauli matrices are also embedded in
$D=2-\eps$-dimensional space. We thus use the following identity
\begin{equation}\label{spatial-trace-pauli}
\sigma_a\sigma_\mu\sigma_a=D\delta_{0\mu}+(2-D)\sigma_a\delta_{a\mu},
\end{equation}
where the sum over the Latin letters $a, b$, used only for the
spatial indices, is assumed. The Greek letters $\mu,\nu$ are
reserved for the spacetime indices. The last term on the right-hand
side turns out to be crucial for the proof of the Ward-Takahashi
identity, guaranteed by the $U(1)$ charge conservation. This is
discussed in later sections. In short, the last term in
Eq.(\ref{spatial-trace-pauli}) yields the last term in Eq.\
(\ref{Cv3}). If the latter were omitted the Ward-Takahashi identity
would be violated. As elaborated on in the discussion section, the
same term also accounts for the discrepancy between the results
found in this work, Eqs. (\ref{constant-C}), and the result we found
previously [Eq.\ (\ref{cond-PRL})] for the Coulomb interaction correction
to the conductivity where the last term was omitted. Details of this
calculation can be found in Appendix E.

\section{Conservation laws, conductivity, and the structure of the polarization tensor}

In the interest of self-containment, in this section we review some
well known results regarding response functions and $U(1)$
conservation laws. Most of these results can be found (scattered) in
many body -- quantum-field theory
textbooks.\cite{Negele-Orland,Peskin}

In order to calculate the response functions to external
electro-magnetic fields, it is useful to define the imaginary time
correlation function
\begin{eqnarray}\label{polarization-tensor}
\Pi_{\mu\nu}(\tau,\br)=\langle
T_{\tau}j_{\mu}(\tau,\br)j_{\nu}(0,0)\rangle,
\end{eqnarray}
where the current "three-vector" $j_{\mu}$ is composed of the
imaginary time density and current as
\begin{eqnarray}\label{eq:3current}
j(\tau,\br)&=&(\rho(\tau,\br),\bj(\tau,\br))\nonumber\\
&=&\left({\psi^{\dagger}(\tau,\br)\psi(\tau,\br),v_F\psi^{\dagger}(\tau,\br)\vec{\sigma}\psi(\tau,\br)}\right).
\end{eqnarray}
In this section we temporarily restore $v_F$ to clearly distinguish
it from the speed of light $c$ used below.

By fluctuation-dissipation theorem,\cite{Negele-Orland} the
expectation value of the electrical current-density operator
$\bJ(t,\br)$, in real time $t$, is related to the imaginary time
correlator $\Pi_{\mu\nu}(i\Omega,\bq)$. The latter is the Fourier
transform (\ref{eq:imSFourier}) of the tensor defined in
Eq.(\ref{polarization-tensor}). The expectation value of the Fourier
transform of the electrical current-density is then
\begin{eqnarray}\label{eq:current}
\langle J_{a}(\Omega,\bq)\rangle &=&
-\frac{e^2}{\hbar}\Pi_{a0}(i\Omega_n\rightarrow
\Omega+i0,\bq)\Phi(\Omega,\bq)\nonumber\\
&+& \frac{e^2}{\hbar}\Pi_{ab}(i\Omega_n\rightarrow
\Omega+i0,\bq)\frac{A_b(\Omega,\bq)}{c}.
\end{eqnarray}
The Fourier components of the electric and magnetic fields are
related to the ones of the scalar and vector potentials as
\begin{eqnarray}\label{eq:EMfieldsAndPotentials}
E_a(\Omega,\bq)&=&i\frac{\Omega}{c}A_a(\Omega,\bq)-iq_a\Phi(\Omega,\bq),\\
B(\Omega,\bq)&=&i\eps_{ab}q_aA_b(\Omega,\bq),
\end{eqnarray}
where $\eps_{ab}$ is completely antisymmetric (Levi-Civita) rank two
tensor. Using Faraday's law of induction we can further relate the
Fourier components of the electric and magnetic fields as
\begin{eqnarray}\label{eq:faraday}
\eps_{ab}q_aE_b(\Omega,\bq)=\frac{\Omega}{c}B(\Omega,\bq).
\end{eqnarray}
In condensed matter systems with massless Dirac particles, propagating with
velocity $v_F$, as the relevant low-energy degrees of freedom
considered here, the (pseudo) Lorentz invariance is violated by
interactions. If we were to consider finite temperature $T$ the
(pseudo) Lorentz invariance would be violated even in the
non-interacting limit. Nevertheless, when spatial $O(2)$ rotational
invariance is preserved, as is the case for problems studied here,
the general structure of the imaginary time polarization tensor
is\cite{Vafek-TesanovicPRL03}
\begin{eqnarray}\label{eq:nonRelFactPi}
\Pi_{\mu\nu}(i\Omega_n,\bq)=\Pi_A(i\Omega_n,|\bq|)A_{\mu\nu}+\Pi_B(i\Omega_n,|\bq|)B_{\mu\nu}
\end{eqnarray}
where the three-tensors are
\begin{eqnarray}\label{eq:nonRelFactAB}
B_{\mu\nu}&=&\delta_{\mu a}\left(\delta_{ab}-\frac{q_{a}q_{b}}{\bq^2}\right)\delta_{b\nu}\\
\label{eq:nonRelFactAB2}
A_{\mu\nu}&=&g_{\mu\nu}-\frac{q_{\mu}q_{\nu}}{q^2}-B_{\mu\nu}.
\end{eqnarray}
The Euclidean three-momenta appearing in the above tensors are
\begin{eqnarray}\label{eq:metric1}
g_{\mu\nu}&=&\mbox{diag}[-1,1,1]_{\mu\nu},\\
\label{eq:metric2}
q_{\mu}&=&g_{\mu\nu}(-i\Omega_n,\bq)_{\nu}=(i\Omega_n,\bq)_{\mu},\\
\label{eq:metric3}
 q^2&=&q_{\mu}g_{\mu\nu}q_{\nu}=\Omega_n^2+\bq^2.
\end{eqnarray}
The real time continuity equation
\begin{equation}\label{eq:continuity}
\frac{\partial}{\partial t}\rho+\nabla\cdot \bJ=0
\end{equation}
requires that, with our choice of the imaginary time "three"-current
$j=(\rho,\vec{j})$, the transversality of the
$\Pi_{\mu\nu}(i\Omega,\bq)$ is equivalent to the condition
\begin{equation}\label{eq:transCondition}
(-i\Omega,\bq)_{\mu}\Pi_{\mu\nu}(i\Omega,\bq)=
\Pi_{\mu\nu}(i\Omega,\bq)\;(-i\Omega,\bq)_{\nu}=0.
\end{equation}
Note that this is explicitly satisfied by the expression
(\ref{eq:nonRelFactPi}). If, in addition, the Lorenz invariance is
satisfied, $\Pi_A=\Pi_B$, and there is no need to separate out the
spatially transverse component of the polarization tensor.

\subsection{Ward-Takahashi identity and vertex functions}

In
addition to the condition (\ref{eq:transCondition}), the continuity
equation (\ref{eq:continuity}) constrains the form of the vertex
function. If we define the four-point matrix function
\begin{equation}
\pi_{\mu}(\br'-\br,\tau'-\tau;\br-\br'',\tau-\tau'')=\langle
T_{\tau}
j_{\mu}(\tau,\br)\psi(\tau',\br')\psi^{\dagger}(\tau'',\br'')\rangle
\end{equation}
where the imaginary time "three"-current was defined in
Eq.(\ref{eq:3current}), then we must have \cite{Peskin}
\begin{eqnarray}
&&\left(\frac{\partial}{\partial \tau},\frac{\nabla}{i}\right)_{\mu}
\pi_{\mu}(\tau'-\tau,\br'-\br;\tau-\tau'',\br-\br'')=\nonumber\\
&&\left(\delta(\tau-\tau'')\delta^{D}(\br-\br'')-\delta(\tau'-\tau)\delta^{D}(\br'-\br)\right)\times\nonumber\\
&&\times\mathcal{G}(\tau'-\tau'',\br'-\br'').
\end{eqnarray}
The above expression relates the exact imaginary time four-point
function to the {\it exact} imaginary time Green's function
\begin{equation}
\mathcal{G}(\tau,\br)=\langle
T_{\tau}\psi(\tau,\br)\psi^{\dagger}(0,0)\rangle.
\end{equation}
If we rewrite the Fourier transform of $\pi_{\mu}$ in terms of the
vertex function $\Lambda_{\mu}$ as
\begin{eqnarray}\label{eq:DefinitionOfVertex}
&&\pi_{\mu}(\bk,i\omega;\bk+\bq,i\omega+i\Omega)=\nonumber\\
&&\mathcal{G}_{\bk}(i\omega)\Lambda_{\mu}(\bk,i\omega;\bk+\bq,i\omega+i\Omega)\mathcal{G}_{\bk+\bq}(i\omega+i\Omega)
\end{eqnarray}
then the Ward-Takahashi identity for the vertex function
$\Lambda_{\mu}$ can be written as
\begin{eqnarray}\label{eq:WardTakahashi}
&&(-i\Omega,\bq)_{\mu}\Lambda_{\mu}(\bk,i\omega;\bk+\bq,i\omega+i\Omega)=\nonumber\\
&&\mathcal{G}^{-1}_{\bk+\bq}(i\omega+i\Omega)-\mathcal{G}^{-1}_{\bk}(i\omega)=\Sigma_{\bk+\bq}(i\omega+i\Omega)-\Sigma_{\bk}(i\omega).\nonumber\\
\end{eqnarray}
This identity has to be satisfied order by order in perturbation
theory. In what follows, we show that this is indeed the case for
the interacting theories studied here when we adopt the dimensional
regularization.

\subsection{Electrical conductivity}

To relate the polarization
tensor to the electrical conductivity, we simply need to relate the
expectation value of the current to the electric field. Since we
have the response to the electromagnetic scalar and vector
potentials, we just need to relate those to the electric and
magnetic fields. Finally, magnetic field can be related to the
electric field using Maxwell's equations. As is well known, at
finite wavevector $\bq$ and frequency $\Omega$, one can define the
logitudinal and transverse conductivity as the proportionality
between the induced current and the longitudinal or transverse
component of the electric field.

Using
Eqs.\ (\ref{eq:current},\ref{eq:nonRelFactPi}-\ref{eq:nonRelFactAB2}),
we find
\begin{eqnarray}
&&\langle J_{a}(\Omega,\bq)\rangle = \frac{e^2}{\hbar}\Pi_{A}(\Omega+i0,|\bq|)\frac{\Omega q_a}{\bq^2-\Omega^2}\Phi(\Omega,\bq)\nonumber\\
&-& \frac{e^2}{\hbar}\Pi_{A}(\Omega+i0,|\bq|)\frac{\Omega^2q_aq_b}{\bq^2(\bq^2-\Omega^2)}\frac{A_b(\Omega,\bq)}{c}\nonumber\\
&+& \frac{e^2}{\hbar}\Pi_{B}(\Omega+i0,|\bq|)\left(\delta_{ab}-\frac{q_aq_b}{\bq^2}\right)\frac{A_b(\Omega,\bq)}{c}.\nonumber\\
\end{eqnarray}
Furthermore, Eqs.\ (\ref{eq:EMfieldsAndPotentials})-(\ref{eq:faraday}) imply
\begin{eqnarray}
&&\langle J_{a}(\Omega,\bq)\rangle = \frac{e^2}{\hbar}\Pi_{A}(
\Omega+i0,|\bq|)\frac{i\Omega}{\bq^2-\Omega^2}\frac{q_aq_b}{\bq^2}E_b(\Omega,\bq)\nonumber\\
&+& \frac{e^2}{\hbar}\Pi_{B}(
\Omega+i0,|\bq|)\frac{1}{i\Omega}\left(\delta_{ab}-\frac{q_aq_b}{\bq^2}\right)E_b(\Omega,\bq).\nonumber\\
\end{eqnarray}
From the above equations we can read off the longitudinal and
transverse electrical conductivity
\begin{eqnarray}\label{eq:condParallel}
\sigma_{\parallel}(\Omega,|\bq|)&=&\frac{e^2}{\hbar}\frac{i\Omega
}{\bq^2-\Omega^2}\Pi_{A}(
\Omega+i0,|\bq|),\\
\label{eq:condPerp}
\sigma_{\perp}(\Omega,|\bq|)&=&\frac{e^2}{\hbar}\frac{\Pi_{B}(
\Omega+i0,|\bq|)}{i\Omega}.
\end{eqnarray}
For $\bq\neq0$ (and $\Omega\neq 0$), $\sigma_{\parallel}$ need not
be equal $\sigma_{\perp}$. However, at $\bq=0$, the a.c.
conductivities
\begin{eqnarray}
\sigma_{\parallel}(\Omega,\bq=0)&=&\sigma_{\perp}(\Omega,\bq=0)
\end{eqnarray}
due to the $O(2)$ spatial rotational symmetry.

In the following, we will work solely in the imaginary time --
Matsubara frequency space, and since we restrict ourselves to $T=0$,
we will drop the subscript $n$ on $i\Omega_n$.

\section{Non-interacting limit: $V(\br)=0$}

For the sake of completeness, and in order to illustrate how the
general results presented in the previous section appear in the
specific solvable problem, we first examine
$\Pi_{\mu\nu}(i\Omega,\bq)$ in the limit of vanishing $V(\br)$. The
Fourier transform (\ref{eq:imSFourier}) of the polarization function
(\ref{polarization-tensor}) in the non-interacting limit is easily
shown to be \begin{eqnarray}
&&\Pi^{(0)}_{\mu\nu}(i\Omega,\bq)=\nonumber\\
&-&N\int_{-\infty}^{\infty}\frac{d\omega}{2\pi}\int\frac{d^D{\bf k}}{(2\pi)^D}
{\rm Tr}[G_{\bk}(i\omega)\sigma_{\mu}G_{\bk+\bq}(i\omega+i\Omega)\sigma_{\nu}]\nonumber\\
\end{eqnarray}
where $\sigma_0$ is the $2\times2$ unit matrix. To this end it is
useful to define the vertex function
\begin{eqnarray}\label{eq:Pdef}
&&\mathcal{P}_{\mu}(\bq,i\Omega)=\nonumber\\
&&\int_{-\infty}^{\infty}\frac{d\omega}{2\pi}\int\frac{d^D{\bf k}}{(2\pi)^D}
G_{\bk}(i\omega)\sigma_{\mu}G_{\bk+\bq}(i\omega+i\Omega),
\end{eqnarray}
in terms of which
\begin{eqnarray}\label{eq:Pi0viaP}
\Pi^{(0)}_{\mu\nu}(i\Omega,\bq)=-N {\rm
Tr}[\mathcal{P}_{\mu}(\bq,i\Omega)\sigma_{\nu}].
\end{eqnarray}
The above expression is divergent at large momenta (UV divergent) as
is easily seen by counting powers. Note that this appears even in
the {\it non-interacting theory} when calculating the response
functions. As is well known in the context of relativistic field
theories, this UV divergence is unphysical and to obtain the correct
answer a regularization is necessary. \cite{Zinn-Justin,Peskin} The
regularization of choice here is dimensional regularization which
leads to finite expressions and which is consistent with $U(1)$ gauge symmetry of the theory.

As shown in detail in the Appendix A, using dimensional
regularization, we obtain
\begin{eqnarray}\label{eq:Pval}
&&\mathcal{P}_{\mu}(\bq,i\Omega)=\frac{\sqrt{\Omega^2+\bq^2}}{64}\nonumber\\
&\times&\left[\sigma_{\mu}-2\delta_{\mu0}-
\frac{(i\Omega+\sigma\cdot\bq)\sigma_{\mu}(i\Omega+\sigma\cdot\bq)
}{\Omega^2+\bq^2} \right].
\end{eqnarray}
Performing the trace in Eq.\ (\ref{eq:Pi0viaP}) we find
\begin{eqnarray}\label{eq:Pi0mat}
&&\Pi^{(0)}_{\mu\nu}(i\Omega,\bq)=\nonumber\\
&-&\frac{N}{16\sqrt{\Omega^2+\bq^2}} \left(\begin{array}{ccc}
-\bq^2 & -i\Omega q_x & -i\Omega q_y \\
-i\Omega q_x & q^2_y+\Omega^2 & -q_xq_y \\
-i\Omega q_y & -q_xq_y & q^2_x+\Omega^2
\end{array}\right)_{\mu\nu}.
\end{eqnarray}
We can write the above matrix more compactly as
\begin{eqnarray}\label{eq:Pi0tens}
\Pi^{(0)}_{\mu\nu}(i\Omega,\bq)&=&-\frac{N}{16}\left(g_{\mu\nu}-\frac{q_{\mu}q_{\nu}}{q^2}\right)\sqrt{\Omega^2+\bq^2}
\end{eqnarray}
where we used definitions from
Eqs.(\ref{eq:metric1}-\ref{eq:metric3}). The correlation function
(\ref{eq:Pi0mat}-\ref{eq:Pi0tens}) is explicitly transverse, as it
should be, and
\begin{equation}
(-i\Omega,\bq)_{\mu}\Pi^{(0)}_{\mu\nu}(i\Omega,\bq)=
\Pi^{(0)}_{\mu\nu}(i\Omega,\bq)\;(-i\Omega,\bq)_{\nu}=0.
\end{equation}

From the above equations we find
\begin{eqnarray}
\Pi^{(0)}_A(i\Omega,\bq)=\Pi^{(0)}_B(i\Omega,\bq)=-\frac{N}{16}\sqrt{\Omega^2+\bq^2}.
\end{eqnarray}
Analytically continuing according to Eqs.
(\ref{eq:condParallel}-\ref{eq:condPerp}), with the branch-cut of
the $\sqrt{z}$-function lying along negative real axis, we find the
well-known expression for the (Gaussian) a.c. conductivity
\begin{eqnarray}\label{Gaussian-cond}
\sigma^{(0)}_{\parallel}(\Omega)=\sigma^{(0)}_{\perp}(\Omega)=\frac{N}{16}\frac{e^2}{\hbar}.
\end{eqnarray}

As a side remark, if we were to define $\tilde{\Pi}^{(0)}_{\mu\nu}$
as a correlation function of a slightly different "three"-current
$(-i\rho,\vec{j})$, the result obtained directly from Eqs.\
(\ref{eq:Pi0mat}-\ref{eq:Pi0tens}) transforms as tensor under
Euclidean $O(3)$ transformations. In real frequencies this is
equivalent to relativistic Lorentz transformations, due to the
invariance of the non-interacting Lagrangian $\mathcal{L}_0$.

Therefore, regulating the UV divergences via dimensional
regularization implemented here leads to finite expressions which
preserve the required $U(1)$ conservation laws. The necessary
regularization of the "integration measure", as done here via
dimensional regularization, is independent of the electron-electron
interaction $V(\br)$, as it must be if the non-interacting theory is
to lead to finite correlation functions. Therefore, as shown already by this
example, regulating only the "momentum transfer" as advocated in
Refs. \onlinecite{MishchenkoEPL08,Sheehy-SchmalianPRB09} is clearly
insufficient.

\section{Short range interactions: $V(\br)=u\delta(\br)$}

While the problem of (2+1)D massless Dirac fermions with the contact
interactions is not exactly solvable, one can calculate the
interaction corrections to the polarization tensor perturbatively in
powers of the interaction strength $u$. Such contact interactions
certainly constitute an idealized special case.\cite{hands} Nevertheless, this
theory has the advantage that one can determine the first correction
in $u$ to the non-interacting (Gaussian) polarization tensor
$\Pi^{(0)}_{\mu\nu}$, found in the previous section, explicitly for
finite $\bq$ and $\Omega$. We can then test the general symmetry
requirements listed before. The technique of choice is again the
(variant of the) dimensional regularization of Veltman and 't Hooft
introduced in section III. Since this interaction violates
Lorentz invariance we can also use this example to study how the
difference between $\Pi_A$ and $\Pi_B$ arises in such theory.

It is straightforward to use the Wick's theorem to show that in this
case, the first order in $u$, and to $\mathcal{O}(N)$, correction to
the polarization tensor is
\begin{eqnarray}\label{eq:deltaPiShort}
&&\delta\Pi_{\mu\nu}(i\Omega,\bq)\!=\!uN\mesk\!\!\mesp\!\!\nonumber\\
&&
\left\{\mbox{Tr}[G_{\bk}(i\omega)\sigma_{\mu}G_{\bk+\bq}(i\omega+i\Omega)G_{\bp}(i\omega')\sigma_{\nu}G_{\bp-\bq}(i\omega'-i\Omega)]\right.\nonumber\\
&+&\left.
\mbox{Tr}[G_{\bk}(i\omega)\sigma_{\mu}G_{\bk+\bq}(i\omega+i\Omega)G_{\bp}(i\omega')G_{\bk+\bq}(i\omega+i\Omega)\sigma_{\nu}]\right.\nonumber\\
&+&\left.
\mbox{Tr}[G_{\bk}(i\omega)\sigma_{\mu}G_{\bk+\bq}(i\omega+i\Omega)\sigma_{\nu}G_{\bk}(i\omega)G_{\bp}(i\omega')].
\right\}
\end{eqnarray}
The last two terms correspond to the self-energy correction, while
the first one is the vertex correction. Because the self-energy for the contact interaction vanishes,
\begin{eqnarray}
\mesk G_{\bk}(i\omega)=0,
\end{eqnarray}
the last two terms in the Eq.(\ref{eq:deltaPiShort}) vanish as well.
The remaining term can be written rather succinctly in terms of
$\mathcal{P}_{\mu}$ defined previously in Eq.(\ref{eq:Pdef}) as
\begin{eqnarray}
\delta\Pi_{\mu\nu}(i\Omega,\bq)
&=&uN\mbox{Tr}\left[\mathcal{P}_{\mu}(\bq,i\Omega)\mathcal{P}_{\nu}(-\bq,-i\Omega)\right].
\end{eqnarray}
The above expression is manifestly transverse, i.e., it satisfies
Eq.(\ref{eq:transCondition}), as can be seen from
Eq.(\ref{eq:Pval}).

\begin{widetext}
Namely, inserting Eq.\ (\ref{eq:Pval}) and performing the traces we find
\begin{equation}
\delta\Pi_{\mu\nu}(i\Omega,\bq)\!\!=\!\!\frac{uN}{512(\Omega^2+\bq^2)}
\left[\begin{array}{ccc} \bq^2(\bq^2-\Omega^2) & i\Omega
q_x(\bq^2-\Omega^2) & i\Omega q_y(\bq^2-\Omega^2)
\\
i\Omega q_x(\bq^2-\Omega^2) & q_x^2(q_y^2-\Omega^2)+(q_y^2+\Omega^2)^2 &-q_xq_y(3\Omega^2+\bq^2)\\
i\Omega q_y(\bq^2-\Omega^2) & -q_xq_y\left(3\Omega^2+\bq^2\right) &
q_x^4-\Omega^2q_y^2+\Omega^4+q^2_x(q^2_y+2\Omega^2)
\end{array}\right]_{\mu\nu}.\nonumber\\
\end{equation}
\end{widetext}
Finally, the above tensor can be factorized as given by Eqs.\ (\ref{eq:nonRelFactPi}-\ref{eq:nonRelFactAB}), and
we find to first non-trivial order in the contact coupling
$u$
\begin{eqnarray}
\Pi_A(i\Omega,|\bq|)&=&-\frac{N}{16}\sqrt{\Omega^2+\bq^2}+\frac{uN}{512}(\Omega^2-\bq^2)+\mathcal{O}(u^2),\nonumber\\
\Pi_B(i\Omega,|\bq|)&=&-\frac{N}{16}\sqrt{\Omega^2+\bq^2}+\frac{uN}{512}(\Omega^2+\bq^2)+\mathcal{O}(u^2).\nonumber\\
\end{eqnarray}
Expectedly, the above expression shows that the interaction
correction to the polarization functions $\Pi_A$ and $\Pi_B$ are
{\em different} (note the sign difference in front of $\bq^2$). As
stated above, the reason for the difference is that the contact
density-density interaction term $u(\psi^{\dagger}(\br)\psi(\br))^2$
breaks the Lorentz invariance of the non-interacting part of the
Lagrangian. Lorentz transformations in general rotate between
density and current and we have purposefully omitted any
current-current interaction.

We can further test the Ward-Takahashi identity
(\ref{eq:WardTakahashi}) for the vertex function
(\ref{eq:DefinitionOfVertex}) in this example with the short range
interactions. It can be readily seen that the first order in $u$
correction to the vertex vector is
\begin{eqnarray}\delta\Lambda_{\mu}(\bk,i\omega;\bk+\bq,i\omega+i\Omega)=-u
\mathcal{P}_{\mu}(\bq,i\Omega).
\end{eqnarray}
It follows from the Eq.\ (\ref{eq:Pval}) that
\begin{eqnarray}
-i\Omega
\mathcal{P}_{0}(\bq,i\Omega)+\bq_a\mathcal{P}_{a}(\bq,i\Omega)=0.
\end{eqnarray}
Therefore the Ward-Takahashi identities (\ref{eq:WardTakahashi}) are
satisfied, since, as mentioned previously in this section, the
self-energy correction vanishes to first order in $u$ for the
short-range interactions.

Finally, from Eqs.\ (\ref{eq:condParallel}-\ref{eq:condPerp}), we
can infer that the above terms correct only the imaginary part of
the a.c. conductivity, but not the real part. At $\bq=0$, correction
is the same for the longitudinal and the transverse components, and
to this order in $u$ we have \begin{eqnarray}
\sigma_{\parallel,\perp}(\Omega)&=&\frac{e^2}{\hbar}\frac{N}{16}\left(1-i\frac{u}{32}\Omega\right).
\end{eqnarray}
Again, the equality between $\sigma_{\parallel}(\Omega)$ and
$\sigma_{\perp}(\Omega)$ is guaranteed due to the $O(2)$ rotational
invariance of this theory. Note also that the fact that the interaction correction
is proportional to the frequency is implied by the power counting at the Gaussian
fixed point of the theory, and is characteristic for any finite-range interaction.\cite{preprint}

\section{Coulomb interaction: $V(\br)=e^2/|\br|$}

Armed with the above results we now focus on the main part of the
paper where we study the effects of the Coulomb interaction. Unlike
in the previous cases, we have been unable to find the explicit
expression for the first order correction to the polarization tensor
at finite $\bq$ and $\Omega$. Nevertheless, we have been able to
show explicitly that the first order correction to the polarization
tensor is transverse, i.e., it satisfies
Eq.(\ref{eq:transCondition}). This is shown using dimensional
regularization in $D=2-\eps$ introduced in section III. Next, we
study the first correction to the Coulomb vertex function which must
also satisfy the Ward-Takahashi identity (\ref{eq:WardTakahashi}).
Since in this case the first order self-energy is known to diverge
logarithmically, the first order correction to the vertex function
should also diverge as $\eps\rightarrow 0$. This can be found
explicitly in terms of elliptic integrals to order $\eps^{-1}$ and
$\eps^0$ and the identity (\ref{eq:WardTakahashi}) is also
explicitly confirmed. Finally, we proceed with the calculation of
the electrical conductivity, first by using the spatial component of
the polarization tensor at $\bq=0$ but finite $\Omega$
(current-current correlation function), and then by using time
component of the polarization tensor at finite but small $\bq$ and
finite $\Omega$. The final results for the conductivity calculated
in both ways are found to be the same. Specifically, we find ${\cal
C}=(11-3\pi)/6$ in Eq.\ (\ref{conductivity-final}).

For unscreened 3D Coulomb interactions $V(\br)=e^2/r$ the effect of
screening due to dielectric medium is easily taken into account by
rescaling $e^2$ in the above formula. The $\mathcal{O}(e^2)$ and
$\mathcal{O}(N)$ correction to the polarization function is then
\begin{eqnarray}\label{eq:deltaPiCoulomb}
&&\delta\Pi^{(c)}_{\mu\nu}(i\Omega,\bq)=N\mesk\!\!\mesp\nonumber\\
&&\left\{V_{\bp-\bk}\right.\nonumber\\
&\times&\left.\mbox{Tr}\left[G_{\bk}(i\omega)\sigma_{\mu}G_{\bk+\bq}(i\omega+i\Omega)G_{\bp+\bq}(i\omega'+i\Omega)\sigma_{\nu}G_{\bp}(i\omega')\right]
\right.\nonumber\\
&+&\left.V_{\bk-\bp}\right.\nonumber\\
&\times&\left.\mbox{Tr}\left[G_{\bk}(i\omega)\sigma_{\mu}G_{\bk+\bq}(i\omega+i\Omega)G_{\bp+\bq}(i\omega'+i\Omega)\right.\right.\nonumber\\
&\times&\left.\left.G_{\bk+\bq}(i\omega+i\Omega)\sigma_{\nu}\right]\right.\nonumber\\
&+&\left.V_{\bk-\bp}\right.\nonumber\\
&\times&\left.\mbox{Tr}\left[G_{\bk}(i\omega)\sigma_{\mu}G_{\bk+\bq}(i\omega+i\Omega)\sigma_{\nu}G_{\bk}(i\omega)G_{\bp}(i\omega')\right]\right\}
\end{eqnarray}
where
\begin{eqnarray}
V_{\bk}=\int d^2\br e^{i\bk\cdot\br}V(\br)=\frac{2\pi e^2}{|\bk|}.
\end{eqnarray}
Just as in the case of contact interactions, the first term in the
expression for $\delta\Pi^{(c)}_{\mu\nu}$ corresponds to the vertex
correction and the last two terms to the self-energy corrections.
Unlike in the case of contact interactions, however, the self energy
correction does not vanish. The expression (\ref{eq:deltaPiCoulomb})
will be used in later sections as a starting point in the
calculation of the Coulomb interaction correction to the a.c.
conductivity in the collisionless regime.

\subsection{Proof of the transversality of $\delta\Pi^{(c)}_{\mu\nu}$
within dimensional regularization}

Because, as mentioned above, the
explicit evaluation of (\ref{eq:deltaPiCoulomb}) at finite $\bq$ and
$\Omega$ yields intractable expressions, we proceed by first
showing that (\ref{eq:deltaPiCoulomb}) is transverse, i.e., that it
satisfies the condition (\ref{eq:transCondition}), when dimensional
regularization employed in this paper is used. As such it therefore
{\it does not} lead to any violation of the charge conservation, a
virtue questioned in Ref.\ \onlinecite{Sheehy-SchmalianPRB09}. By 2D
rotational invariance, this in turn implies that the Coulomb
polarization tensor can be written in the form
(\ref{eq:nonRelFactPi}).

To prove (\ref{eq:transCondition}) we follow Ref.\
\onlinecite{Sheehy-SchmalianPRB09} and use
\begin{eqnarray}
-i\Omega \sigma_0+\bq\cdot\sigma
&=&G^{-1}_{\bk+\bq}(i\omega+i\Omega)-G^{-1}_{\bk}(i\omega)
\end{eqnarray}
to find
\begin{eqnarray}
&&(-i\Omega,\bq)_{\mu}\delta\Pi^{(c)}_{\mu\nu}(i\Omega,\bq)=
N\int\frac{d^D\bk}{(2\pi)^D}\frac{d^D\bp}{(2\pi)^D}\int_{-\infty}^{\infty}\frac{d\omega}{2\pi}\frac{d\omega'}{2\pi}\nonumber\\
&&V_{\bp-\bk}\left\{\mbox{Tr} \left[
G_{\bk}(i\omega)\sigma_{\nu}G_{\bk}(i\omega)G_{\bp}(i\omega')\right]-\nonumber\right.\\
&&\left.\mbox{Tr} \left[
G_{\bk+\bq}(i\omega+i\Omega)\sigma_{\nu}G_{\bk+\bq}(i\omega+i\Omega)G_{\bp+\bq}(i\omega'+i\Omega)\right]\right\}.\nonumber
\end{eqnarray}
At this point it is not immediately obvious that we can shift the
integration variables $\bk$ and $\bp$ in the second term by $\bq$,
which if true would readily yield the desired relation
(\ref{eq:transCondition}), since the frequency integral can be
shifted. We therefore define a function of frequency and {\it two}
momentum variables
\begin{equation}\label{self-energy-definition}
{\Sigma}_{\bp,\bq}(i\Omega)=
\int\frac{d^D\bk}{(2\pi)^D}\int_{-\infty}^{\infty}\frac{d\omega'}{2\pi}V_{\bk-\bp}G_{\bk+\bq}(i\omega'+i\Omega)
\end{equation}
in terms of which we have unambiguously
\begin{eqnarray}
&&(-i\Omega,\bq)_{\mu}\delta\Pi^{(c)}_{\mu\nu}(i\Omega,\bq)\nonumber\\
&=&N\int\frac{d^D\bk}{(2\pi)^D}\int_{-\infty}^{\infty}\frac{d\omega}{2\pi}\left\{\mbox{Tr}
\left[
G_{\bk}(i\omega)\sigma_{\nu}G_{\bk}(i\omega){\Sigma}_{\bk,0}(0)\right]\nonumber\right.\\
&-&\left.\mbox{Tr} \left[
G_{\bk+\bq}(i\omega+i\Omega)\sigma_{\nu}G_{\bk+\bq}(i\omega+i\Omega){\Sigma}_{\bk,\bq}(i\Omega)\right]\right\}.\nonumber
\end{eqnarray}
To continue, we need to find an explicit expression for
${\Sigma}_{\bp,\bq}(i\Omega)$. Using the identity
(\ref{dim-reg-int}), Feynman parametrization
\begin{equation}\label{FP1}
\frac{1}{A^\alpha B^\beta}=\frac{\Gamma(\alpha+\beta)}{\Gamma(\alpha)\Gamma(\beta)}
\int_0^1dy\frac{y^{\alpha-1}(1-y)^{\beta-1}}{[y A+(1-y)B]^{\alpha+\beta}},
\end{equation}
 for $\alpha=\beta=1/2$, and
\begin{equation}\label{Beta}
\int_0^1dy
y^{\alpha-1}(1-y)^{\beta-1}=\frac{\Gamma(\alpha)\Gamma(\beta)}{\Gamma(\alpha+\beta)},
\end{equation}
we find
\begin{equation}\label{self-energy-integral}
{\Sigma}_{\bp,\bq}(i\Omega)=\frac{e^2}{(4\pi)^{\frac{D}{2}}}\frac{\sigma\cdot(\bp+\bq)}{|\bp+\bq|^{2-D}}
\frac{\Gamma\left(1-\frac{D}{2}\right)\Gamma\left(\frac{D+1}{2}\right)\Gamma\left(\frac{D-1}{2}\right)}{\Gamma(D)},
\end{equation}
which agrees with Eq.(12) of Ref.\ \onlinecite{Vafek-Case}. Note
that this identity shows that within dimensional regularization,
${\Sigma}_{\bp,\bq}(i\Omega)={\Sigma}_{\bp+\bq,0}(i\Omega).$
Moreover, in what follows, there is no need to shift the integration
variable. Rather, since the commutator of the self-energy and the
Green's function vanishes,
\begin{eqnarray}
\left[G_{\bk+\bq}(i\omega+i\Omega),{\Sigma}_{\bp,\bq}(i\Omega)\right]=0,
\end{eqnarray}
after a straightforward use of the cyclic property of the trace and
the identity
\begin{eqnarray}
\int_{-\infty}^{\infty}\frac{d\omega}{2\pi}G_{\bk+\bq}(i\omega+i\Omega)G_{\bk+\bq}(i\omega+i\Omega)=0,
\end{eqnarray}
we prove that
\begin{eqnarray}
(-i\Omega,\bq)_{\mu}\delta\Pi^{(c)}_{\mu\nu}(i\Omega,\bq)=0.
\end{eqnarray}
The same procedure as the one used above also leads to
\begin{eqnarray}
\delta\Pi^{(c)}_{\mu\nu}(i\Omega,\bq)(-i\Omega,\bq)_{\nu}=0.
\end{eqnarray}
This proof holds to all orders of $\eps$. The regularization
technique implemented here is therefore perfectly adequate and does
not lead to violation of the charge conservation.

\subsection{Coulomb vertex and the proof of the Ward-Takahashi
identity}

Next, we will demonstrate that the dimensional
regularization used here preserves the Ward-Takahashi identity for
the Coulomb vertex function. This proof is technically more involved
than the proof in the previous section, but nevertheless, we find it
important to present its details since our technique is not widely
used in the community. We show the desired identity to order
$\eps^{-1}$ and $\eps^{0}$. Most of the technical details are
presented in the appendices, and in this section we just present
the main steps of the derivation.

The Coulomb vertex function to the first order in the coupling
constant is
\begin{eqnarray}\label{eq:CV-def}
&&\delta\Lambda_{\mu}(\bp,i\nu;\bp+\bq,i\nu+i\Omega)=\mathcal{P}^c_{\mu}(\bq,\bp,i\Omega)=\nonumber\\
&&-\int\frac{d^D\bk}{(2\pi)^D}\int_{-\infty}^{\infty}\frac{d\omega}{2\pi}
V_{\bp-\bk}G_{\bk}(i\omega)\sigma_{\mu}G_{\bk+\bq}(i\omega+i\Omega).\nonumber\\
\end{eqnarray}
The integrals on the right are logarithmically divergent in $D=2$ as
can be easily seen by powercounting. This divergence is related to
the divergence of the electron self-energy, calculated in the
previous section,
\begin{eqnarray}\label{eq:SelfEnergy}
\Sigma_{\bk}(i\omega)&\equiv&\Sigma_{\bk,0}(i\omega)=
\frac{e^2}{8}\sigma\cdot\bk\nonumber\\
&\times&\left(\frac{2}{\eps}-\gamma+\ln64\pi-\ln\bk^2+{\cal
O}(\eps)\right),
\end{eqnarray}
where the Euler-Mascheroni constant $\gamma=0.577$ and, as before,
$\eps=2-D$. Indeed, if the Ward-Takahashi identity,
\begin{eqnarray}
(-i\Omega,\bq)_{\mu}\mathcal{P}^{c}_{\mu}(\bq,\bp,i\Omega)=\Sigma_{\bp+\bq}(i\nu+i\Omega)-\Sigma_{\bp}(i\nu),
\end{eqnarray}
is to be satisfied, the vertex function must diverge
logarithmically, which manisfests in the dimensional regularization as the first-order pole in Laurent expansion in the parameter $\eps$.

In the second part of the Appendix A we use dimensional
regularization to determine $\mathcal{P}^{c}(\bq,\bp,i\Omega)$ to
orders $\eps^{-1}$ and $\eps^0$. Our final expression for finite
$\bq$,$\bp$ and $i\Omega$, Eq.\ (\ref{eq:CoulombVertexFinal}), is
left as an integral over a Feynman parameter $x$. We wish to stress
that all of the integrals in this equation can be performed in the
closed form in terms of elliptic integrals. However, we found that
doing so leads to intractable and unrevealing expressions. We
therefore chose to work with the expression
(\ref{eq:CoulombVertexFinal}) and in effect manipulate the integral
representation of the elliptic integrals. In the limiting case of
$\bq=0$, the vertex function is determined in the closed form in
Appendix A up to, and including, $\eps^0$.

In Appendix B we in turn find that the vertex function
(\ref{eq:CoulombVertexFinal}) satisfies
\begin{eqnarray}
&&(-i\Omega,\bq)_{\mu}\mathcal{P}^{c}_{\mu}(\bq,\bp,i\Omega)=N(\bp,\bq)-\frac{e^2}{4}\sqrt{\Omega^2+\bq^2}\nonumber\\
&\times&\sigma\cdot\left(\bp(\Omega^2+\bq^2) L(\bp,\bq,\Omega) +\bq
M(\bp,\bq,\Omega)\right).
\end{eqnarray}
Using the dimensional regularization, we then show that the function
\begin{eqnarray}
N(\bp,\bq)&=&\frac{e^2}{8}\left(\frac{2}{\eps}\sigma\cdot\bq+(\ln64\pi-\gamma)\sigma\cdot\bq\right.\nonumber\\
&-&\left.\sigma\cdot(\bp+\bq)\ln(\bp+\bq)^2
+\sigma\cdot\bp\ln\bp^2\right)\nonumber\\
&=&\Sigma_{\bp+\bq}(i\nu+i\Omega)-\Sigma_{\bp}(i\nu),
\end{eqnarray}
and, in Appendix C, that $L(\bp,\bq,\Omega)=M(\bp,\bq,\Omega)=0$.
This proves the Ward-Takahashi identity to the first order in
perturbation theory.

\subsection{Calculation of the a.c. conductivity from the current-current correlator (Kubo formula)}

In this section we calculate the diagonal spatial component of the
Coulomb interaction correction of the polarization tensor,
$\delta\Pi^{(c)}_{xx}$, at $\bq=0$ and finite $i\Omega$. We then use
this result to calculate the corresponding correction to the
electrical conductivity. Given the decomposition
(\ref{eq:nonRelFactPi}), one should in principle specify the
direction in the $\bq$-plane along which the limit $\bq\rightarrow
0$ is taken. For example, if $q_x$ is taken to $0$ before $q_y$,
$\delta\Pi^{(c)}_{xx}(i\Omega,0)$ is proportional to
$\Pi_B(i\Omega,0)$. On the other hand, if $q_y$ is taken to $0$
before $q_x$, then $\delta\Pi^{(c)}_{xx}(i\Omega,0)$ is proportional
to $\Pi_A(i\Omega,0)$. Similarly, if the limit is taken along a line
that forms an angle $\theta$ with the $q_x$ axis, then
$\delta\Pi^{(c)}_{xx}(i\Omega)$ is proportional to $\cos^2\theta
\Pi_A(i\Omega,0)+\sin^2\theta\Pi_B(i\Omega,0)$. However, due to the
$O(2)$ rotational invariance, $\Pi_A(i\Omega,0)=\Pi_B(i\Omega,0)$
and the result is independent of $\theta$. We can therefore use
either Eq.\ (\ref{eq:condParallel}) or Eq.\ (\ref{eq:condPerp})
along with diagonal spatial part of the polarization tensor
(\ref{eq:deltaPiCoulomb}) to calculate the a.c. conductivity.

We start by showing that
\begin{eqnarray}\label{eq:deltaPiCoulombAtZero}
\delta\Pi^{(c)}_{\mu\nu}(\bq=0,i\Omega=0)=0.
\end{eqnarray}
This is expected, since a space and time independent vector and
scalar potential correspond to a pure gauge, and as such have no
effect on the physics of the problem. Within our formalism, this
identity can be shown to the first order in the Coulomb interaction
by first performing the integral over the frequencies in Eq.
(\ref{eq:deltaPiCoulomb}), which, as  can be easily seen, yields
\begin{eqnarray}
&&\delta\Pi^{(c)}_{\mu\nu}(i\Omega=0,\bq=0)
=\frac{N}{4}\int\frac{d^D\bp}{(2\pi)^D}\frac{1}{|\bp|}\nonumber\\
&\times&{\rm Tr}\left[
\mathcal{P}_{\mu}^{(c)}(0,\bp,0)\left(\sigma_{\nu}-\frac{\sigma\cdot\bp\sigma_{\nu}\sigma\cdot\bp}{\bp^2}\right)
\right]+\frac{N}{8}\nonumber\\
&\times&\int\frac{d^D\bk}{(2\pi)^D}\frac{1}{|\bk|^3}
\mbox{Tr}\left[\left(\sigma\cdot\bk\sigma_{\mu}-\sigma_{\mu}\sigma\cdot\bk\right)\left(
\Sigma_{\bk}
\sigma_{\nu}-\sigma_{\nu}\Sigma_{\bk}\right)\right]\nonumber.
\end{eqnarray}
Substituting the expression for the self-energy
(\ref{eq:SelfEnergy}) and the static vertex
(\ref{eq:CoulombVertexAtZero}), performing the traces and using
$k_ak_b\rightarrow\delta_{ab}\bk^2/D$, we conclude that
Eq.(\ref{eq:deltaPiCoulombAtZero}) holds, as it should. We are
therefore free to subtract it from the expression for
$\delta\Pi^{(c)}_{\mu\nu}$ at either finite $\Omega$ and/or finite
$\bq$.

Next, we set $\bq=0$ and consider finite $\Omega$ in Eq.\
(\ref{eq:deltaPiCoulomb}). The polarization tensor can be written as
sum of the contributions from the self-energy correction and the
vertex correction
\begin{equation}\label{polarization-tensor-coulomb}
\delta\Pi_{\mu\nu}^{(c)}(i\Omega,0)=\delta\Pi_{\mu\nu}^{(a)}(i\Omega,0)+\delta\Pi_{\mu\nu}^{(b)}(i\Omega,0),
\end{equation}
where the self-energy is given by
\begin{eqnarray} \label{pol-self-energy}
&&\delta\Pi_{\mu\nu}^{(a)}(i\Omega,0)=2N\int_{-\infty}^\infty\frac{{d\omega}}{2\pi}\frac{d\omega'}{2\pi}\int\frac{d^D{\bf
k}}{(2\pi)^D}\frac{d^D{\bf p}}{(2\pi)^D}V_{{\bf k}-{\bf p}}
\nonumber\\
&&{\rm Tr}\left[G_{{\bf k}}(i\omega)\sigma_\mu G_{{\bf
k}}(i\omega+i\Omega)\sigma_\nu G_{{\bf k}}(i\omega) G_{{\bf
p}}(i\omega')\right],
\end{eqnarray}
and the vertex correction reads
\begin{eqnarray}\label{pol-vertex}
&&\delta\Pi_{\mu\nu}^{(b)}(i\Omega,0)=N\int_{-\infty}^\infty\frac{{d\omega}}{2\pi}\frac{d\omega'}{2\pi}\int\frac{d^D{\bf
k}}{(2\pi)^D}\frac{d^D{\bf p}}{(2\pi)^D}V_{{\bf k}-{\bf p}}
\nonumber\\
&& {\rm Tr}\left[G_{{\bf k}}(i\omega)\sigma_\mu G_{{\bf
k}}(i\omega+i\Omega)G_{{\bf p}}(i\omega'+i\Omega)\sigma_\nu G_{{\bf
p}}(i\omega')\right].
\end{eqnarray}
Both of these expressions need to be regulated due to the UV
divergence.

In appendix E we calculate both of these contributions to the
electrical conductivity. The contribution to the conductivity coming
from the self-energy part expanded up to the order $\epsilon^0$ is
found to be
\begin{equation}
\sigma_a=\frac{\sigma_0
e^2}{2}\left(-\frac{1}{\epsilon}+\frac{3}{2}+\gamma-\ln(64\pi)\right).
\end{equation}
The corresponding vertex part is found to be
\begin{eqnarray}
\sigma_b
&=&\frac{\sigma_0e^2}{2}\left[\left(\frac{1}{\eps}-\frac{1}{2}-\gamma+\ln64\pi\right)+\frac{8-3\pi}{3}\right].
\end{eqnarray}
Adding these two terms we obtain the first order correction to the
a.c. conductivity due to the Coulomb interaction
\begin{equation}\label{cond-perpendicular}
\delta\sigma^{(c)}=\sigma_a+\sigma_{b}=\frac{11-3\pi}{6}\sigma_0e^2,
\end{equation}
which corresponds to the value
\begin{equation}\label{constant-C}
{\cal C}=\frac{11-3\pi}{6}
\end{equation}
in Eq.\ (\ref{conductivity-final}). We discuss this result in light
of previous work as well as present day experiments in the
concluding section.

\subsection{Calculation of the a.c. conductivity using the density-density correlator}

To show that our previous result for the conductivity is consistent,
we now calculate the longitudinal conductivity given by Eq.\
(\ref{eq:condParallel}) and show that it yields the same value of
the constant ${\cal C}$ as in Eq.\ (\ref{constant-C}). This must be
the case if the Ward-Takahashi identity and the $O(2)$ rotational
invariance hold. The longitudinal correction to the conductivity can
be calculated by focusing on $\delta\Pi^{(c)}_{00}$, since
$B_{00}=0$. Unlike for Kubo formula, this component of the
polarization tensor must be calculated at finite $\Omega$ and finite
$\bq$, since at $\bq=0$ it vanishes. Fortunately, we need only the
leading order term in the expansion in small $\bq^2$, from which we
can extract the conductivity.

According to Eq.\ (\ref{eq:deltaPiCoulomb}), the Coulomb interaction
correction to the density-density correlator reads
\begin{equation}\label{definePi00}
\delta\Pi^{(c)}_{00}(i\Omega,\bq)=\delta\Pi^{(a)}_{00}(i\Omega,\bq)+\delta\Pi^{(b)}_{00}(i\Omega,\bq),
\end{equation}
where, just as before, we have separated the self-energy
contribution
\begin{eqnarray}\label{density-density-self-energy}
&&\delta\Pi^{(a)}_{00}(i\Omega,\bq)=N
\int_{-\infty}^\infty\frac{d\omega}{2\pi}\frac{d\omega'}{2\pi}
\int\frac{d^2{\bf k}}{(2\pi)^2}\frac{d^2{\bf p}}{(2\pi)^2}V_{\bk-\bp}\nonumber\\
&\times&\left\{{\rm
Tr}\left[G_{\bk}(i\omega)G_{\bp}(i\omega')G_{\bk}(i\omega)
G_{\bk+\bq}(i\omega+i\Omega)\right]\right.\nonumber\\
&+&\left.{\rm Tr}\left[G_{\bk}(i\omega)G_{\bp}(i\omega')G_{\bk}(i\omega)
G_{\bk-\bq}(i\omega-i\Omega)\right]\right\},
\end{eqnarray}
and the vertex contribution
\begin{eqnarray}\label{density-density-vertex}
&&\delta\Pi^{(b)}_{00}(i\Omega,\bq)=N
\int_{-\infty}^\infty\frac{d\omega}{2\pi}\frac{d\omega'}{2\pi}
\int\frac{d^2{\bf k}}{(2\pi)^2}\frac{d^2{\bf p}}{(2\pi)^2}V_{\bk-\bp}\nonumber\\
&\times&{\rm Tr}\left[G_{\bk}(i\omega)G_{\bp}(i\omega')G_{\bp-\bq}(i\omega'-i\Omega)G_{\bk-\bq}(i\omega-i\Omega)\right].\nonumber\\
\end{eqnarray}
The details of the calculations are presented in the Appendix F.
Here we just state the final result
\begin{equation}\label{cond-parallel}
\delta\sigma^{(c)}_\parallel=\frac{11-3\pi}{6}\sigma_0e^2,
\end{equation}
in agreement with the result (\ref{cond-perpendicular}) obtained
from the current-current correlator. Such agreement is expected
since, as we have shown to this order in the Coulomb interaction,
the dimensional regularization explicitly preserves the $U(1)$ gauge
symmetry of the theory of the Coulomb interacting Dirac fermions.

\section{Discussion and connection with previous work}

Let us now discuss the result (\ref{constant-C}) for the correction to the a.c. conductivity due to the long-range Coulomb interaction
in light of the ones previously reported in the literature.\cite{HerbutJuricicVafekPRL08,MishchenkoEPL08,Sheehy-SchmalianPRB09}

In Ref.\ \onlinecite{HerbutJuricicVafekPRL08}, the Coulomb correction to the conductivity is shown to have the form given
by Eq.\ (\ref{conductivity-final}), consistent with the renormalizability of the quantum field theory of the Coulomb interacting Dirac fermions.
Moreover, the value of the constant ${\mathcal C}=(25-6\pi)/12$ has been calculated from the current-current correlator, and using hard-cutoff regularization. In Appendix D we show that in general hard cutoff violates the Ward-Takahashi identity. Nevertheless, it can be shown that  the integral for   ${\mathcal C}$ is, despite appearances, in fact UV convergent, but sensible to the order of integration. A correct way of performing the integral is to integrate both momenta up to finite cutoffs, and take the cutoff to infinity after all the integrals are done first.
This, as shown in the Appendix H, corrects the value of  the constant precisely down to the ${\mathcal C}=(11-3\pi)/6$. As we showed in Appendix G, the previuos result ${\mathcal C}=(25-6\pi)/12$
 is also obtained when using a version of the dimensional regularization in which the Pauli matrices are treated as embedded in strictly two spatial dimensions, and which also violates the Ward-Takahashi identity, as we argued in Appendix D. Technically, the origin of the discrepancy between the results for the Coulomb correction to conductivity within the two versions of the dimensional regularization
may be traced if we consider the self-energy correction to the conductivity in Eq.\ (\ref{a-tilde-full}) and its counterpart with
Pauli matrices in $D=2-\eps$, given by Eq. (\ref{a-complete}). The difference arises from the factor
$D-1=1-\eps$ which is a consequence of the different treatment of Pauli matrices within the two schemes. The
self-energy piece has a singular part proportional to $1/\eps$, and when multiplied  by a term linear in $\eps$
coming from $D-1$, it gives rise to a finite contribution to the self-energy correction. Analogous situation
occurs in the vertex part, and in that case the last three terms in the integrand of Eq.\ (\ref{sigmab1-1})
account for the difference. Namely, when the trace over spatial indices of Pauli matrices is taken in
$D=2$, these three terms cancel out, as it may be seen from the term proportional to $({\bk}\cdot{\bp})^2$ in
the integrand in Eq.\ (\ref{D6}), but, in fact, when Pauli matrices are embedded in $D=2-\eps$, these terms
yield a finite contribution to the conductivity, which may be directly checked following the steps in Eqs.\
(\ref{b11-xx})-(\ref{b11}).

On the other hand, in Ref.\ \onlinecite{MishchenkoEPL08},  the result for the constant
${\cal C}=(19-6\pi)/12$ has been calculated using three different methods, namely, the density-density correlator, the current-current correlator and the kinetic equation, and it has been argued that in order to obtain the unique value for the constant ${\mathcal C}$ a short-distance cutoff on the long-range Coulomb interaction has to be imposed. This regularization is an analogue of the Pauli-Villars regularization in QED, but without the additional Pauli-Villars fermions introduced, that are, in fact, necessary to render it consistent.\cite{BjorkenDrell}
This value for the constant ${\mathcal C}$ has also been obtained in Ref.\ \onlinecite{Sheehy-SchmalianPRB09} by regulating the short-distance behavior of the Coulomb interaction in the same manner as in Ref.\ \onlinecite{MishchenkoEPL08}. Although it has been shown that the same regularization preserves the Ward-Takahashi identity, besides lacking the Pauli-Villars fermions, this regularization cannot be applied to the theory of free Dirac fermions. Namely, the latter needs to be regularized when calculating the polarization bubble. Clearly, this cannot be achieved by imposing a short-distance cutoff on the long-range Coulomb interaction. Therefore, this regularization cannot serve as a consistent regularization of the entire field theory.

\section{acknowledgements}
V. J. wishes to acknowledge the support of the Netherlands Organisation for Scientific Research (NWO). O. V. was supported in part by NSF
grant DMR-00-84173 and NSF CAREER award grant DMR-0955561.
I. F. H. is supported by the NSERC of Canada. The authors also wish to thank the Aspen Center for Physics where a part of this work was preformed.
\appendix

\section{Vertex integrals}
The quantity of interest, which enters into the evaluation of the
bare bubble and the leading order correction to the polarization
tensor for short range interactions $u$ is
\begin{eqnarray}
\mathcal{P}_{\mu}(\bq,i\Omega)=\int\frac{d^D\bk}{(2\pi)^D}\int_{-\infty}^{\infty}\frac{d\omega}{2\pi}
G_{\bk}(i\omega)\sigma_{\mu}G_{\bk+\bq}(i\omega+i\Omega).\nonumber\\
\end{eqnarray}
Substituting (\ref{GreensFxn}), using Feynman parametrization (\ref{FP1})
for $\alpha=\beta=1$, and interchanging the order of integrations, we obtain
\begin{eqnarray}
&&\mathcal{P}_{\mu}(\bq,i\Omega)=\int_0^1dx\nonumber\\
&\times&
\mesk\frac{1}{\left[(\omega+x\Omega)^2+(\bk+x\bq)^2+\Delta\right]^2}
\nonumber\\
&\times&[i\omega+\sigma\cdot\bk]\sigma_{\mu}[i\omega+i\Omega+\sigma\cdot(\bk+\bq)]
\end{eqnarray}
where
\begin{eqnarray}\label{Delta}
\Delta=x(1-x)(\Omega^2+\bq^2).
\end{eqnarray}
The standard next step when working in dimensional regularization is
to define new integration variables $\ell_{\omega}=\omega+x\Omega$
and $\ell=\bk+x\bq$. We then perform the integral over
$\ell_{\omega}$ to obtain
\begin{eqnarray}
&&\mathcal{P}_{\mu}(\bq,i\Omega)=\int_0^1dx\int\frac{d^D\ell}{(2\pi)^D}\left[\frac{-\sigma_{\mu}}{4\sqrt{\ell^2+\Delta}}+\right.\nonumber\\
&&\left.
\frac{(\sigma\cdot\ell-xS)\sigma_{\mu}
(\sigma\cdot\ell+(1-x)S)}{4(\ell^2+\Delta)^{\frac{3}{2}}}\right],
\end{eqnarray}
where we defined
\begin{equation}\label{S}
S\equiv i\Omega+\sigma\cdot\bq.
\end{equation}
Since the integration measure is $O(2)$-symmetric, only the terms even in $\ell$ in numerator give a non-trivial contribution, and we find
\begin{eqnarray}
&&\mathcal{P}_{\mu}(\bq,i\Omega)=\frac{1}{4}\int_0^1dx\int\frac{d^D\ell}{(2\pi)^D}\left[\frac{-\sigma_{\mu}}
{\sqrt{\ell^2+\Delta}}\right.\nonumber\\
&+&\left. \sigma_a\sigma_{\mu}\sigma_a\frac{\ell^2
}{D(\ell^2+\Delta)^{\frac{3}{2}}}-\frac{x(1-x)S\sigma_{\mu}S
}{(\ell^2+\Delta)^{\frac{3}{2}}}\right].
\end{eqnarray}
We next use  the dimensional regularization integrals (\ref{dim-reg-int}) and (\ref{dim-reg-int1}), as well as the identity (\ref{spatial-trace-pauli})
to find
\begin{eqnarray}
&&\mathcal{P}_{\mu}(\bq,i\Omega)=\frac{1}{8\pi}\int_0^1dx\left[\sigma_{\mu}
\sqrt{\Delta}-2\delta_{\mu0}\sqrt{\Delta}\right.\nonumber\\
&&\left. -
(i\Omega+\sigma\cdot\bq)\sigma_{\mu}(i\Omega+\sigma\cdot\bq)
\frac{x(1-x)}{\sqrt{\Delta}}\right].
\end{eqnarray}
Using Eq.(\ref{Delta}),
we finally have
\begin{eqnarray}
&&\mathcal{P}_{\mu}(\bq,i\Omega)=\frac{\sqrt{\Omega^2+\bq^2}}{64}\nonumber\\
&\times&\left[\sigma_{\mu}-2\delta_{\mu0}-
\frac{(i\Omega+\sigma\cdot\bq)\sigma_{\mu}(i\Omega+\sigma\cdot\bq)
}{\Omega^2+\bq^2} \right].\nonumber\\
\end{eqnarray}

\begin{widetext}
\subsection{Coulomb vertex}
The Coulomb vertex function to the first order in the coupling
constant is
\begin{equation}\label{CV-def}
\mathcal{P}^c_{\mu}(\bq,\bp,i\Omega)=-\int\frac{d^D\bk}{(2\pi)^D}\int_{-\infty}^{\infty}\frac{d\omega}{2\pi}
\frac{2\pi e^2}{|\bp-\bk|}G_{\bk}(i\omega)\sigma_{\mu}G_{\bk+\bq}(i\omega+i\Omega),
\end{equation}
with the free fermion Green's function given by Eq.\ (\ref{GreensFxn}).
After introducing Feynman parameters using Eq.\ (\ref{FP1}), we obtain
\begin{eqnarray}\label{Coulomb-vertex}
\mathcal{P}^c_{\mu}(\bq,\bp,i\Omega)&=&-\int_0^1dx\int\frac{d^D\ell}{(2\pi)^D}\frac{2\pi
e^2}{|\bp+x\bq-\ell|}\left[\frac{-\sigma_{\mu}}{4\sqrt{\ell^2+\Delta}}+
\frac{(\sigma\cdot\ell-x(i\Omega+\sigma\cdot\bq))\sigma_{\mu}
(\sigma\cdot\ell+(1-x)(i\Omega+\sigma\cdot\bq))}{4(\ell^2+\Delta)^{\frac{3}{2}}}\right].\nonumber\\
\end{eqnarray}

Now, we consider two terms in the above form of the Coulomb vertex separately.
Using the Feynman parametrization (\ref{FP1}) and the $D$-dimensional integral (\ref{dim-reg-int}),
the first term in the last equation acquires the form
\begin{eqnarray}\label{Cv1}
\int\frac{d^D\ell}{(2\pi)^D}\frac{1}{|\bp+x\bq-\ell|}
\frac{1}{\sqrt{\ell^2+\Delta}}
&=&\frac{1}{\pi}\int_0^1\!\!\frac{dy}{\sqrt{y(1-y)}}\int\!\!\frac{d^D\ell}{(2\pi)^D}
\frac{1}{\ell^2+(1-y)\left(y(\bp+x\bq)^2+\Delta\right)}\nonumber\\
&=&\frac{1}{\pi}\frac{\Gamma[1-\frac{D}{2}]}{(4\pi)^{\frac{D}{2}}}\int_0^1\!\!\frac{dy}{\sqrt{y}(1-y)^{\frac{3-D}{2}}}
\frac{1}{\left(y(\bp+x\bq)^2+\Delta\right)
^{1-\frac{D}{2}}}.
\end{eqnarray}
After expanding the integrand to the first order in the parameter $\eps=2-D$ and integrating over $y$, we have
\begin{eqnarray}
\int\frac{d^D\ell}{(2\pi)^D}\frac{1}{|\bp+x\bq-\ell|}
\frac{1}{\sqrt{\ell^2+\Delta}}
=\frac{\Gamma[1-\frac{D}{2}]}{(4\pi)^{\frac{D}{2}}}
\left(1-\frac{\eps}{2}\ln\frac{(\bp+x\bq)^2}{16}
-\eps\tanh^{-1}\sqrt{\frac{x(1-x)(\Omega^2+\bq^2)}{(\bp+x\bq)^2+x(1-x)(\Omega^2+\bq^2)}}\right).\nonumber\\
\end{eqnarray}
The second term in Eq.\ (\ref{Coulomb-vertex}), after introducing the Feynman parameter, shifting the variable $\ell-y(\bp+x\bq)\rightarrow\ell$,
and integrating over $\ell$, becomes
\begin{eqnarray}\label{Cv3}
&&\int\frac{d^D\ell}{(2\pi)^D}
\frac{(\sigma\cdot\ell-xS)\sigma_{\mu}(\sigma\cdot\ell+(1-x)S)}{|\bp+x\bq-\ell|(\ell^2+\Delta)^{\frac{3}{2}}}=
\frac{2}{\pi}\frac{\Gamma[1-\frac{D}{2}]}{(4\pi)^{\frac{D}{2}}}\int_0^1dy\frac{(1-y)^{\frac{D-1}{2}}}{\sqrt{y}}\frac{\frac{D}{2}\delta_{\mu0}
}{\left(y(\bp+x\bq)^2+\Delta\right)^{1-\frac{D}{2}}}\nonumber\\
&+&\frac{2}{\pi}\frac{\Gamma[2-\frac{D}{2}]}{(4\pi)^{\frac{D}{2}}}\int_0^1dy\frac{(1-y)^{\frac{D-3}{2}}}{\sqrt{y}}
\frac{(y\sigma\cdot(\bp+x\bq)-xS)\sigma_{\mu}(y\sigma\cdot(\bp+x\bq)+(1-x)S)}{\left(y(\bp+x\bq)^2+\Delta\right)^{2-\frac{D}{2}}}\nonumber\\
&+&\frac{2}{\pi}\frac{\Gamma[2-\frac{D}{2}]}{(4\pi)^{\frac{D}{2}}}\int_0^1dy\frac{(1-y)^{\frac{D-1}{2}}}{\sqrt{y}}
\frac{\delta_{\mu
a}\sigma_a}{\left(y(\bp+x\bq)^2+\Delta\right)^{1-\frac{D}{2}}},
\end{eqnarray}
where $S=i\Omega+\sigma\cdot\bq$, and we also used the identity (\ref{spatial-trace-pauli}) for the trace of the Pauli matrices over spatial indices. Note that the last term in the previous equation arises from the term proportional to $\eps=2-D$ in Eq.\ (\ref{spatial-trace-pauli}). If we treated Pauli matrices strictly in $D=2$, this term would be omitted what would thus lead to the violation of the Ward identity, as it may be directly checked in Eq.\ (\ref{N}). Therefore, in order to preserve the gauge invariance of the theory, it is crucial to treat Pauli matrices, as well as the momentum integrals, in a general spatial dimension, and only at the end of the calculation to take $D=2-\eps$, and expand in $\eps$.
After expanding in $\eps$, keeping
terms up to order $\eps^0$, and performing the integral over $y$ in
the first term, we have
\begin{eqnarray}\label{Cv2}
&&\int\frac{d^D\ell}{(2\pi)^D}
\frac{(\sigma\cdot\ell-xS)\sigma_{\mu}(\sigma\cdot\ell+(1-x)S)}{|\bp+x\bq-\ell|(\ell^2+\Delta)^{\frac{3}{2}}}=
\frac{\Gamma[1-\frac{D}{2}]}{(4\pi)^{\frac{D}{2}}}\delta_{\mu0}
\left(
1-\frac{\eps}{2}\ln\frac{(\bp+x\bq)^2}{16}-\eps\tanh^{-1}\sqrt{\frac{\Delta}{(\bp+x\bq)^2+\Delta}}\right.\nonumber\\
&-&\left.\eps\frac{\sqrt{\Delta((\bp+x\bq)^2+\Delta)}-\Delta}{(\bp+x\bq)^2}
-\frac{\eps}{2}\right)+\frac{1}{4\pi}\delta_{\mu a}\sigma_a\nonumber\\
&+&\frac{1}{2\pi^2}\int_0^1dy\frac{1}{\sqrt{y(1-y)}}
\frac{(y\sigma\cdot(\bp+x\bq)-xS)\sigma_{\mu}(y\sigma\cdot(\bp+x\bq)+(1-x)S)}{y(\bp+x\bq)^2+\Delta}\nonumber\\
&=&\frac{\Gamma[1-\frac{D}{2}]}{(4\pi)^{\frac{D}{2}}}\delta_{\mu0}
\left(1-\frac{\eps}{2}\ln\frac{(\bp+x\bq)^2}{16}-\eps\tanh^{-1}\sqrt{\frac{\Delta}{(\bp+x\bq)^2+\Delta}}\right.\nonumber\\
&-&\left.\eps\frac{\sqrt{\Delta((\bp+x\bq)^2+\Delta)}-\Delta}{(\bp+x\bq)^2}
-\frac{\eps}{2}\right)+\frac{1}{4\pi}\delta_{\mu
a}\sigma_a
+\frac{1}{2\pi^2} \left(
-x(1-x)S\sigma_{\mu}S\;\mathcal{I}_0[(\bp+x\bq)^2,\Delta]\right.\nonumber\\
&+&\left.(-xS\sigma_{\mu}\sigma\cdot(\bp+x\bq)+(1-x)(\bp+x\bq)\cdot\sigma
\sigma_{\mu}S)\mathcal{I}_1[(\bp+x\bq)^2,\Delta]+\sigma\cdot(\bp+x\bq)\sigma_{\mu}\sigma\cdot(\bp+x\bq)\mathcal{I}_2[(\bp+x\bq)^2,\Delta]\right)\nonumber\\
\end{eqnarray}
with $\Delta$ defined in Eq.\ (\ref{Delta}). The remaining integrals
over $y$ read
\begin{eqnarray}
\mathcal{I}_0(a,\Delta)=\int_0^1\frac{dy}{\sqrt{y(1-y)}}\frac{1}{y a+\Delta}&=&\frac{\pi}{\sqrt{\Delta(a+\Delta)}}\label{mathcalI-0}\label{calI-0}\\
\mathcal{I}_1(a,\Delta)=\int_0^1\frac{dy}{\sqrt{y(1-y)}}\frac{y}{y a+\Delta}&=&\frac{\pi}{a}\left(1-\sqrt{\frac{\Delta}{a+\Delta}}\right)\label{mathcalI-1}
\label{calI-1}\\
\mathcal{I}_2(a,\Delta)=\int_0^1\frac{dy}{\sqrt{y(1-y)}}\frac{y^2}{y
a+\Delta}&=&\frac{\pi}{2a^2}\left(a-2\Delta\left(1-\sqrt{\frac{\Delta}{a+\Delta}}\right)\right)\label{mathcalI-2}.\label{calI-2}
\end{eqnarray}
Therefore, using Eqs.\ (\ref{Cv1}) and (\ref{Cv2}), we can write the Coulomb vertex in the form
\begin{eqnarray}\label{eq:CoulombVertexFinal}
{\mathcal P}_\mu^c(\bq,\bp,i\Omega)&=&-\frac{\pi e^2}{2}\int_0^1
dx\,\,\left\{\frac{\Gamma{[1-\frac{D}{2}]}}{(4\pi)^{D/2}}\eps\delta_{\mu0}\left(\frac{\Delta-\sqrt{\Delta(\Delta+(\bp+x\bq)^2)}}{(\bp+x\bq)^2}
-\frac{1}{2}\right)+\frac{1}{4\pi}\sigma_a\delta_{\mu a}\right.\nonumber\\
&-&\left.\sigma_a\delta_{\mu a}\frac{\Gamma[1-\frac{D}{2}]}{(4\pi)^{D/2}}\left(
1-\frac{\eps}{2}\ln\frac{(\bp+x\bq)^2}{16}-\eps\tanh^{-1}\sqrt{\frac{x(1-x)(\Omega^2+\bq^2)}{(\bp+x\bq)^2+x(1-x)(\Omega^2+\bq^2)}}\right)\right.
\nonumber\\
&+&\left.\frac{1}{2\pi} \left[
\;\frac{- x(1-x)(i\Omega+\sigma\cdot\bq)\sigma_{\mu}(i\Omega+\sigma\cdot\bq)}{\sqrt{\Delta((\bp+x\bq)^2+\Delta)}}\right.\right.\nonumber\\
&+&\left.\left.\frac{\left(-x(i\Omega+\sigma\cdot\bq)\sigma_{\mu}\sigma\cdot(\bp+x\bq)+(1-x)(\bp+x\bq)\cdot\sigma
\sigma_{\mu}(i\Omega+\sigma\cdot\bq)\right)}{(\bp+x\bq)^2}\left(1-\sqrt{\frac{\Delta}{(\bp+x\bq)^2+\Delta}}\right)\right.\right.\nonumber\\
&+&\left.\left.\frac{
\sigma\cdot(\bp+x\bq)\sigma_{\mu}\sigma\cdot(\bp+x\bq)}{2(\bp+x\bq)^4}\left((\bp+x\bq)^2-2\Delta
\left(1-\sqrt{\frac{\Delta}{(\bp+x\bq)^2+\Delta}}\right)\right)
\right]\right\},
\end{eqnarray}
where again $ \Delta=x(1-x)(\Omega^2+\bq^2)$, and we explicitly wrote the functions
${\mathcal I}_0$, ${\mathcal I}_1$, and ${\mathcal I}_2$, defined in Eqs.(\ref{calI-0})-(\ref{calI-2}).
The above expression
diverges as $D\rightarrow 2$ from below, or equivalently as
$\eps\rightarrow 0^+$. As discussed in the main text, this
divergence is tied to the divergence of the self-energy and is a
consequence of the Ward-Takahashi identity proved below. All the
integrals over $x$ in the above expression can be performed in the
closed form in terms of elliptic integrals. (The expressions
involving $\tanh^{-1}$ need to be integrated by parts first to bring
them to the form easily expressible in terms of the elliptic
integrals.) However, we found that doing so leads to intractable
expressions and we thus chose to work with the above form of the Coulomb vertex, in which the integrals over the variable $x$
can be thought of as the integral representation of the elliptic
integrals.

At $\bq=0$ the above expressions simplify significantly and we have
\begin{eqnarray}\label{eq:CoulombVertexFinalq0}
{\mathcal P}_\mu^c(0,\bp,i\Omega)&=&-\delta_{\mu a}\frac{
e^2}{8}\left[ \sigma_a\left(1-\left( \frac{2}{\eps}-\gamma+\ln
64\pi-\ln\bp^2
\right)+\frac{2|\Omega|}{\sqrt{\Omega^2+4\bp^2}}K\right)\right.
\nonumber\\
&+&\left. \; [\bp\cdot\sigma
,\sigma_{a}]\frac{i\Omega}{\bp^2}\left(1-\frac{|\Omega|}{\sqrt{\Omega^2+4\bp^2}}K-\frac{\sqrt{\Omega^2+4\bp^2}}{|\Omega|}
(E-K)
\right)\right.\nonumber\\
&+&\left.\frac{
\sigma\cdot\bp\sigma_{a}\sigma\cdot\bp}{\bp^2}\left(1-\frac{\Omega^2}{3\bp^2}+\frac{(\Omega^4-16\bp^4)E+(16\bp^4-2\Omega^2\bp^2)K}{3|\Omega|\bp^2\sqrt{\Omega^2+4\bp^2}}
\right) \right],
\end{eqnarray}
where the arguments of the complete elliptic integrals of the first
and second kind, respectively, are
$$
K\equiv K\left(\frac{|\Omega|}{\sqrt{\Omega^2+4\bp^2}}\right);\;\;\;
E\equiv E\left(\frac{|\Omega|}{\sqrt{\Omega^2+4\bp^2}}\right).
$$
Note that to this order, at $\bq=0$ (and at any $\Omega$), only the
spatial components of $\mathcal{P}^c_{\mu}$ are finite. This is a
consequence of the Ward-Takahashi identity, since to this order the
self-energy is frequency-independent.

Finally, at $\bq=0$ and $\Omega=0$ the integrals in
Eq.(\ref{Coulomb-vertex}) can be performed for arbitrary $D$ without
the necessity of expanding in powers of $\eps$. The Coulomb vertex
then becomes
\begin{eqnarray}\label{eq:CoulombVertexAtZero}
{\mathcal
P}_\mu^c(0,\bp,0)&=&\frac{e^2}{(4\pi)^{\frac{D}{2}}}\frac{\delta_{\mu
a}}{|\bp|^{2-D}}
\frac{\Gamma[\frac{D}{2}+\frac{1}{2}]\Gamma[\frac{D}{2}-\frac{1}{2}]}{\Gamma[D]}
\left(\sigma_a\Gamma\left[1-\frac{D}{2}\right]
-\left(\sigma_a+\frac{\sigma\cdot\bp\sigma_a\sigma\cdot\bp}{\bp^2}\right)\Gamma\left[2-\frac{D}{2}\right]
\right).
\end{eqnarray}
This form of the vertex function is used to show that $\delta
\Pi^{(c)}_{\mu\nu}(0,0)=0$ for any $D$, a fact which is in turn used
in the calculation of the electrical conductivity.

\section{Coulomb vertex and the Ward-Takahashi identities in dimensional regularization}

In this appendix we show in detail that to the leading order in the
Coulomb interaction coupling constant $e^2$ and to $\mathcal{O}(N)$,
the Ward-Takahashi identity, questioned to hold in
Ref.\ \onlinecite{Sheehy-SchmalianPRB09}, is satisfied. As a first step, we
define the contraction $q^\mu{\mathcal P}_{\mu}^c\equiv-i\Omega{\mathcal
P}_0^c+q_a{\mathcal P}_a^c$. Using Eq.\ (\ref{eq:CoulombVertexFinal}), a
straightforward calculation shows that all the terms in the
contraction proportional to $i\Omega$ cancel out, and the contraction simplifies to
\begin{eqnarray}\label{contraction}
q^\mu{\mathcal P}_\mu^c(\bq,\bp,i\Omega)&=&-\frac{\pi
e^2}{2}\int_0^1dx\,\,\left\{-{\bf\sigma}\cdot\bq\frac{\Gamma[1-\frac{D}{2}]}{(4\pi)^{D/2}}\left(
1-\frac{\eps}{2}\ln\frac{(\bp+x\bq)^2}{16}-\eps\tanh^{-1}\sqrt{\frac{\Delta}{(\bp+x\bq)^2+\Delta}}\right)
+\frac{1}{4\pi}{\bf\sigma}\cdot\bq\right.\nonumber\\
&+&\left.\frac{1}{2\pi}\left[-{\bf\sigma}\cdot\bq\sqrt{\frac{\Delta}{(\bp+x\bq)^2+\Delta}}+
{\bf\sigma}\cdot(\bp+x\bq)\frac{(1-2x)(\Omega^2+\bq^2)}{(\bp+x\bq)^2}\left(1-\sqrt{\frac{\Delta}{(\bp+x\bq)^2+\Delta}}\right)\right.\right.\nonumber\\
&-&\left.\left.\frac{\Delta}{(\bp+x\bq)^4}\sigma\cdot(\bp+x\bq)\sigma\cdot\bq\sigma\cdot(\bp+x\bq)+
\frac{\sigma\cdot(\bp+x\bq)\sigma\cdot\bq\sigma\cdot(\bp+x\bq)}{2(\bp+x\bq)^2}\right.\right.\nonumber\\
&+&\left.\left.\frac{\sigma\cdot(\bp+x\bq)\sigma\cdot\bq\sigma\cdot(\bp+x\bq)}{(\bp+x\bq)^4}\sqrt{\frac{\Delta^3}{(\bp+x\bq)^2+\Delta}}\right]\right\}.
\end{eqnarray}

In order to show the Ward-Takahashi identity, we first note that the self-energy to the first-order in the Coulomb coupling is independent of the frequency. Thus all the terms in the contraction (\ref{contraction}) that contain  frequency have to vanish if the Ward-Takahashi identity holds.
In fact, as we will show in what follows, the contraction can be written in the form
\begin{equation}\label{contraction-decomposition}
q^\mu{\mathcal P}_\mu^c(\bq,\bp,i\Omega)=N(\bp,\bq)-\frac{e^2}{4}(\Omega^2+\bq^2)W(\bp,\bq)-\frac{e^2}{4}\sigma\cdot\bp\sqrt{(\Omega^2+\bq^2)^3}
L(\bp,\bq,\Omega)-\frac{e^2}{4}\sigma\cdot\bq\sqrt{\Omega^2+\bq^2}M(\bp,\bq,\Omega),
\end{equation}
where the functions $N(\bp,\bq)$, $W(\bp,\bq)$, $L(\bp,\bq,\Omega)$, and $M(\bp,\bq,\Omega)$ are defined in Eqs.\ (\ref{N}), (\ref{W}), (\ref{L}), and (\ref{M1}), respectively. This condition is, therefore, satisfied if $W(\bp,\bq)=0$, $M(\bp,\bq,\Omega)=0$, and $L(\bp,\bq,\Omega)=0$. Finally, to complete the proof of the identity, we will show that $N(\bp,\bq)=\Sigma_{\bp+\bq}(i\nu+i\omega)-\Sigma_\bp(i\nu)$.

Let us first show that term proportional to $\Omega^2+\bq^2$ vanishes, i.e., that
\begin{eqnarray}\label{W}
W(\bp,\bq)\equiv\int_0^1dx\,\,\left({\bf\sigma}\cdot(\bp+x\bq)\frac{(1-2x)}{(\bp+x\bq)^2}
-\frac{x(1-x)}{(\bp+x\bq)^4}\sigma\cdot(\bp+x\bq)\sigma\cdot\bq\sigma\cdot(\bp+x\bq)\right)=0.
\end{eqnarray}
Using the identity
\begin{equation}\label{id1}
\sigma\cdot\bp\,\sigma\cdot\bq\,\sigma\cdot\bp=2\bp\cdot\bq\,\sigma\cdot\bp-\bp^2\,\sigma\cdot\bq,
\end{equation}
we can rewrite
the above integral as
\begin{eqnarray}
W(\bp,\bq)&=&\sigma\cdot\bp\int_0^1dx\,\,\frac{\bp^2(1-2x)-(2\bp\cdot\bq+\bq^2)x^2}{(\bp+x\bq)^4}\nonumber\\
&+&\frac{\sigma\cdot\bq}{\bq^2}\int_0^1dx\,\,\left(\frac{x^2[2\bq^2\bp\cdot\bq-\bp^2\bq^2+4(\bp\cdot\bq)^2]+2x\bp^2[\bq^2+2\bp\cdot\bq]+\bp^4}
{(\bp+x\bq)^4}-1\right)\nonumber\\
&\equiv&\sigma\cdot\bp\,\, W_1(\bp,\bq)
+\frac{\sigma\cdot\bq}{\bq^2}\,\, W_2(\bp,\bq).
\end{eqnarray}
When $\bp=\bq$, it is easy to show that both $W_1$ and $W_2$ vanish, and we thus concentrate on the case $\cde\equiv\bp^2\bq^2-(\bp\cdot\bq)^2>0$.
In order to calculate the integrals $W_1$ and $W_2$, we use the following identities
\begin{eqnarray}
K_0&\equiv&\int_0^1\,\,\frac{dx}{({\bf p}+x{\bf q})^4}=\frac{\bp^2\bq^2-2(\bp\cdot\bq)^2-(\bp\cdot\bq)\bq^2}{2\bp^2(\bp+\bq)^2\cde}+
\frac{\bq^2}{2\cde}\int_0^1\,\,
\frac{dx}{({\bf p}+x{\bf q})^2},\\
K_1&\equiv&\int_0^1dx\,\,\frac{x}{({\bf p}+x{\bf q})^4}=\frac{\bq^2+\bp\cdot\bq}{2(\bp+\bq)^2\cde}-\frac{\bp\cdot\bq}{2\cde}
\int_0^1\frac{dx}{({\bf p}+x{\bf q})^2},\\
K_2&\equiv&\int_0^1dx\,\,\frac{x^2}{({\bf p}+x{\bf q})^4}=-\frac{\bp^2+\bp\cdot\bq}{2(\bp+\bq)^2\cde}+
\frac{\bp^2}{2\cde}\int_0^1\,\,\frac{dx}{({\bf p}+x{\bf q})^2}.
\end{eqnarray}
Straightforward calculation yields $W_1(\bp,\bq)=0$ and $W_2(\bp,\bq)=0$, and thus $W(\bp,\bq)=0$.
The contraction given by Eq.\ (\ref{contraction}) then simplifies to
\begin{eqnarray}\label{contraction1}
q^\mu{\mathcal P}_\mu^c(\bq,\bp,i\Omega)&=&\frac{\pi e^2}{2}
{\bf\sigma}\cdot\bq\,\,{\mathcal J}_0
-\frac{ e^2}{8}\int_0^1dx\,\,\left[{\bf\sigma}\cdot\bq+\frac{\sigma\cdot(\bp+x\bq)\sigma\cdot\bq\sigma\cdot(\bp+x\bq)}{(p+x\bq)^2}\right.\nonumber\\
&+&\left.2
\left(-\sigma\cdot\bq-\sigma\cdot(\bp+x\bq)\frac{(1-2x)(\Omega^2+\bq^2)}{(\bp+x\bq)^2}
+x(1-x)(\Omega^2+\bq^2)\frac{\sigma\cdot(\bp+x\bq)\sigma\cdot\bq\sigma\cdot(\bp+x\bq)}{(\bp+x\bq)^4}\right)\right.\nonumber\\
&\times&\left.\sqrt{\frac{x(1-x)(\Omega^2+\bq^2)}{(\bp+x\bq)^2+x(1-x)(\Omega^2+\bq^2)}}\right],
\end{eqnarray}
where
\begin{equation}
{\mathcal J}_0\equiv\frac{\Gamma[1-\frac{D}{2}]}{(4\pi)^{D/2}}\int_0^1dx\,\,\left(
1-\frac{\eps}{2}\ln\frac{(\bp+x\bq)^2}{16}-\eps\tanh^{-1}\sqrt{\frac{x(1-x)(\Omega^2+\bq^2)}{(\bp+x\bq)^2+x(1-x)(\Omega^2+\bq^2)}}\right).
\end{equation}
After a partial integration in the last term, ${\mathcal J}_0$ becomes
\begin{equation}
{\mathcal J}_0=\frac{\Gamma[1-\frac{D}{2}]}{(4\pi)^{D/2}}\int_0^1dx\,\,\left(1+\eps\ln4-\eps\ln\left[\sqrt{(\bp+x\bq)^2+x(1-x)(\Omega^2+\bq^2)}+\sqrt{x(1-x)(\Omega^2+\bq^2)}\right]\right).
\end{equation}
Finally, after performing another partial integration in the last term of the previous equation, we obtain
\begin{eqnarray}\label{mcJ0}
{\mathcal J}_0&=&\frac{\Gamma[1-\frac{D}{2}]}{(4\pi)^{D/2}}\int_0^1dx\,\,\left\{1+\eps\ln4-\frac{\eps}{2}\ln[(\bp+\bq)^2]+\eps\frac{x(x\bq^2+\bp\cdot\bq)}
{(\bp+x\bq)^2}\right.\nonumber\\
&+&\left.\frac{\eps}{2}\frac{1}{(\bp+x\bq)^2}\left[\frac{1-2x}{1-x}(\bp+x\bq)^2-2x(x\bq^2+\bp\cdot\bq)\right]\sqrt{\frac{x(1-x)(\Omega^2+\bq^2)}{(\bp+x\bq)^2+x(1-x)(\Omega^2+\bq^2)}}
\right\}.
\end{eqnarray}
Expansion of the prefactor in $\eps$ has the form
\begin{equation}
\frac{\Gamma[1-\frac{D}{2}]}{(4\pi)^{D/2}}=\frac{1}{2\pi}\left(\frac{1}{\eps}+\frac{1}{2}[-\gamma+\ln4\pi]\right)+{\mathcal O}(\eps),
\end{equation}
which, after keeping the terms up to the order $\eps^0$ in Eq.\ (\ref{mcJ0}), yields
\begin{eqnarray}\label{mcJ01}
{\mathcal J}_0&=&\frac{1}{2\pi}\int_0^1dx\,\,\left\{\frac{1}{\eps}+\frac{1}{2}[-\gamma+\ln64\pi]-\frac{1}{2}\ln[(\bp+\bq)^2]+\frac{x(x\bq^2+\bp\cdot\bq)}
{(\bp+x\bq)^2}\right.\nonumber\\
&+&\left.\frac{1}{2(\bp+x\bq)^2}\left[\frac{1-2x}{1-x}(\bp+x\bq)^2-2x(x\bq^2+\bp\cdot\bq)\right]\sqrt{\frac{x(1-x)(\Omega^2+\bq^2)}{(\bp+x\bq)^2+x(1-x)(\Omega^2+\bq^2)}}\right\}.
\end{eqnarray}
Now, after substituting Eq.\ (\ref{mcJ01}) into Eq.\ (\ref{contraction1}), the contraction reads
\begin{equation}
q^\mu{\mathcal
P}_\mu^c(\bq,\bp,i\Omega)=N(\bp,\bq)-\frac{e^2}{4}\sigma\cdot\bp\sqrt{(\Omega^2+\bq^2)^3}
L(\bp,\bq,\Omega)-\frac{e^2}{4}\sigma\cdot\bq\sqrt{\Omega^2+\bq^2}M(\bp,\bq,\Omega),
\end{equation}
which is, in fact, the form (\ref{contraction-decomposition}) of the contraction, since we have already shown that $W(\bp,\bq)=0$.
The frequency-independent part in the above equation reads
\begin{equation}\label{N}
N(\bp,\bq)=\frac{e^2}{8}\int_0^1dx\,\,\left\{\sigma\cdot\bq\left[\frac{2}{\eps}-\gamma+\ln64\pi-\ln[(\bp+\bq)^2]+
\frac{2x(x\bq^2+\bp\cdot\bq)}{(\bp+x\bq)^2}-1\right]-\frac{\sigma\cdot(\bp+x\bq)\,\sigma\cdot\bq\,\sigma\cdot(\bp+x\bq)}{(\bp+x\bq)^2}\right\},
\end{equation}
while the remaining terms are
\begin{equation}\label{L}
L({\bf p},{\bf q},\Omega)=\int_0^1dx\frac{x^2({\bf p}+{\bf q})^2-(1-x)^2{\bf p}^2}{({\bf p}+x{\bf q})^4}\sqrt{\frac{x(1-x)}{({\bf p}+x{\bf q})^2+x(1-x)(\Omega^2+{\bf q}^2)}},
\end{equation}
and
\begin{eqnarray}\label{M1}
M({\bf p},{\bf q},\Omega)&=&\frac{1}{2}\int_0^1 dx \left( \frac{x^2({\bf p}+{\bf q})^2-(1-x)^2{\bf p}^2}{(1-x)({\bf p}+x{\bf q})^2}-2 \right)
\sqrt{\frac{x(1-x)}{({\bf p}+x{\bf q})^2+x(1-x)(\Omega^2+\bq^2)}}\nonumber\\
&+&\int_0^1 dx\left(\frac{x(\Omega^2+{\bf q}^2)[(3x-2){\bf p}^2+2x(2x-1){\bf p}\cdot{\bf q}+x^3{\bf q}^2]}{({\bf p}+x{\bf q})^4}\right)\sqrt{\frac{x(1-x)}{({\bf p}+x{\bf q})^2+x(1-x)(\Omega^2+{\bf q}^2)}}.\nonumber\\
\end{eqnarray}
The frequency-independent term (\ref{N}), after using the identity (\ref{id1}), and performing the remaining integral, has the form
\begin{equation}
N(\bp,\bq)=\frac{e^2}{8}\left(\frac{2}{\eps}\sigma\cdot\bq+(-\gamma+\ln64\pi)\sigma\cdot\bq-\sigma\cdot(\bp+\bq)\ln[(\bp+\bq)^2]
+\sigma\cdot\bp\ln\bp^2\right)=\Sigma_{\bp+\bq}(i\nu+i\Omega)-\Sigma_{\bp}(i\nu).
\end{equation}
Here, $\Sigma_\bp(i\nu)$ is the self-energy defined in Eq.\ (\ref{eq:SelfEnergy}) in the main text.

\section{Evaluation of the functions $L(\bp,\bq,\Omega)$ and $M({\bf p},{\bf
q},\Omega)$} In this Appendix we show that the
functions $L(\bp,\bq,\Omega)$, Eq.\ (\ref{L}), and $M({\bf p},{\bf
q},\Omega)$, Eq.\ (\ref{M1}), vanish identically, and therefore, the
Ward-Takahashi identity is, indeed, satisfied within the dimensional
regularization used here.

In order to evaluate integrals in Eqs.\ (\ref{L}) and (\ref{M1}), we introduce new variables
\begin{equation}
p'=\frac{p}{q}e^{-i\varphi},\,\,w=\frac{\Omega}{q},
\end{equation}
with $p=|\bp|$, $q=|\bq|$, and $\cos\varphi=\bp\cdot\bq/(pq)$,
and thus we can write
\begin{equation}
(\bp+x\bq)^2=q^2(x+p')(x+p'^*),
\end{equation}
with $p'^*$ denoting complex conjugate of $p'$.
The integral (\ref{L}) can now be rewritten in terms of the new variables as
\begin{eqnarray}\label{L1}
L=\frac{1}{q^2}\sqrt{1+\frac{q^2}{\Omega^2}}\int_0^1dx\frac{(1+p')(1+p'^*)x^3(1-x)-x(1-x)^3p'p'^*}{(x+p')^2(x+p'^*)^2}
\frac{1}{\sqrt{x(x-1)(x-x_+)(x-x_-)}},
\end{eqnarray}
where $x_\pm$ are the roots of the quadratic equation $({\bf p}+x{\bf q})^2+x(1-x)(\Omega^2+q^2)=0$,
\begin{equation}\label{x+}
x_\pm=\frac{({\bf p}+{\bf q})^2-p^2+\Omega^2\pm\sqrt{(({\bf p}+{\bf q})^2-p^2+\Omega^2)^2+4\Omega^2p^2}}{2\Omega^2},
\end{equation}
or in terms of the variables $p'$ and $w$
\begin{equation}
x_\pm=\frac{1+2Re(p')+w^2\pm\sqrt{(1+2Re(p')+w^2)^2+4w^2p'p'^*}}{2w^2}.
\end{equation}
Assuming that $x_+>1$, we have the folowing sequence
\begin{equation}
x_+>1>0>x_-,
\end{equation}
which is important for expressing the function $L(\bp,\bq,\Omega)$ in terms of the elliptic integrals, as we will see below.
We use partial fractions to calculate the integral in Eq.\ (\ref{L1}).  The first term reads
\begin{eqnarray}
\frac{x^3(1-x)}{(x+p')^2(x+p'^*)^2}&=&-1+\left(\frac{ip'^2[p'(1+2p')-(3+4p')p'^*]}{8[Im(p')]^3(x+p')}
+\frac{p'^3(1+p')}{4[Im(p')]^2(x+p')^2}+C.c.\right)\nonumber\\
&\equiv&-1+\left(\frac{A}{x+p'}+\frac{B}{(x+p')^2}+C.c.\right),
\end{eqnarray}
while the second term is
\begin{eqnarray}
\frac{x(1-x)^3}{(x+p')^2(x+p'^*)^2}&=&-1+\left(\frac{i(1+p')^2[p'^2-2p'p'^*-Re(p')]}{4[Im(p')]^3(x+p')}
+\frac{p'(1+p')^3}{4[Im(p')]^2(x+p')^2}+C.c.\right)\nonumber\\
&\equiv&-1+\left(\frac{A_1}{x+p'}+\frac{B_1}{(x+p')^2}+C.c.\right).
\end{eqnarray}
We therefore reduced the problem of evaluating the integral (\ref{L1}) to the calculation of the following integrals
\begin{equation}\label{Im}
I_m=\int_0^1dx\frac{1}{(x+p)^m}\frac{1}{\sqrt{x(x-1)(x-x_+)(x-x_-)}},
\end{equation}
with $m=0,1,2$, and $x_+>1>0>x_-$. In terms of the integrals $I_m$, the integral $L$ reads
\begin{eqnarray}
L&=&-[1+2Re(p')]I_0+\left[((1+p')(1+p'^*)A-p'p'^*A_1)I_1
+((1+p')(1+p'^*)B-p'p'^*B_1)I_2+C.c.\right]\nonumber\\
&=&-[1+2Re(p')]I_0+\left[\frac{-ip'(1+p')(1+2p')}{2Im(p')}I_1+\frac{ip'^2(1+p')^2}{2Im(p')}I_2+C.c.\right].\nonumber\\
\end{eqnarray}
The integrals $I_m$ with $m=0,1,2$ and $x_+>1>0>x_-$ have the form (Eqs. (255.00), (255.38), (340.01), and (340.02) in Ref. \onlinecite{Byrd})
\begin{eqnarray}
I_0&=&gF(k),\label{I0}\\
I_1&=&\frac{g}{(1+p')\alpha_1^2}\left[(\alpha_1^2-\alpha^2)\Pi(\alpha_1^2,k)+\alpha^2F(k)\right],\label{I1}\\
I_2&=&\frac{g}{(1+p')^2\alpha_1^4}\left[\alpha^4F(k)+2\alpha^2(\alpha_1^2-\alpha^2)\Pi(\alpha_1^2,k)
+\frac{(\alpha_1^2-\alpha^2)^2}{2(\alpha_1^2-1)(k^2-\alpha_1^2)}\left(\alpha_1^2E(k)+(k^2-\alpha_1^2)F(k)\right.\right.\nonumber\\
&+&\left.\left.(2\alpha_1^2k^2+2\alpha_1^2-\alpha_1^4-3k^2)\Pi(\alpha_1^2,k)\right)\right],\label{I2}
\end{eqnarray}
where $F(k)$, $E(k)$, and $\Pi(\alpha^2,k)$ are the complete elliptic integrals of the first, second, and the third kind, respectively, defined in
terms of the corresponding incomplete integrals as $F(k)\equiv F(\pi/2,k)$, $E(k)\equiv E(\pi/2,k)$, and $\Pi(\alpha^2,k)\equiv\Pi(\pi/2,\alpha^2,k)$, with incomplete elliptic integrals defined as in Ref. \onlinecite{Byrd}. Here,
\begin{equation}\label{k}
k^2=\frac{x_+-x_-}{x_+(1-x_-)},\,\,\,\,\,
g=\frac{1}{\sqrt{x_+(1-x_-)}},
\end{equation}
and
\begin{equation}\label{alpha}
\alpha^2=\frac{1}{x_+},\,\,\,\,\,
\alpha_1^2=\frac{x_++p'}{x_+(1+p')}.
\end{equation}
Therefore, the integral $L$ has the form
\begin{equation}
L=R(p',w) F(k)+P(p',w)E(k)+Im[G(p',w)\Pi(\alpha_1^2,k)].
\end{equation}
Note that imaginary part of $\alpha_1^2$ is non-vanishing.
In the following, we will show that the coefficient $P(p',w)=0$. Then, by expressing $Im[G(p',w)\Pi(p',w)]$ in terms of the functions $F(k)$ and  $\Pi(\alpha_2^2,k)$, with $\alpha_2^2$ defined below purely real, we obtain that the function $L$ vanishes.
The coefficient $P(p',w)$ reads
\begin{eqnarray}
P(p',w)&=&\frac{ip'^2(\alpha_1^2-\alpha^2)^2}{4Im(p')(\alpha_1^2-1)\alpha_1^2(k^2-\alpha_1^2)}+C.c
=-\frac{i p'x_+(1-x_+)(1+p')}{4Im(p')(p'+x_+)(p'+x_-)}+C.c\nonumber\\
&=&\frac{1}{2Im(p')}Im\left[\frac{p'x_+(1-x_+)(1+p')}{(x_++p')(x_-+p')}\right],
\end{eqnarray}
but
\begin{eqnarray}
&&\frac{p'x_+(1-x_+)(1+p')}{(x_++p')(x_-+p)}=-\frac{1}{4w^2(1+4w^2)}\left[1+2Re(p')+w^2
+\sqrt{(1+2Re(p')
+w^2)^2+4p'p'^*w^2}\right]\nonumber\\
&\times&\left[1+2Re(p')-w^2+\sqrt{(1+2Re(p')+w^2)^2+4p'p'^*w^2}\right]
\end{eqnarray}
is purely real, and thus $P(p',w)$ vanishes identically. The coefficient $R(p',w)$ reads
\begin{equation}\label{R}
R(p',w)=-g(1+2Re(p'))+g\left[\left(\frac{-ip'(1+p')(1+2p')}{2Im(p')(x_++p')}+\frac{ip'(1+p')}{4Im(p')(x_++p')^2}[2p'(1+p')+x_+(1-x_+)]
\right)+C.c.\right],
\end{equation}
while
\begin{equation}
G(p',w)=g\frac{p'(\alpha_1^2-\alpha^2)}{Im(p')\alpha_1^2}\left[1+2p'+\frac{p'}{\alpha_1^2}
\left(\frac{\alpha^2(3\alpha_1^4+k^2-2\alpha_1^2(1+k^2))+\alpha_1^2(\alpha_1^4+3k^2-2\alpha_1^2(1+k^2))}{2(\alpha_1^2-1)(k^2-\alpha_1^2)}
\right)\right].
\end{equation}
Imaginary part of the product $G(p',w)\Pi(\alpha_1^2,k)$  can be expressed in terms of the complete elliptic function of the first kind and
the third kind, as given by Eq.\ (419.00) in Ref.\ \onlinecite{Byrd},
\begin{eqnarray}
Im[G(p',w)\Pi(\alpha_1^2,k)]&=&\frac{g}{Im(p')}\frac{1}{m_2r^2(s_2t_1-s_1t_2)}\left\{[a_1(s_1r^2-k^2s_2m_2)+b_1(k^2t_2m_2-t_1r^2)]F(k)\right.\nonumber\\
&+&\left.n_2m_2r^2(a_1s_1-b_1t_1)\Pi(\alpha_2^2,k)\right\},
\end{eqnarray}
where
\begin{eqnarray}\label{G}
G&=&a_1+i b_1,\,\,\alpha_1^2=-\gamma_1-i\gamma_2,\,\,r^2=\gamma_1^2+\gamma_2^2,\nonumber\\
m_2&=&-\frac{2\gamma_1+k^2}{\gamma_1},\,\,\alpha_2^2=\frac{k^2m_2^2}{r^2},\nonumber\\
s_1&=&1-\frac{k^2}{r^2},\,\, n_2=\frac{m_2[\alpha_2^4-(2+m_2)\alpha_2^2+(1-2\alpha_2^2)k^2]-r^2}{m_2(\gamma_2^2+(\gamma_1+\alpha_2^2)^2)}\nonumber\\     t_1&=&\frac{2(k^2+2\gamma_1+\gamma_1^2)}{\gamma_2(2\gamma_1+k^2)},\,\,\,\, t_2=\frac{m_2^2+(\gamma_1+2-\alpha_2^2)m_2+n_2m_2(\gamma_1+\alpha_2^2)}{m_2\gamma_2},\,\,\,
s_2=1-n_2-\frac{1}{m_2},
\end{eqnarray}
since $r^2+2\gamma_1+k^2=0$ in our case, with $k^2$ given by Eq.\ (\ref{k}), and $a_1$, $b_1$, $\gamma_1$, and $\gamma_2$ real.
Thus, the integral $L$ acquires the form
\begin{equation}
L=[R(p',w)+R_1(p',w)]F(k)+S_1(p',w)\Pi(\alpha_2^2,k),
\end{equation}
where
\begin{equation}\label{R1}
R_1(p',w)=\frac{g}{Im(p')}\frac{a_1(s_1r^2-k^2s_2m_2)+b_1(k^2t_2m_2-t_1r^2)}{m_2r^2(s_2t_1-s_1t_2)}
\end{equation}
and
\begin{equation}
S_1(p',w)=\frac{g}{Im(p')}\frac{n_2(a_1s_1-b_1t_1)}{s_2t_1-s_1t_2},
\end{equation}
while $R(p',w)$ is given by Eq. (\ref{R}).

Straightforward calculation shows that $a_1s_1-b_1t_1=0$ and $s_2t_1-s_1t_2\neq0$, and thus the coefficient $S_1(p',w)$ vanishes identically.
Furthermore, using that $a_1s_1-b_1t_1=0$, the form of the  coefficient $R_1(p',w)$, given by Eq.\ (\ref{R1}), can be simplified to
\begin{equation}
R_1(p',w)=-\frac{g}{Im(p')}\frac{a_1k^2}{r^2t_1}.
\end{equation}
Finally, the previous equation together with  Eqs. (\ref{R}) and (\ref{G}), since  $w^2$ is purely real, yields $R(p',w)+R_1(p',w)=0$, and, therefore, the integral $L({\bf p},{\bf q},\Omega)$ given by Eq.\ (\ref{L}) vanishes. When the two vectors are (anti)collinear, i.e., when $Im(p')=0$, the continuity implies $L({\bf p},{\bf q},\Omega)=0$. When $x_+<1$, the analogous calculation shows that $L({\bf p},{\bf q},\Omega)=0$, as well.

Let us now turn to the integral $M({\bf p},{\bf q},\Omega)$, and consider the case $Im(p')\neq0$, and $x_+>1$. In terms of partial
fractions, this integral reads
\begin{eqnarray}
M&=&\frac{1+w^2}{w}\int_0^1dx \left\{\left[\frac{p'^3(1+p')^2}{2iIm(p')(x+p')^2}+\frac{ip'^2(1+p')(1+2p')}{2Im(p')(x+p')}+C.c.\right]
+ p'^{*2}+2(1+p')Re(p')+x-x^2\right\}\nonumber\\
&\times&\frac{1}{\sqrt{x(x-1)(x-x_+)(x-x_-)}}
+\frac{1}{2w}\int_0^1dx\left\{\left[\frac{p'^2(1+p')}{x+p'}+C.c.\right]-2Re[p'(1+p')]+[2Re(p')-1]x+2x^2\right\}\nonumber\\
&\times&\frac{1}{\sqrt{x(x-1)(x-x_+)(x-x_-)}}.
\end{eqnarray}
It follows from Eq. (255.17) in Ref.\ \onlinecite{Byrd} that
\begin{eqnarray}
J_1&\equiv&\int_0^1\frac{xdx}{\sqrt{x(x-1)(x-x_+)(x-x_-)}}=\frac{g}{\alpha^2}\left[(\alpha^2-1)\Pi(\alpha^2,k)+F(k)\right]\\
J_2&\equiv&\int_0^1\frac{x^2dx}{\sqrt{x(x-1)(x-x_+)(x-x_-)}}=\frac{g}{\alpha^4}\left\{F(k)+2(\alpha^2-1)\Pi(\alpha^2,k)\right.\nonumber\\
&+&\left.\frac{\alpha^2-1}{2(k^2-\alpha^2)}\left[\alpha^2E(k)+(k^2-\alpha^2)F(k)+(2\alpha^2k^2+2\alpha^2-\alpha^4-3k^2)\Pi(\alpha^2,k)\right] \right\},
\end{eqnarray}
which together with Eqs.\ (\ref{I0}), (\ref{I1}) and (\ref{I2}) allows us to express the function $M$ in terms of the elliptic integrals as
\begin{equation}
M=\frac{g}{w}\left[A(p',w)\Pi(\alpha_1^2,k)+A(p',w)^*\Pi(\alpha_1^2,k)^*+B(p',w)F(k)
+ X(p',w)\Pi(\alpha^2,k)+Y(p',w)E(k)\right],
\end{equation}
with the coefficients in the above equation of the form
\begin{eqnarray}
A(p',w)&=&\frac{p'^2}{2}\left(1-\frac{\alpha^2}{\alpha_1^2}\right)\left\{1+\frac{i(1+2p')(1+w^2)}{Im(p')}\right.\nonumber\\
&-&\left.\frac{ip'(1+w^2)}{\alpha_1^2Im(p')}\left[2\alpha^2+\frac{(\alpha_1^2-\alpha^2)(2\alpha_1^2k^2+2\alpha_1^2-\alpha_1^4-3k^2)}
{2(\alpha_1^2-1)(k^2-\alpha_1^2)}\right]\right\},\\
B(p',w)&=&\frac{1+w^2}{Im(p')}Im\left\{\frac{p'^3}{\alpha_1^4}\left[\alpha^4+\frac{(\alpha_1^2-\alpha^2)^2}{2(\alpha_1^2-1)}\right]\right\}
+Re\left[\frac{\alpha^2p'^2}{\alpha_1^2}\left(1+\frac{i(1+2p')(1+w^2)}{Im(p')}\right)\right]\nonumber\\
&+&(1+w^2)[p'^{*2}+2(1+p')Re(p')]-Re[p'(1+p')]+\frac{1}{\alpha^2}\left[\frac{1}{2}+w^2+Re(p')\right]\\
&-&\frac{w^2(1+\alpha^2)}{2\alpha^4},\\
X(p',w)&=&\left(1-\frac{1}{\alpha^2}\right)\left[\frac{1}{2}+Re(p')+w^2-\frac{w^2(2\alpha^2k^2-\alpha^4+k^2-2\alpha^2)}
{2\alpha^2(k^2-\alpha^2)}\right],\nonumber\\
Y(p',w)&=&\frac{1+w^2}{2Im(p')}Im\left[\frac{p'^3(\alpha_1^2-\alpha^2)^2}{\alpha_1^2(\alpha_1^2-1)(k^2-\alpha_1^2)}\right]-
\frac{w^2(\alpha^2-1)}{2\alpha^2(k^2-\alpha^2)}.
\end{eqnarray}
Here, $k$, $\alpha$, and $\alpha_1$ are defined by Eqs.\ (\ref{k}) and (\ref{alpha}). Since $w^2$ is purely real, the functions $A(p',w)$, $B(p',w)$, $X(p',w)$, and $Y(p',w)$ vanish, and thus the integral $M({\bf p},{\bf q},\Omega)$ given by Eq.\ (\ref{M1}) is identically equal to zero. This result also implies that, because of the continuity, in the  case of (anti)collinear vectors ${\bf p}$ and ${\bf q}$, i.e., when $Im(p')=0$, the function $M({\bf p},{\bf q},\Omega)$ also vanishes.    When the root $x_+$ in Eq,\ (\ref{x+}) is less than one, the analogous calculation shows that the function $M=0$, as well.
\end{widetext}

\section{Violation of the Ward-Takahashi identity within hard cutoff regularization }

In this Appendix we show that Ward-Takahashi identity does not hold within the hard cutoff regularization.
In order to show that, we will follow the same steps as in the previous two appendices.
Let us first calculate the self-energy, given by Eq.\ (\ref{self-energy-definition}),
\begin{eqnarray}
\Sigma_{\bp}(i\omega)&=&\int\frac{d\omega'}{2\pi}\int\frac{d^2{\bk}}{(2\pi)^2}
\frac{2\pi e^2}{|\bp-\bk|}\frac{i\omega'+\sigma\cdot\bk}{\omega'^2+k^2}\nonumber\\
&=&\pi e^2\int\frac{d^2{\bk}}{(2\pi)^2}\frac{\sigma\cdot\bk}{p|\bp-\bk|},
\end{eqnarray}
which after using the Feynman parametrization (\ref{FP1}), shifting the momentum variable,
and performing angular integration becomes
\begin{eqnarray}
\Sigma_{\bp}(i\omega)&=&\frac{e^2}{2\pi}\sigma\cdot\bp\int_0^1dy\sqrt{\frac{y}{1-y}}\nonumber\\
&\times&\int_0^\Lambda dk\frac{k}{k^2+y(1-y)p^2},
\end{eqnarray}
where $\Lambda$ is the momentum cutoff regulating the ultraviolet divergence of the integral. Remaining integrations then yield
\begin{equation}\label{self-energy-UV}
\Sigma_{\bp}(i\omega)=\frac{e^2}{4}\sigma\cdot\bp\left(\ln\Lambda-\ln p+2\ln2\right).
\end{equation}
Therefore, the divergent part of the momentum integral appears as the logarithm of the
cutoff which corresponds to $1/\eps$ pole in the dimensional regularization, see Eq.\ (\ref{eq:SelfEnergy}).
Note that the divergent parts appear with precisely the same coefficients,
but the finite parts are different within the two regularizations.

We now consider the Coulomb vertex function defined in Eq.\ (\ref{eq:CV-def})
which after introducing the Feynman parameters may be written in
the form given by Eq.\  (\ref{Coulomb-vertex}) with the integral performed in $D=2$.
After following the same steps as in Appendix A, and performing straightforward integrations, we obtain
\begin{eqnarray}
&&\int\frac{d^2\ell}{(2\pi)^2}\frac{1}{|\bp+x\bq-\ell|}
\frac{1}{\sqrt{\ell^2+x(1-x)(\Omega^2+q^2)}}\nonumber\\
&=&\frac{1}{2\pi}\left(\ln\Lambda-\frac{1}{2}\ln\frac{(\bp+x\bq)^2}{16}\right.\nonumber\\
&-&\left.\tanh^{-1}
\sqrt{\frac{x(1-x)(\Omega^2+q^2)}{(\bp+x\bq)^2+x(1-x)(\Omega^2+q^2)}}\right).
\end{eqnarray}
Analogously, we have
\begin{widetext}
\begin{eqnarray}
&&\int\frac{d^2\ell}{(2\pi)^2}\frac{(\sigma\cdot\ell-xS)\sigma_{\mu}(\sigma\cdot\ell+(1-x)S)}{|\bp+x\bq-\ell|(\ell^2+\Delta)^{\frac{3}{2}}}
=\frac{\delta_{\mu0}}{2\pi}\left(\ln\Lambda-\frac{1}{2}\ln\frac{(\bp+x\bq)^2}{16}
-\tanh^{-1}
\sqrt{\frac{x(1-x)(\Omega^2+q^2)}{(\bp+x\bq)^2+x(1-x)(\Omega^2+q^2)}}\right.\nonumber\\
&-&\left.\frac{\sqrt{\Delta((\bp+x\bq)^2+\Delta)}-\Delta}{(\bp+x\bq)^2}\right)
+\frac{1}{2\pi^2} \left\{
-x(1-x)S\sigma_{\mu}S\;\mathcal{I}_0[(\bp+x\bq)^2,\Delta]+\left[-xS\sigma_{\mu}\sigma\cdot(\bp+x\bq)\right.\right.\nonumber\\
&+&\left.\left.(1-x)(\bp+x\bq)\cdot\sigma
\sigma_{\mu}S\right]\mathcal{I}_1[(\bp+x\bq)^2,\Delta]+\sigma\cdot(\bp+x\bq)\sigma_{\mu}\sigma\cdot(\bp+x\bq)\mathcal{I}_2[(\bp+x\bq)^2,\Delta]\right\},
\end{eqnarray}
\end{widetext}
with ${\mathcal I}_0$, ${\mathcal I}_1$, and ${\mathcal I}_2$ defined in Eqs.\
(\ref{mathcalI-0})-(\ref{mathcalI-2}). Note that the term $\sim\delta_{\mu a}\sigma_a$ is not present in the
above equation, since the Pauli matrices here are treated strictly in $D=2$, and thus
$\sigma_a\sigma_\mu\sigma_a=2\delta_{\mu 0}$. Taking the contraction $q^\mu{\cal P}_\mu^{c}$, and following the
steps in Appendix B, its frequency-independent part reads
\begin{eqnarray}
{\tilde N}(\bp,\bq)&=&\frac{e^2}{4}\int_0^1dx\,\,\left\{\sigma\cdot\bq\left[\ln\Lambda+\ln4-\ln[|\bp+\bq|]\right.\right.\nonumber\\
&+&\left.\left.
\frac{x(x\bq^2+\bp\cdot\bq)}{(\bp+x\bq)^2}\right]\right.\nonumber\\
&-&\left.\frac{\sigma\cdot(\bp+x\bq)\,\sigma\cdot\bq\,\sigma\cdot(\bp+x\bq)}{2(\bp+x\bq)^2}\right\},
\end{eqnarray}
which after performing straightforward integrals yields
\begin{equation}\label{WT-violation}
{\tilde N}(\bp,\bq)=\Sigma_{\bp+\bq}(i\nu+i\Omega)-\Sigma_\bp(i\nu)+\frac{e^2}{8}\sigma\cdot\bq,
\end{equation}
and therefore the Ward-Takahashi identity is violated within the hard-cutoff regularization scheme. The other frequency-dependent terms, actually, vanish,
since they have the same form as within the dimensional regularization, but even if this were not the case they could not cancel purely momentum-dependent term on the right-hand side of Eq.\ (\ref{WT-violation}) that spoils the Ward-Takahashi identity.
In fact,
using exactly the same procedure, one can show that within dimensional regularization with Pauli matrices
treated in strictly $D=2$ the Ward-Takahashi identity is also violated precisely because of the last term on
the right-hand side in the frequency-independent part of the contraction $q^\mu{\cal P}_\mu^{c}$.

\section{Kubo formula and the a.c. conductivity within dimensional regularization with Pauli matrices in $D=2-\eps$}

In this Appendix we perform explicit calculation of the Coulomb correction to the conductivity within the dimensional regularization, in which both the momentum integrals and the Pauli matrices are treated in $D=2-\eps$, which, as we demonstrated, is consistent with the $U(1)$ gauge symmetry of the theory, and show that this regularization yields ${\mathcal C}=(11-3\pi)/6$ in Eq.\ (\ref{conductivity-final}).

Let us first calculate  contribution coming from the self-energy
part. Using Eq.\ (\ref{self-energy-integral}) and the identity
\begin{equation}
\sigma_x {\bf\sigma}\cdot{\bf k}\sigma_x=2\sigma_x
k_x-{\bf\sigma}\cdot{\bf k},
\end{equation}
the self-energy part of $\delta\Pi_{\mu\nu}(i\Omega,0)$ for
$\mu=\nu=x$, $\delta\Pi_{xx}^{(a)}(i\Omega,0)$, can be written as
\begin{eqnarray}
&&\delta\Pi_{xx}^{(a)}(i\Omega,{0})=2N
e^2\frac{\Gamma\left(1-\frac{D}{2}\right)\Gamma\left(\frac{D+1}{2}\right)
\Gamma\left(\frac{D-1}{2}\right)}
{(4\pi)^{D/2}\Gamma\left(D\right)}\nonumber\\
&\times&\int\frac{d\omega}{2\pi}\int\frac{d^D{\bf k}}{(2\pi)^D}
\frac{k^{D-2}}{(\omega^2+k^2)^2[(\omega+\Omega)^2+k^2]}
\nonumber\\
&\times&{\rm Tr}\{(i\omega+\sigma\cdot{\bf k})[i(\omega+\Omega)+2\sigma_x k_x-\sigma\cdot{\bf k}]\nonumber\\
&&(i\omega+\sigma\cdot{\bf k})\sigma\cdot{\bf k}\}.
\end{eqnarray}
Performing the trace and the frequency integral, we obtain
\begin{eqnarray}
&&\delta\Pi_{xx}^{(a)}(i\Omega,{0})=-4N
e^2\frac{\left(1-\frac{1}{D}\right)\Gamma\left(1-\frac{D}{2}\right)
\Gamma\left(\frac{D+1}{2}\right)}{(4\pi)^{D/2}
\Gamma\left(D\right)}\nonumber\\
&\times&\Gamma\left(\frac{D-1}{2}\right)\int\frac{d^D{\bf
k}}{(2\pi)^D}\frac{k^{D-1}(4k^2-\Omega^2)}{(\Omega^2+4k^2)^2}.
\end{eqnarray}
After subtracting the zero-frequency part of the Coulomb correction
to the polarization tensor, and using Eq.\ (\ref{eq:condPerp}), the
self-energy part of the Coulomb interaction correction to the
conductivity reads
\begin{eqnarray}
\sigma_a&=&-32\sigma_0\Omega
e^2\frac{\left(1-\frac{1}{D}\right)\Gamma\left(1-\frac{D}{2}\right)
\Gamma\left(\frac{D+1}{2}\right)\Gamma\left(\frac{D-1}{2}\right)}{(4\pi)^{D}
\Gamma\left(D\right)\Gamma\left(\frac{D}{2}\right)}\nonumber\\
&\times&\int_0^\infty dk\,\,
k^{2D-4}\frac{\Omega^2+12k^2}{(\Omega^2+4k^2)^2},
\end{eqnarray}
where $\sigma_0$ is the Gaussian conductivity of the Dirac fermions
given by Eq.\ (\ref{Gaussian-cond}). The remaining integral then
yields
\begin{eqnarray}\label{a-complete}
\sigma_a&=&-\sigma_0
e^2\Omega^{2D-4}\frac{\left(1-\frac{1}{D}\right)\Gamma\left(1-\frac{D}{2}\right)
\Gamma\left(\frac{D+1}{2}\right)
\Gamma\left(\frac{D-1}{2}\right)}{(4\pi)^{D}
\Gamma\left(D\right)\Gamma\left(\frac{D}{2}\right)}\nonumber\\
&\times&\frac{2^{8-2D}(D-1)\pi}{\cos(D\pi)}.
\end{eqnarray}
Taking $D=2-\epsilon $ and expanding up to the order $\epsilon^0$,
we obtain the self-energy part of the Coulomb contribution to the
conductivity
\begin{equation}\label{a}
\sigma_a=\frac{1}{2}\sigma_0
e^2\left(-\frac{1}{\epsilon}+\frac{3}{2}+\gamma-\ln(64\pi)+\mathcal{O}(\epsilon)\right).
\end{equation}

Let us now concentrate on the vertex part of the Coulomb correction
to the conductivity. Taking the trace in $\delta\Pi_{xx}^{(b)}$
given by Eq.\ (\ref{pol-vertex}), we have
\begin{eqnarray}
&&\delta\Pi_{xx}^{(b)}(i\Omega,0)=2N\int_{-\infty}^\infty\frac{{d\omega}}{2\pi}\int\frac{d^D{\bf
k}}{(2\pi)^D}\int_{-\infty}^\infty\frac{d\omega'}{2\pi} \nonumber\\
&\times&\int\frac{d^D{\bf p}}{(2\pi)^D}V_{{\bf k}-{\bf p}}\frac{1}{[(\omega+\Omega)^2+k^2][(\omega'+\Omega)^2+p^2]}\nonumber\\
&\times&\frac{1}{(\omega^2+k^2)(\omega'^2+p^2)}\{\omega(\omega+\Omega)\omega'(\omega'+\Omega)\nonumber\\
&-&[\omega\omega'+(\omega+\Omega)(\omega'+\Omega)]{\bf k}\cdot{\bf
p}
\nonumber\\
&+& 4{\bf k}\cdot{\bf p}k_xp_x-2k_x^2p^2
-2p_x^2k^2+p^2k^2\nonumber\\
&-&[\omega(\omega'+\Omega)
+\omega'(\omega+\Omega)](2k_xp_x-{\bf k}\cdot{\bf p})\nonumber\\
&-&\omega(\omega+\Omega)(2p_x^2-p^2)-\omega'(\omega'+\Omega)(2k_x^2-k^2)\}.
\end{eqnarray}
Integration over the frequencies then yields
\begin{eqnarray}
&&\delta\Pi_{xx}^{(b)}(i\Omega,{ 0})=2N\int\frac{d^D{\bf
k}}{(2\pi)^D}\int\frac{d^D{\bf p}}{(2\pi)^D} V_{{\bf
k}-{\bf p}}\nonumber\\
&\times&\frac{1}
{kp(\Omega^2+4k^2)(\Omega^2+4p^2)}[\Omega^2(k_xp_x-{\bf k}\cdot{\bf p})\nonumber\\
&+&4({\bf k}\cdot{\bf p}k_xp_x+k^2p^2-p_x^2k^2-p^2k_x^2)].
\end{eqnarray}
After subtracting  the zero-frequency part of
$\delta\Pi_{xx}(i\Omega,{\bf 0})$, we obtain the vertex part of the
Coulomb correction to the conductivity
\begin{eqnarray}
\sigma_b&=&8\sigma_0\Omega\int\frac{d^D{\bf
k}}{(2\pi)^D}\int\frac{d^D {\bf p}}{(2\pi)^D}\,\, V_{{\bf
k}-{\bf p}}\,\,\frac{1}{k^3p^3(\Omega^2+4k^2)}\nonumber\\
&\times&\frac{1}{\Omega^2+4p^2}\left\{({\bf k}\cdot{\bf
p}k_xp_x+k^2p^2-p_x^2k^2-p^2k_x^2)
[\Omega^2\right.\nonumber\\
&+&\left.4(k^2+p^2)]+4k^2p^2({\bf k}\cdot{\bf p}-k_xp_x)\right \}\nonumber\\
&=&\sigma_{b1}+\sigma_{b2}
\end{eqnarray}
where
\begin{eqnarray}\label{sigmab1}
\sigma_{b1}&=&8\sigma_0\Omega\int\frac{d^D{\bf
k}}{(2\pi)^D}\int\frac{d^D {\bf p}}{(2\pi)^D}\,\, V_{{\bf
k}-{\bf p}}\,\,\frac{\Omega^2+4(k^2+p^2)}{k^3p^3(\Omega^2+4k^2)}\nonumber\\
&\times&\frac{1}{\Omega^2+4p^2}\left\{({\bf k}\cdot{\bf p}k_xp_x+k^2p^2-p_x^2k^2-p^2k_x^2)\right\},\nonumber\\
\end{eqnarray}
and
\begin{eqnarray}\label{sigmab2}
\sigma_{b2}&=&32\sigma_0\Omega\int\frac{d^D{\bf
k}}{(2\pi)^D}\int\frac{d^D {\bf p}}{(2\pi)^D}\,\, V_{{\bf
k}-{\bf p}}\nonumber\\
&\times&\frac{{\bf k}\cdot{\bf
p}-k_xp_x}{kp(\Omega^2+4k^2)(\Omega^2+4p^2)}.
\end{eqnarray}
The contribution $\sigma_{b1}$, by adding and subtracting $\Omega^2$
in the first term in the numerator, may  be rewritten as
\begin{equation}\label{sigmab1-decomposition}
\sigma_{b1}=\sigma_{b1}^{(1)}+\sigma_{b1}^{(2)}+\sigma_{b1}^{(3)},
\end{equation}
where
\begin{widetext}
\begin{eqnarray}\label{sigmab1-1}
\sigma_{b1}^{(1)}&=&2\sigma_0\Omega\int\frac{d^D{\bf
k}}{(2\pi)^D}\int\frac{d^D {\bf p}}{(2\pi)^D}\,\, V_{{\bf k}-{\bf
p}}\frac{{\bf k}\cdot{\bf
p}k_xp_x+k^2p^2-p_x^2k^2-p^2k_x^2}{k^3p^3\left[k^2+\left(\frac{\Omega}{2}
\right)^2\right]\left[p^2+\left(\frac{\Omega}{2}\right)^2\right]}\left[k^2+\left(\frac{\Omega}{2}\right)^2+p^2+\left(\frac{\Omega}{2}\right)^2\right]\nonumber\\
&=&4\sigma_0\Omega\int\frac{d^D{\bf k}}{(2\pi)^D}\int\frac{d^D {\bf
p}}{(2\pi)^D}\,\, V_{{\bf k}-{\bf p}}\frac{{\bf k}\cdot{\bf
p}k_xp_x+k^2p^2-p_x^2k^2-p^2k_x^2}{k^3p^3\left[k^2+\left(\frac{\Omega}{2}
\right)^2\right]},
\end{eqnarray}

\begin{equation}\label{sigmab1-2-xx}
\sigma_{b1}^{(2)}=-\frac{1}{2}\sigma_0\Omega^3\int\frac{d^D{\bf
k}}{(2\pi)^D}\int\frac{d^D {\bf p}}{(2\pi)^D}\,\, V_{{\bf k}-{\bf
p}}\frac{1}{kp\left[k^2+\left(\frac{\Omega}{2}\right)^2\right]\left[p^2+
\left(\frac{\Omega}{2}\right)^2\right]},
\end{equation}
and
\begin{equation}\label{sigmab1-3}
\sigma_{b1}^{(3)}=-\frac{1}{2}\sigma_0\Omega^3\int\frac{d^D{\bf
k}}{(2\pi)^D}\int\frac{d^D {\bf p}}{(2\pi)^D}\,\, V_{{\bf k}-{\bf
p}}\frac{{\bf k}\cdot{\bf p}k_xp_x-2p^2k_x^2}{k^3p^3
\left[k^2+\left(\frac{\Omega}{2}\right)^2\right]\left[p^2+\left(\frac{\Omega}{2}\right)^2\right]}.
\end{equation}
\end{widetext}
The advantage of this decomposition of the integral $\sigma_b$ is
that its diverging part is now isolated, and it is contained in the
integral $\sigma_{b1}^{(1)}$, whereas all the other integrals are
finite in $D=2$.

We first consider the term $\sigma_{b1}^{(1)}$. Using the Feynman
parametrization
\begin{eqnarray}\label{FP2}
&&\frac{1}{A^\alpha B^\beta
C^\gamma}=\frac{\Gamma[\alpha+\beta+\gamma]}{\Gamma[\alpha]\Gamma[\beta]\Gamma[\gamma]}
\int_0^1 dx \int_0^{1-x} dy
\nonumber\\
&\times&
\frac{(1-x-y)^{\alpha-1}x^{\beta-1}y^{\gamma-1}}{\left\{(1-x-y)A+xB+yC\right\}^{\alpha+\beta+\gamma}},
\end{eqnarray}
we write
\begin{eqnarray}\label{aux-xx}
&&\frac{1}{k^3|{\bf k}-{\bf
p}|\left[(\frac{\Omega}{2})^2+k^2\right]}=\frac{4}{\pi}\int_0^1dx\int_0^{1-x}dy
\nonumber\\
&&\frac{(1-x-y)^{1/2}x^{-1/2}}{\left[({\bf k}-x{\bf
p})^2+x(1-x)p^2+y\left(\frac{\Omega}{2}\right)^2\right]^3}.
\end{eqnarray}
Shifting the momentum ${\bf k}-x{\bf p}\rightarrow{\bf k}$, and
retaining terms even in ${\bf k}$ as these are the only
non-vanishing ones due to the rotational invariance of the
integrand, we obtain
\begin{eqnarray}\label{b11-xx}
&&\sigma_{b1}^{(1)}=32\sigma_0\Omega
e^2\int_0^1dx\int_0^{1-x}dy\,\,(1-x-y)^{-1/2}x^{1/2}
\nonumber\\
&\times&\int\frac{d^D{\bf p}}
{(2\pi)^D}\frac{p^2-p_x^2}{p^3}\int\frac{d^D {\bf k}}{(2\pi)^D}\,\,\left[\left(1-\frac{1}{D}\right)k^2+x^2p^2\right]\nonumber\\
&\times&\frac{1}{\left[k^2+x(1-x)p^2+y\left(\frac{\Omega}{2}\right)^2\right]^3}.\nonumber\\
\end{eqnarray}
After performing the integral over ${\bf k}$, we have
\begin{eqnarray}
&&\sigma_{b1}^{(1)}=\frac{32\sigma_0\Omega
e^2}{(4\pi)^D\Gamma\left(\frac{D}{2}\right)}
\left(1-\frac{1}{D}\right)
\int_0^1dx\int_0^{1-x}dy\,\, x^{-1/2}\nonumber\\
&\times&(1-x-y)^{1/2} \int_0^\infty dp
\frac{p^{D-2}}{\left[x(1-x)p^2+y\left(\frac{\Omega}{2}\right)^2\right]^{2-\frac{D}{2}}}
\nonumber\\
&\times&\left\{\frac{1}{2}(D-1)
\Gamma\left(2-\frac{D}{2}\right)+\frac{x^2p^2\Gamma\left(3-\frac{D}{2}\right)}{x(1-x)p^2+y\left(\frac{\Omega}{2}\right)^2}\right\}.\nonumber\\
\end{eqnarray}
Integration over $p$ then yields
\begin{eqnarray}
\sigma_{b1}^{(1)}&=&\sigma_0e^2\Omega^{2D-4}\frac{2^{9-2D}\Gamma\left(\frac{5}{2}-D\right)
\Gamma\left(\frac{D+1}{2}\right)\left(1-\frac{1}{D}\right)} {(4\pi)^D\Gamma\left(\frac{D}{2}\right)}\nonumber\\
&\times&\int_0^1dx\,\, x^{-D/2}(1-x)^{-\frac{D+1}{2}}\int_0^{1-x}dy \,\, y^{D-\frac{5}{2}}\nonumber\\
&\times&(1-x-y)^{1/2}.
\end{eqnarray}
Using Eq.\ (\ref{Beta}) and the identity
\begin{equation}
\int_0^{1-x}dy\,\,(1-x-y)^{1/2}y^{D-\frac{5}{2}}=\frac{\sqrt{\pi}}{2}\frac{\Gamma\left(D-\frac{3}{2}\right)}
{\Gamma(D)}(1-x)^{D-1},
\end{equation}
after integration over $y$ and $x$, $\sigma_{b1}^{(1)}$ acquires the
form
\begin{eqnarray}
\sigma_{b1}^{(1)}&=&\sigma_0e^2\Omega^{2D-4}\frac{4^{4-D}\left(1-\frac{1}{D}\right)\Gamma\left(\frac{5}{2}-D\right)
\Gamma\left(1-\frac{D}{2}\right)}{(4\pi)^D
\Gamma\left(\frac{D}{2}\right)\Gamma(D)}\nonumber\\
&\times&\Gamma\left(D-\frac{3}{2}\right)\Gamma\left(\frac{D+1}{2}\right)\Gamma\left(\frac{D-1}{2}\right).
\end{eqnarray}
Finally, expanding the previous result in the parameter $\epsilon$,
we obtain
\begin{equation}\label{b11}
\sigma_{b1}^{(1)}=\frac{1}{2}\sigma_0e^2\left[\frac{1}{\epsilon}-\frac{1}{2}\left(1+2\gamma-12\ln2-2\ln\pi\right)+
{\mathcal{O}(\epsilon)}\right].
\end{equation}
Therefore, poles coming from the self-energy and the vertex parts
cancel out, as it should be, since the theory of Coulomb interacting
Dirac fermions is renormalizable, at least to the second order in
the Coulomb coupling.\cite{Vafek-Case}

Let us turn to  the remaining contributions which are all finite in
$D=2$. We first consider the term $\sigma_{b1}^{(2)}$ in Eq.\
(\ref{sigmab1-2-xx}). Using the Feynman parametrization (\ref{FP2}), we have
\begin{eqnarray}\label{aux1-xx}
&&\frac{1}{p|{\bf k}-{\bf
p}|\left[(\frac{\Omega}{2})^2+p^2\right]}=\frac{1}{\pi}\int_0^1dx\int_0^{1-x}dy
\nonumber\\
&\times&\frac{(1-x-y)^{-1/2}x^{-1/2}}{\left[({\bf p}-x{\bf
k})^2+x(1-x)k^2+y\left(\frac{\Omega}{2}\right)^2\right]^2}.
\end{eqnarray}
After shifting the momentum ${\bf p}-x{\bf k}\rightarrow{\bf p}$,
and integrating over ${\bf p}$, the term $\sigma_{b1}^{(2)}$ in Eq.\ (\ref{sigmab1-2-xx}) becomes
\begin{eqnarray}
\sigma_{b1}^{(2)}&=&-\frac{1}{4\pi}\sigma_0 e^2\Omega^3\int_0^1dx\int_0^{1-x}dy\,\,(1-x-y)^{-1/2}\nonumber\\
&\times&\int\frac{d^2{\bf
k}}{(2\pi)^2}\frac{x^{-1/2}}{k\left[x(1-x)k^2+y\left(\frac{\Omega}{2}\right)^2\right]\left[k^2+
\left(\frac{\Omega}{2}\right)^2 \right] }.\nonumber\\
\end{eqnarray}
Integration over the remaining momentum variable then yields
\begin{eqnarray}
\sigma_{b1}^{(2)}&=&-\frac{1}{2\pi}\sigma_0e^ 2\int_0^1dx\int_0^{1-x}dy\,\,\frac{1}{\sqrt{x(1-x-y)}}\nonumber\\
&\times&\frac{1}{y+\sqrt{xy(1-x)}}.
\end{eqnarray}
After integrating out the Feynman parameter $y$, the term
$\sigma_{b1}^{(2)}$ is
\begin{equation}\label{b12-xx}
\sigma_{b1}^{(2)}=\frac{i}{\pi}\sigma_0e^2\int_0^1dx\,\,\frac{\sec^{-1}\sqrt{x}}{\sqrt{x}(1-x)}=
-\frac{\pi}{2}\sigma_0e^2.
\end{equation}

We now evaluate the term $\sigma_{b1}^{(3)}$ in Eq.\
(\ref{sigmab1-3}). Using Eq.\ (\ref{aux-xx}), after shifting the
momentum variable, and retaining only terms even in ${\bf p}$,  we have
\begin{eqnarray}
\sigma_{b1}^{(3)}&=&-4\sigma_0\Omega^3e^2\int_0^1dx\int_0^{1-x}dy\,\,(1-x-y)^{1/2}x^{-1/2}\nonumber\\
&\times&\int\frac{d^D {\bf k}}{(2\pi)^D}\frac{k_x^2}{k^3\left[k^2+\left(\frac{\Omega}{2}\right)^2 \right]}\nonumber\\
&\times&\int\frac{d^D {\bf
p}}{(2\pi)^D}\frac{\left(\frac{1}{D}-2\right)p^2-x^2k^2}{\left[p^2+x(1-x)k^2+
y\left(\frac{\Omega}{2}\right)^2\right]^3}.
\end{eqnarray}
After integrating  over ${\bf p}$ and setting $D=2$, we obtain
\begin{eqnarray}
\sigma_{b1}^{(3)}&=&\sigma_0\Omega^3e^2\frac{1}{8\pi^2}\int_0^1dx\int_0^{1-x}dy\,\,(1-x-y)^{1/2}x^{-1/2}\nonumber\\
&\times&\int_0^\infty
\frac{dk}{\left[k^2+\left(\frac{\Omega}{2}\right)^2\right]\left[x(1-x)k^2+
y\left(\frac{\Omega}{2}\right)^2\right]}\nonumber\\
&\times&\left[\frac{3}{2}+\frac{x^2k^2}{x(1-x)k^2+y\left(\frac{\Omega}{2}\right)^2}\right].
\end{eqnarray}
Integration over $k$ then gives
\begin{eqnarray}
\sigma_{b1}^{(3)}&=&-\frac{1}{4\pi}\sigma_0e^ 2\int_0^1dx\int_0^{1-x}dy\,\,(1-x-y)^{1/2}\nonumber\\
&\times&(1-x)^{-1/2}\frac{3(x-1)\sqrt{\frac{y}{x(1-x)}}+2x-3}
{\sqrt{y}[\sqrt{y}+\sqrt{x(1-x)}]^2},
\end{eqnarray}
which, after integrating over the remaining variables, yields
\begin{equation}\label{b13}
\sigma_{b1}^{(3)}=\frac{1}{12}(4+3\pi)\sigma_0e^2.
\end{equation}

Let us now calculate the term $\sigma_{b2}$ given by Eq.\
(\ref{sigmab2}). Using Eq.\ (\ref{aux1-xx}), shifting the momentum
${\bf p}-x{\bf k}\rightarrow {\bf p}$, and retaining only terms even
in ${\bf p}$, we have
\begin{eqnarray}
\sigma_{b2}&=&4\sigma_0e^2\Omega\int_0^1dx\int_0^{1-x}dy\,\,(1-x-y)^{-1/2}x^{-1/2}\nonumber\\
&\times&\int\frac{d^D{\bf k}}{(2\pi)^D}
\frac{x(k^2-k_x^2)}{k\left[k^2+\left(\frac{\Omega}{2}\right)^2\right]}\nonumber\\
&\times&\int\frac{d^D{\bf
p}}{(2\pi)^D}\frac{1}{\left[p^2+x(1-x)k^2+y\left(\frac{\Omega}{2}\right)^2\right]^2}.
\end{eqnarray}
After the integration over ${\bf p}$ and setting $D=2$, we obtain
\begin{eqnarray}
\sigma_{b2}&=&\frac{\sigma_0e^2\Omega}{(2\pi)^2}\int_0^1dx\int_0^{1-x}dy\,\,(1-x-y)^{-1/2}x^{1/2}\nonumber\\
&\times&\int_0^\infty dk
\frac{k^2 }{\left[k^2+\left(\frac{\Omega}{2}\right)^2\right]\left[x(1-x)k^2+y\left(\frac{\Omega}{2}\right)^2\right]}.\nonumber\\
\end{eqnarray}
Integration over $k$ then gives
\begin{eqnarray}
&&\sigma_{b2}=\frac{\sigma_0e^2}{4\pi}\int_0^1dx\int_0^{1-x}dy\,\,\frac{[(1-x)(1-x-y)]^{-\frac{1}{2}}}
{\sqrt{y}+\sqrt{x(1-x)}}\nonumber\\
&=&\sigma_0e^2\frac{1}{4\pi}\int_0^1dx\,\,\frac{1}{\sqrt{1-x}}\left[\pi+2i\sqrt{\frac{x}{1-x}}\sec^{-1}
\sqrt{x}\right],\nonumber\\
\end{eqnarray}
where $\sec^{-1}x$ is the inverse function of $\sec x\equiv1/\cos
x$. Finally, integration over $x$ yields
\begin{equation}\label{b2-xx}
\sigma_{b2}=\frac{4-\pi}{4}\sigma_0e^2.
\end{equation}
Therefore, using Eqs.\ (\ref{a}), (\ref{b11}), (\ref{b12-xx}),
(\ref{b13}), and (\ref{b2-xx}) we obtain the first order correction
to the a.c. conductivity due to the Coulomb interaction
\begin{equation}
\delta\sigma^{(c)}=\sigma_a+\sigma_{b1}^{(1)}+\sigma_{b1}^{(2)}+\sigma_{b1}^{(3)}+\sigma_{b2}=\frac{11-3\pi}{6}\sigma_0e^2,
\end{equation}
which corresponds to the value
\begin{equation}
{\cal C}=\frac{11-3\pi}{6}
\end{equation}
in Eq.\ (\ref{conductivity-final}).

\section{Longitudinal conductivity using density-density correlator}

In order to obtain the longitudinal conductivity, we expand
$\delta\Pi_{00}^{(c)}(i\Omega,{\bf q})$ in Eq.\ (\ref{definePi00}), which is the time component ($\mu=\nu=0$)
of the Coulomb correction to the polarization tensor (\ref{eq:deltaPiCoulomb}), to the order ${\bf q}^2$.

Let us first consider the self-energy part
(\ref{density-density-self-energy}) which may be written as
\begin{equation}\label{Pia00sum}
\delta\Pi^{(a)}_{00}(i\Omega,\bq)=\delta\Pi^{(a1)}_{00}(i\Omega,\bq)+\delta\Pi^{(a1)}_{00}(-i\Omega,-\bq),
\end{equation}
where
\begin{eqnarray}
&&\delta\Pi^{(a1)}_{00}(i\Omega,\bq)=N \int
\frac{d^2{\bf k}}{(2\pi)^2}\int\frac{d\omega}{2\pi}\int \frac{d^2{\bf p}}{(2\pi)^2}\int\frac{d\omega'}{2\pi}\nonumber\\
&&V_{\bk-\bp}{\rm
Tr}\left[G_{\bk}(i\omega)G_{\bp}(i\omega')G_{\bk}(i\omega)
G_{\bk+\bq}(i\omega+i\Omega)\right].\nonumber\\
\end{eqnarray}
Using Eq.\ (\ref{self-energy-integral}) to integrate over the momentum $\bp$, and taking the trace over Pauli matrices, we
obtain
\begin{eqnarray}
&&\delta\Pi^{(a1)}_{00}(i\Omega,\bq)=2Ne^2\frac{\Gamma\left(1-\frac{D}{2}\right)\Gamma\left(\frac{D+1}{2}\right)
\Gamma\left(\frac{D-1}{2}\right)}
{(4\pi)^{D/2}\Gamma\left(D\right)}\nonumber\\
&\times&\int\frac{d\omega}{2\pi}\int\frac{d^D{\bf
k}}{(2\pi)^D}\frac{k^{D-2}}
{(\omega^2+{k}^2)^2[(\omega+\Omega)^2+({\bf k}+{\bf q})^2]}\nonumber\\
&\times&[(-\omega^2+{k}^2){\bf k}\cdot({\bf k}+{\bf q})-2\omega(\omega+\Omega){k}^2].\nonumber\\
\end{eqnarray}
Note that above expression contains a part divergent in $D=2$ that arises
from the self-energy (\ref{self-energy-integral}), and when multiplied with the remaining terms of the order $\eps$ gives a finite result,
i.e., the final result does not have a pole in $\eps$ .
Therefore, in order not to overlook this subtle cancellation, we have to perform the integrations in $D$-dimensions first, and only at the
end of the calculation to take $D=2-\eps$, with $\eps\rightarrow0$.
Expanding the ${\bf q}$-dependent term in the denominator to the
quadratic order in ${\bf q}$, and keeping only the terms quadratic in $\bq$ in the expression for $\delta\Pi_{00}^{(a1)}(i\Omega,\bq)$, we find
\begin{eqnarray}\label{Pia1-app}
&&\delta\Pi^{(a1)}_{00}(i\Omega,\bq)=2Ne^2\frac{\Gamma\left(1-\frac{D}{2}\right)\Gamma\left(\frac{D+1}{2}\right)
\Gamma\left(\frac{D-1}{2}\right)}
{(4\pi)^{D/2}\Gamma\left(D\right)}\nonumber\\
&\times&\int\frac{d\omega}{2\pi}\int\frac{d^D{\bf
k}}{(2\pi)^D}\frac{k^{D-2}}
{(\omega^2+{k}^2)^2[(\omega+\Omega)^2+{k}^2]^2}\nonumber\\
&\times&[(-\omega^2+{k}^2){\bf k}\cdot({\bf k}+{\bf q})-2\omega(\omega+\Omega){k}^2]\nonumber\\
&\times&\left(-2{\bf k}\cdot{\bf q}-{ q}^2+\frac{4({\bf k}\cdot{\bf q})^2}{(\omega+\Omega)^2+{k}^2}\right)\nonumber\\
&=&\delta\Pi^{(a11)}_{00}(i\Omega,\bq)+\delta\Pi^{(a12)}_{00}(i\Omega,\bq),
\end{eqnarray}
where
\begin{eqnarray}\label{Pia11}
&&\delta\Pi^{(a11)}_{00}(i\Omega,\bq)=-4Ne^2\frac{\Gamma\left(1-\frac{D}{2}\right)\Gamma\left(\frac{D+1}{2}\right)
\Gamma\left(\frac{D-1}{2}\right)}{(4\pi)^{D/2}\Gamma\left(D\right)}\nonumber\\
&\times&\int\frac{d\omega}{2\pi}\int\frac{d^D{\bf
k}}{(2\pi)^D}\frac{(-\omega^2+{k}^2)({\bf k}\cdot{\bf q})^2}
{k^{2-D}(\omega^2+{ k}^2)^2[(\omega+\Omega)^2+{k}^2]^2},\nonumber\\
\end{eqnarray}
and
\begin{eqnarray}\label{Pia12}
&&\delta\Pi^{(a12)}_{00}(i\Omega,\bq)=2Ne^2\frac{\Gamma\left(1-\frac{D}{2}\right)\Gamma\left(\frac{D+1}{2}\right)
\Gamma\left(\frac{D-1}{2}\right)}{(4\pi)^{D/2}\Gamma\left(D\right)}\nonumber\\
&\times&\int\frac{d\omega}{2\pi}\int\frac{d^D{\bf
k}}{(2\pi)^D}\frac{(-\omega^2+{k}^2){
k}^2-2\omega{k}^2(\omega+\Omega)}
{k^{2-D}(\omega^2+{k}^2)^2[(\omega+\Omega)^2+{ k}^2]^2}\nonumber\\
&\times&\left(-{ q}^2+\frac{4({\bf k}\cdot{\bf
q})^2}{(\omega+\Omega)^2+{k}^2}\right).
\end{eqnarray}
We first consider the term $\delta\Pi_{00}^{(a11)}$ in Eq.\ (\ref{Pia11}). After integrating over $\omega$ and using the
rotational symmetry of  the integrand, we have
\begin{eqnarray}
&&\delta\Pi^{(a11)}_{00}(i\Omega,\bq)=-Ne^2\frac{\Gamma\left(1-\frac{D}{2}\right)\Gamma\left(\frac{D+1}{2}\right)
\Gamma\left(\frac{D-1}{2}\right)}{(4\pi)^{D/2}\Gamma\left(D\right)}\nonumber\\
&\times&\frac{{ q}^2}{D}\int\frac{d^D{\bf
k}}{(2\pi)^D}k^{D-3}\frac{32k^4-12k^2\Omega^2-\Omega^4}{(\Omega^2+k^2)^3}.
\end{eqnarray}
After integrating over ${\bf k}$, and expanding the result in $\eps$, we obtain to the order $\eps^0$
\begin{equation}
\delta\Pi^{(a11)}_{00}(i\Omega,\bq)=\frac{3}{64}Ne^2\frac{q^2}{\omega}=\frac{3}{4}\sigma_0e^2\frac{{q}^2}{|\Omega|}.
\end{equation}
The term $\delta\Pi^{(a12)}_{00}(i\Omega,\bq)$ given by Eq.\
(\ref{Pia12}), after integration over the frequency and using the
rotational symmetry of the integrand, acquires the form
\begin{eqnarray}
&&\delta\Pi^{(a12)}_{00}(i\Omega,\bq)=-Ne^2\frac{\Gamma\left(1-\frac{D}{2}\right)\Gamma\left(\frac{D+1}{2}\right)
\Gamma\left(\frac{D-1}{2}\right)}{(4\pi)^{D/2}\Gamma\left(D\right)}\nonumber\\
&\times&\frac{q^2}{2D}\int\frac{d^D{\bf k}}{(2\pi)^D}\frac{k^{D-3}}{(\Omega^2+4k^2)^3}\nonumber\\
&\times&[16(D-5)k^4+24k^2\Omega^2+(3-D)\Omega^4].
\end{eqnarray}
Integration over ${\bf k}$ in the last expression, and expansion in $\eps$ then yield
\begin{equation}
\delta\Pi^{(a12)}_{00}(i\Omega,\bq)=-\frac{1}{32}Ne^2\frac{q^2}{|\Omega|}=-\frac{1}{2}\sigma_0e^2\frac{q^2}{|\Omega|}.
\end{equation}
Therefore, using Eq.\ (\ref{Pia1-app}), we have
\begin{equation}
\delta\Pi^{(a1)}_{00}(i\Omega,\bq)=\frac{1}{64}Ne^2\frac{q^2}{|\Omega|}=\frac{1}{4}\sigma_0e^2\frac{q^2}{|\Omega|},
\end{equation}
which together with Eq.\ (\ref{Pia00sum}) gives for the self-energy part
\begin{equation}\label{Pia00-final}
\delta\Pi^{(a)}_{00}(i\Omega,\bq)=\frac{1}{2}\sigma_0e^2\frac{q^2}{|\Omega|}.
\end{equation}

Let us now concentrate on the vertex part of the density-density
correlator given by Eq.\ (\ref{density-density-vertex}). In order to
calculate this contribution, we need the following integral over the
frequency
\begin{widetext}
\begin{eqnarray}
&&\int_{-\infty}^{\infty}\frac{d\omega'}{2\pi}G_{\bp}(i\omega')G_{\bp-\bq}(i\omega'-i\Omega)
=\frac{1}{2}\frac{1}{\left(|\bp-\bq|+p\right)^2+\Omega^2}\nonumber\\
&\times&\left(-|\bp-\bq|-p+i\Omega\sigma\cdot\frac{2\bp-\bq}{|\bp-\bq|}+(-i\Omega\sigma\cdot\bp+p^2-\bp\cdot\bq-i\sigma_3\bp\times\bq)
\frac{|\bp-\bq|+p}{|\bp-\bq|p}\right).
\end{eqnarray}
Since the frequency integrals in Eq.\ (\ref{density-density-vertex}) are decoupled, we use the above equation to perform them separately, and then
take the trace using the identity
\begin{equation}\label{trace-identity-appendix}
{\rm Tr}[(a+{\bf b}\cdot\sigma)(c+{\bf d}\cdot\sigma)]=2ac+2{\bf
b}\cdot{\bf d}.
\end{equation}
Integrations over $\omega$ and $\omega'$ in the vertex part, given by Eq.\ (\ref{density-density-vertex}), yield the coefficients
\begin{eqnarray}\label{abcd}
a&=&-(|\bp-\bq|+p)\left(1-\frac{\bp\cdot(\bp-\bq)}{p|\bp-\bq|}\right),\,\,
c=-(|\bk-\bq|+k)\left(1-\frac{\bk\cdot(\bk-\bq)}{k|\bk-\bq|}\right),\nonumber\\
{\bf
b}&=&\left\{i\Omega\left(\frac{\bp-\bq}{|\bp-\bq|}-\frac{\bp}{p}\right),-i(\bp\times\bq)_z\frac{|\bp-\bq|+p}{|\bp-\bq|p}\right\},\,\,
{\bf
d}=\left\{i\Omega\left(\frac{\bk-\bq}{|\bk-\bq|}-\frac{\bk}{k}\right),i(\bk\times\bq)_z\frac{|\bk-\bq|+k}{|\bk-\bq|k}\right\}.
\end{eqnarray}
\end{widetext}
in Eq.\ (\ref{trace-identity-appendix}), where $(\bp\times\bq)_z=p_x q_y-p_y q_x$.
Notice that ${\bf b}$ and ${\bf d}$ are three-dimensional vectors.
Expanding the expressions in Eq.\ (\ref{abcd}) to the order $q^2$,
we obtain
\begin{eqnarray}
a&=&-\frac{q^2p^2-(\bq\cdot\bp)^2}{p^3},\nonumber\\
c&=&-\frac{q^2k^2-(\bq\cdot\bk)^2}{k^3}.\nonumber\\
{\bf b}&=&\left\{i\Omega\frac{\bp(\bp\cdot\bq)-\bq
p^2}{p^3},-2i\frac{(\bp\times\bq)_z}{p}\right\},
\nonumber\\
{\bf d}&=&\left\{i\Omega\frac{\bk(\bk\cdot\bq)-\bq
k^2}{k^3},2i\frac{(\bk\times\bq)_z}{k}\right\},
\end{eqnarray}
thus $ac={\cal O} (q^4)$, and therefore does not contribute to the conductivity, whereas
\begin{eqnarray}
{\bf b}\cdot{\bf d}&=&-\Omega^2\frac{[\bp(\bp\cdot\bq)-\bq
p^2]\cdot[\bk(\bk\cdot\bq)-\bq
k^2]}{p^3k^3}\nonumber\\
&+&4\frac{(\bp\times\bq)_z(\bk\times\bq)_z}{pk}.
\end{eqnarray}
Setting $D=2$ in the momentum integrals, since there are no
divergent subintegrals in the vertex part as it was the case in the
self-energy term, the vertex part of the density-density correlator
to the order $q^2$ has the form
\begin{equation}
\delta\Pi^{(b)}_{00}(i\Omega,\bq)
=\delta\Pi^{(b1)}_{00}(i\Omega,\bq)+\delta\Pi^{(b2)}_{00}(i\Omega,\bq),
\nonumber
\end{equation}
where
\begin{eqnarray}\label{Pib100}
&&\delta\Pi^{(b1)}_{00}(i\Omega,\bq)=-N\frac{\Omega^2}{2}\int
\frac{d^2{\bf k}}{(2\pi)^2}\int\frac{d^2{\bf p}}{(2\pi)^2}V_{{\bf k}-{\bf p}}\nonumber\\
&\times&\frac{\left(\bp(\bp\cdot\bq)-\bq
p^2\right)\cdot\left(\bk(\bk\cdot\bq)-\bq
k^2\right)}{p^3k^3\left[\left(2p\right)^2+
\Omega^2\right]\left[\left(2k\right)^2+\Omega^2\right]}\nonumber\\
&\equiv&N\int \frac{d^2{\bf
k}}{(2\pi)^2}\frac{\left(\bk(\bk\cdot\bq)-\bq k^2\right)\cdot{\bf
I}_1({\bf k},{\bf q},\Omega)}{k^3((2k)^2+\Omega^2)},
\end{eqnarray}
and
\begin{eqnarray}\label{Pib200}
&&\delta\Pi^{(b2)}_{00}(i\Omega,\bq)=2N\int
\frac{d^2{\bf k}}{(2\pi)^2}\int\frac{d^2{\bf p}}{(2\pi)^2}V_{{\bf k}-{\bf p}}\nonumber\\
&\times&\frac{\bp\cdot\bk q^2-\bp\cdot\bq \,\,\bq\cdot
\bk}{pk\left[\left(2k\right)^2+\Omega^2\right]\left[(2p)^2+
\Omega^2\right]}\nonumber\\
&\equiv&N\int \frac{d^2{\bf k}}{(2\pi)^2}\frac{I_2({\bf k},{\bf
q},\Omega)}{k[(2k)^2+\Omega^2]}.
\end{eqnarray}
Here, we defined
\begin{equation}\label{Pib00I1}
{\bf I}_1({\bf k},{\bf
q},\Omega)\equiv-\frac{\Omega^2}{8}\int\frac{d^2{\bf p}}{(2\pi)^2}V_{{\bf
k}-{\bf p}}\frac{\bp(\bp\cdot\bq)-\bq
p^2}{p^3\left[p^2+\left(\frac{\Omega}{2}\right)^2\right]},
\end{equation}
and
\begin{equation}\label{Pib00I2}
I_2({\bf k},{\bf q},\Omega)\equiv\frac{1}{2}\int\frac{d^2{\bf
p}}{(2\pi)^2}V_{{\bf k}-{\bf p}}\frac{\bp\cdot\bk q^2-\bp\cdot\bq
\,\,\bq\cdot
\bk}{p\left[p^2+\left(\frac{\Omega}{2}\right)^2\right]}.
\end{equation}
We consider the term $\delta\Pi_{00}^{(b1)}(i\Omega,\bq)$, given by Eq.\ (\ref{Pib100}), and, as a first step, calculate the
integral ${\bf I}_1({\bf k})$. Using the Feynman
parametrization (\ref{FP2}), this integral can be written as
\begin{eqnarray}
&&{\bf I}_1(\bk,\bq,\Omega)=-e^2\Omega^2\int_0^1dx\int_0^{1-x}dy\frac{\sqrt{1-x-y}}{\sqrt{x}}\nonumber\\
&\times&\int\frac{d^2{\bf p}}{(2\pi)^2}\frac{\bp(\bp\cdot\bq)-\bq
p^2}{\left[(\bp-x\bk)^2+x(1-x)\bk^2+\frac{y}{4}\Omega^2\right]^3},\nonumber\\
\end{eqnarray}
which after shifting the momentum variable ${\bf p}-x{\bf
k}\rightarrow{\bf p}$, and integrating over ${\bf p}$, acquires the
form
\begin{eqnarray}
&&{\bf I}_1(\bk,\bq,\Omega)=-e^2\frac{\Omega^2}{8\pi}\int_0^1dx\int_0^{1-x}dy\frac{\sqrt{1-x-y}}{\sqrt{x}}\nonumber\\
&\times&\left(\frac{-\frac{1}{2}\bq}{x(1-x)\bk^2+\frac{y}{4}\Omega^2}+\frac{x^2\bk(\bk\cdot\bq)-\bq
x^2\bk^2}{(x(1-x)\bk^2+\frac{y}{4}\Omega^2)^2}\right).\nonumber\\
\end{eqnarray}
Using the previous result for the integral ${\bf I}_1({\bf
k},\bq,\Omega)$, after integration over ${\bf k}$, we obtain
\begin{eqnarray}\label{Pib100-final}
&&\delta\Pi^{(b1)}_{00}(i\Omega,\bq)= -\frac{Ne^2\bq^2}{64\pi|\Omega|}\int_0^1dx\int_0^{1-x}dy\frac{\sqrt{1-x-y}}{\sqrt{xy}}\nonumber\\
&\times&\left(\frac{1}{\sqrt{y}+\sqrt{(1-x)x}}+\frac{
x^{\frac{3}{2}}}{\sqrt{1-x}\left(\sqrt{(1-x)x}+\sqrt{y}\right)^2}\right)\nonumber\\
&=&-N\frac{e^2q^2}{64\pi|\Omega|}\left[\frac{4\pi}{3}+\pi\left(\pi-\frac{8}{3}\right)\right]\nonumber\\
&=&N\frac{e^2q^2}{|\Omega|}\frac{1}{16}\left(\frac{1}{3}-\frac{\pi}{4}\right)=\frac{e^2q^2}{|\Omega|}\sigma_0\left(\frac{1}{3}-\frac{\pi}{4}\right).
\end{eqnarray}

We now evaluate the term $\delta\Pi^{(b2)}_{00}(i\Omega,\bq)$ in
Eq.\ (\ref{Pib200}). First, we compute the integral
$I_2({\bf k},\bq,\Omega)$ in Eq.\ (\ref{Pib00I2}) using the Feynman
parametrization (\ref{FP2})
\begin{eqnarray}
&&I_2({\bf k},\bq,\Omega)=e^2\int_0^1dx\int_0^{1-x}dy\frac{1}{\sqrt{x}\sqrt{1-x-y}}\nonumber\\
&\times&\int\frac{d^2{\bf p}}{(2\pi)^2}\frac{\bp\cdot\bk
q^2-\bp\cdot\bq \bq\cdot
\bk}{((\bp-x\bk)^2+x(1-x)\bk^2+\frac{y}{4}\Omega^2)^2}\nonumber\\
&=&\frac{e^2q^2}{8\pi}\int_0^1dx\int_0^{1-x}dy\frac{\sqrt{x}}{\sqrt{1-x-y}}
\frac{k^2}{[x(1-x)k^2+\frac{y}{4}\Omega^2]}.\nonumber\\
\end{eqnarray}
We use this result to calculate integral over ${\bf k}$ in Eq.\ (\ref{Pib200}). After performing straightforward integrations, we find
\begin{eqnarray}\label{Pib200-final}
&&\delta\Pi^{(b2)}_{00}(i\Omega,\bq)=\frac{Ne^2q^2}{64\pi|\Omega|}\int_0^1dx\int_0^{1-x}dy
\frac{\sqrt{x}}{\sqrt{1-x-y}}\nonumber\\
&\times&\frac{1}{x(1-x)+\sqrt{x(1-x)y}}=N\frac{e^2q^2}{64|\Omega|}(4-\pi)\nonumber\\
&=&\frac{e^2q^2}{|\Omega|}\frac{4-\pi}{4}.
\end{eqnarray}
Using Eqs.\ (\ref{Pib100-final}) and (\ref{Pib200-final}), the
vertex part of $\delta\Pi_{00}^{(c)}$ reads
\begin{equation}
\delta\Pi^{(b)}_{00}(i\Omega,\bq)=\frac{e^2q^2}{|\Omega|}\frac{8-3\pi}{6},
\end{equation}
which together with the self-energy part (\ref{Pia00-final}) yields
up to the order $q^2$
\begin{equation}
\delta\Pi_{00}^{(c)}(i\Omega,\bq)=\frac{e^2q^2}{|\Omega|}\frac{11-3\pi}{6}.
\end{equation}
Finally, using Eq.\ (\ref{eq:condParallel}), we obtain the Coulomb
interaction correction to the longitudinal conductivity
\begin{equation}
\sigma^{(c)}_\parallel=\frac{11-3\pi}{6}\sigma_0e^2,
\end{equation}
in agreement with the result (\ref{cond-perpendicular}) obtained
from the current-current correlator, which is expected, since the
dimensional regularization explicitly preserves the $U(1)$ gauge
symmetry of the theory of the Coulomb interacting Dirac fermions.

\section{Kubo formula and the a.c. conductivity within dimensional regularization with Pauli matrices in $D=2$}

In this Appendix we present the calculation of the Coulomb
correction to the a.c. conductivity using dimensional
regularization, but treating the Pauli matrices in $D=2$ spatial
dimensions strictly . As we commented earlier, this leads to the
violation of the Ward identity, and thus it is incompatible with the
$U(1)$ gauge symmetry of the theory, but yields the number obtained
in Ref.\ \onlinecite{HerbutJuricicVafekPRL08}. We use Eq.\
(\ref{eq:condPerp}) with the Coulomb correction to the polarization
tensor given by Eq.\ (\ref{polarization-tensor-coulomb}). Since the
system of Dirac fermions interacting only via the long-range Coulomb
interaction is isotropic, translationally and time-reversal
invariant, the trace over spatial indices of the polarization tensor
at zero momentum and a finite frequency is
$\Pi_{aa}(i\Omega,0)=D\Pi_B(i\Omega,0)$, which we use to calculate
the Coulomb correction to the conductivity.

Let us first consider the self-energy part obtained from Eq.\ (\ref{pol-self-energy}). Using that in $D=2$,
$\sigma_a\sigma_\mu\sigma_a=2\delta_{0\mu}$, taking the trace over the Pauli matrices, integrating over the
frequencies, and subtracting the zero-frequency part, the self-energy contribution reads
\begin{eqnarray}
&&{\tilde\sigma}_{a}=-\frac{8}{D}\sigma_0\Omega \int\frac{d^D{\bf k}}{(2\pi)^D}\int\frac{d^D{\bf p}}{(2\pi)^D}\frac{2\pi e^2}{|{\bf
k}-{\bf p}|}\nonumber\\
&\times&\frac{{\bf k}\cdot{\bf p}}{kp}\frac{\Omega^2+12k^2}{k^2(\Omega^2+4k^2)^2}.
\end{eqnarray}
Performing the momentum integrals, we have
\begin{eqnarray}\label{a-tilde-full}
&&{\tilde\sigma}_{a}=-\sigma_0e^2\Omega^{2D-4}\frac{2^{8-4D}
\Gamma(\frac{D+1}{2})(1-\frac{1}{D})} {\pi^{D-1}\Gamma(D/2)\Gamma(D){\rm Cos}{(\pi D)}}\nonumber\\
&\times&\Gamma\left(1-\frac{D}{2}\right)\Gamma\left(\frac{D-1}{2}\right)=\frac{1}{D-1}\sigma_a,
\end{eqnarray}
with $\sigma_a$ given by Eq.\  (\ref{a-complete}). This result, after expanding in $\eps$,
reads
\begin{equation}\label{a-tilde}
{\tilde\sigma}_{a}=\frac{1}{2}\sigma_0e^2\left(\frac{1}{\eps}+\frac{1}{2}+\gamma-\ln64\pi+{\cal O}(\eps)\right).
\end{equation}

The vertex part of the Coulomb correction to the conductivity is obtained from Eq.\ (\ref{pol-vertex}).  The trace over spatial
indices of $\delta\Pi_{\mu\nu}^{(b)}$ in Eq.\ (\ref{pol-vertex}) at the momentum ${\bf q}={ 0}$, using the
standard anticommutation and trace relations for the Pauli matrices in $D=2$,  assumes the form
\begin{widetext}
\begin{eqnarray}
&&\delta\Pi_{aa}^{(b)}(i\Omega,{\bf 0})=-4N\int_{-\infty}^\infty\frac{{d\omega}}{2\pi}\int\frac{d^D{\bf
k}}{(2\pi)^D}\int_{-\infty}^\infty\frac{d\nu}{2\pi}\int\frac{d^D{\bf p}}{(2\pi)^D}\nonumber\\
&\times& V_{{\bf k}-{\bf
p}}\frac{((\omega+\Omega)(\nu+\Omega)-2{\bf k}\cdot{\bf p})(-\omega\nu+{\bf k}\cdot{\bf p})-\omega\nu{\bf
k}\cdot{\bf p}+k^2 p^2}{(\omega^2+k^2)((\omega+\Omega)^2+k^2)(\nu^2+p^2)((\nu+\Omega)^2+p^2)}.
\end{eqnarray}
Performing the integrals over the frequencies $\omega$ and $\nu$ in the above equation, we have
\begin{equation}
\delta\Pi_{aa}^{(b)}(i\Omega,{\bf 0})=2N\int\frac{d^D{\bf k}}{(2\pi)^D}\int\frac{d^D{\bf p}}{(2\pi)^D} V_{{\bf
k}-{\bf p}}\frac{{\bf k}\cdot{\bf p}(4{\bf k}\cdot{\bf p}-\Omega^2)}{kp(\Omega^2+4k^2)(\Omega^2+4p^2)}.
\end{equation}
After subtracting  the zero-frequency part of $\delta\Pi_{aa}(i\Omega,{ 0})$, we obtain the vertex part of
the Coulomb correction to the conductivity
\begin{equation}\label{D6}
{\tilde\sigma}_{b}=\frac{8}{D}\sigma_0\Omega\int\frac{d^D{\bf k}}{(2\pi)^D}\int\frac{d^D {\bf p}}{(2\pi)^D}\,\,\frac{2\pi
e^2}{|{\bf k}-{\bf p}|}\,\,\frac{{\bf k}\cdot{\bf p}\left[{\bf k}\cdot{\bf
p}(\Omega^2+4k^2+4p^2)+4k^2p^2\right]}{k^3p^3(\Omega^2+4k^2)(\Omega^2+4p^2)}={\tilde\sigma}_{b1}+{\tilde\sigma}_{b2}
+{\tilde\sigma}_{b3},
\end{equation}
where
\begin{equation}
{\tilde\sigma}_{b1}=\frac{4}{D}\sigma_0\Omega\int\frac{d^D{\bf k}}{(2\pi)^D}\int\frac{d^D {\bf p}}{(2\pi)^D}\,\,\frac{2\pi
e^2}{|{\bf k}-{\bf p}|}\,\,\frac{\Omega^2+4(k^2+p^2)}{k p(\Omega^2+4k^2)(\Omega^2+4p^2)},
\end{equation}
 diverges in $D=2$, as we will show in the following, and the remaining integrals
\begin{equation}\label{sigmaa2}
{\tilde\sigma}_{b2}=\frac{32}{D}\sigma_0\Omega\int\frac{d^D{\bf k}}{(2\pi)^D}\int\frac{d^D {\bf p}}{(2\pi)^D}\,\,\frac{2\pi
e^2}{|{\bf k}-{\bf p}|}\,\,\frac{{\bf k}\cdot{\bf p}}{k p (\Omega^2+4k^2)(\Omega^2+4p^2)},
\end{equation}
and
\begin{equation}\label{sigmab3}
{\tilde\sigma}_{b3}=\frac{4}{D}\sigma_0\Omega\int\frac{d^D{\bf k}}{(2\pi)^D}\int\frac{d^D {\bf p}}{(2\pi)^D}\,\,\frac{2\pi
e^2}{|{\bf k}-{\bf p}|}\,\,\frac{[2({\bf k}\cdot{\bf p})^2-k^2p^2][\Omega^2+8k^2]}{k^3
p^3(\Omega^2+4k^2)(\Omega^2+4p^2)}
\end{equation}
are finite in two dimensions. The above decomposition is obtained from Eq.\ (\ref{D6}) by adding and subtracting $\Omega^2$ in the term in the numerator multiplying $(\bk\cdot\bp)^2$.
We now further decompose the integral ${\tilde\sigma}_{b1}$ in order to isolate its diverging part
\begin{equation}
{\tilde\sigma}_{b1}={\tilde\sigma}_{b1}^{(1)}+{\tilde\sigma}_{b1}^{(2)}.
\end{equation}
Here, the term
\begin{equation}\label{tilde-sigmab11-appendix}
{\tilde\sigma}_{b1}^{(1)}=\frac{8}{D}\sigma_0\Omega\int\frac{d^D{\bf k}}{(2\pi)^D}\int\frac{d^D {\bf p}}{(2\pi)^D}\,\,\frac{2\pi
e^2}{|{\bf k}-{\bf p}|}\frac{1}{kp(\Omega^2+4k^2)},
\end{equation}
has the pole in Laurent expansion in $\eps$,  while the remaining one
\begin{equation}\label{sigmab1-2}
{\tilde\sigma}_{b1}^{(2)}=-\frac{4}{D}\sigma_0\Omega^3\int\frac{d^D{\bf k}}{(2\pi)^D}\int\frac{d^D {\bf p}}{(2\pi)^D}\,\,
\frac{2\pi e^2}{|{\bf k}-{\bf p}|}\frac{1}{kp(\Omega^2+4k^2)(\Omega^2+4p^2)}
\end{equation}
is finite in $D=2$.

 We first consider the term divergent in $D=2$, namely, ${\tilde\sigma}_{b1}^{(1)}$. Integration over ${\bf p}$
in Eq.\ (\ref{tilde-sigmab11-appendix}), after performing the standard steps, yields
\begin{equation}
\int\frac{d^D{\bf p}}{(2\pi)^D}\frac{1}{p|{\bf k}-{\bf
p}|}=\frac{k^{D-2}}{\pi}\frac{\Gamma\left(1-\frac{D}{2}\right)\left[\Gamma\left(\frac{D-1}{2}\right)
\right]^2}{(4\pi)^{D/2} \Gamma(D-1)},
\end{equation}
while after integrating over ${\bf k}$, we find
\begin{equation}
{\tilde\sigma}_{b1}^{(1)}=\sigma_0 e^2\Omega^{2D-4}\frac{\Gamma\left(1-\frac{D}{2}\right)
\left[\Gamma\left(\frac{D-1}{2}\right)\right]^2\Gamma\left(D-\frac{3}{2}\right)\Gamma\left(\frac{5}{2}-D\right)
}{2^{4D-7}\pi^{D}D\Gamma(D-1)\Gamma\left(\frac{D}{2}\right)}.
\end{equation}
The previous expression, after expanding in $\eps$, reads
\begin{equation}
{\tilde\sigma}_{b1}^{(1)}=\frac{1}{2}\sigma_0e^2\left[\frac{1}{\epsilon}+\frac{1}{2}-\gamma+\ln(64\pi)+O(\epsilon)\right].
\end{equation}
Thus the poles in ${\tilde\sigma}_{a}$, given by Eq.\ (\ref{a-tilde}),  and ${\tilde\sigma}_{b1}^{(1)}$ cancel out, and
\begin{equation}\label{a+b11}
{\tilde\sigma}_{b1}^{(1)}+{\tilde\sigma}_{a}=\frac{1}{2}\sigma_0e^2+{\cal O}(\epsilon).
\end{equation}

The remaining integrals are finite in $D=2$, and can be calculated as follows. Consider the term ${\tilde\sigma}_{b2}$
given by Eq.\ (\ref{sigmaa2}). Using Feynman parametrization (\ref{FP2}),
we write
\begin{equation}\label{aux}
\frac{1}{p|{\bf k}-{\bf p}|\left[(\frac{\Omega}{2})^2+p^2\right]}=\frac{1}{\pi}\int_0^1dx\int_0^{1-x}dy
\frac{(1-x-y)^{-1/2}x^{-1/2}}{\left[({\bf p}-x{\bf k})^2+x(1-x)k^2+y\left(\frac{\Omega}{2}\right)^2\right]^2}.
\end{equation}
As usual, we now shift the momentum ${\bf p}-x{\bf k}\rightarrow{\bf p}$, and integrate over ${\bf p}$,  to obtain
\begin{eqnarray}
{\tilde\sigma}_{b2}&=&\frac{1}{2\pi}\sigma_0 e^2\Omega\int_0^1dx\int_0^{1-x}dy\,\,(1-x-y)^{-1/2}x^{1/2}\nonumber\\
&\times&\int\frac{d^2{\bf k}}{(2\pi)^2}\frac{k}{\left[x(1-x)k^2+y\left(\frac{\Omega}{2}\right)^2\right]\left[k^2+\left(\frac{\Omega}{2}\right)^2 \right] }.
\end{eqnarray}
The integral over ${\bf k}$ in the previous equation then yields
\begin{equation}
{\tilde\sigma}_{b2}=\frac{1}{4\pi}\sigma_0e^ 2\int_0^1dx\int_0^{1-x}dy\,\,\frac{1}{\sqrt{(1-x)(1-x-y)}\left(\sqrt{x(1-x)}+\sqrt{y}\right)},
\end{equation}
while the integration over the variable $y$ gives
\begin{equation}
{\tilde\sigma}_{b2}=\frac{1}{4\pi}\sigma_0e^2\int_0^1dx\frac{1}{\sqrt{1-x}}\left[\pi+2i\sqrt{\frac{x}{1-x}}\sec^{-1}\sqrt{x}\right].
\end{equation}
Finally, after performing the remaining integral over $x$, we obtain
\begin{equation}\label{b2}
{\tilde\sigma}_{b2}=\frac{4-\pi}{4}\sigma_0e^2.
\end{equation}
Similarly, using Eq.\ (\ref{aux}) after shifting ${\bf p}-x{\bf k}\rightarrow{\bf p}$, and integrating over ${\bf p}$,
the term ${\tilde\sigma}_{b1}^{(2)}$ in Eq.\ (\ref{sigmab1-2}) becomes
\begin{eqnarray}
{\tilde\sigma}_{b1}^{(2)}&=&-\frac{1}{16\pi}\sigma_0 e^2\Omega^3\int_0^1dx\int_0^{1-x}dy\,\,(1-x-y)^{-1/2}x^{-1/2}\nonumber\\
&\times&\int\frac{d^2{\bf k}}{(2\pi)^2}\frac{1}{k\left[x(1-x)k^2+y\left(\frac{\Omega}{2}\right)^2\right]\left[k^2+\left(\frac{\Omega}{2}\right)^2 \right] }.
\end{eqnarray}
Integration over the remaining momentum variable yields
\begin{equation}
{\tilde\sigma}_{b1}^{(2)}=-\frac{1}{8\pi}\sigma_0e^ 2\int_0^1dx\int_0^{1-x}dy\,\,\frac{1}{\sqrt{x(1-x-y)}\left(y+\sqrt{xy(1-x)}\right)}.
\end{equation}
After integrating out the Feynman parameter $y$, the term ${\tilde\sigma}_{b1}^{(2)}$ is
\begin{equation}\label{b12}
{\tilde\sigma}_{b1}^{(2)}=\frac{i}{4\pi}\sigma_0e^2\int_0^1dx\,\,\frac{\sec^{-1}\sqrt{x}}{\sqrt{x}(1-x)}=-\frac{\pi}{8}\sigma_0e^2.
\end{equation}
Let us now calculate the term $\sigma_{b3}$ in Eq.\ (\ref{sigmab3}). Using Feynman parametrization (\ref{FP2}), we can write
\begin{equation}\label{aux1}
\frac{1}{p^3|{\bf k}-{\bf p}|\left[(\frac{\Omega}{2})^2+p^2\right]}=\frac{4}{\pi}\int_0^1dx\int_0^{1-x}dy
\frac{(1-x-y)^{1/2}x^{-1/2}}{\left[({\bf p}-x{\bf k})^2+x(1-x)k^2+y\left(\frac{\Omega}{2}\right)^2\right]^3}.
\end{equation}
After shifting ${\bf p}-x{\bf k}\rightarrow{\bf p}$, and integrating over ${\bf p}$, we have
\begin{eqnarray}
{\tilde\sigma}_{b3}&=&\frac{1}{8\pi}\sigma_0 e^2\Omega\int_0^1dx\int_0^{1-x}dy\,\,(1-x-y)^{1/2}x^{3/2}\nonumber\\
&\times&\int\frac{d^2{\bf k}}{(2\pi)^2}\frac{k(\Omega^2+8k^2)}{\left[x(1-x)k^2+y\left(\frac{\Omega}{2}\right)^2\right]^2\left[k^2+\left(\frac{\Omega}{2}\right)^2 \right] },
\end{eqnarray}
while integration over ${\bf k}$ then gives
\begin{equation}
{\tilde\sigma}_{b3}=\frac{1}{8\pi}\sigma_0e^ 2\int_0^1dx\int_0^{1-x}dy\,\,\frac{\sqrt{1-x-y}\left[x(1-x)+4\sqrt{xy(1-x)}+2y\right]}
{(1-x)^{3/2}\left(y+\sqrt{xy(1-x)}\right)\left(\sqrt{y}+\sqrt{x(1-x)}\right)},
\end{equation}
which after calculating the integral over the variable $y$ becomes
\begin{eqnarray}\label{b3}
{\tilde\sigma}_{b3}&=&\frac{1}{8\pi}\sigma_0e^2\int_0^1dx\,\,\ (1-x)^{-3/2}\left[2(x-1)\sqrt{x}+\pi(1-x^2)+2i\sqrt{x^3(1-x)}\sec^{-1}\sqrt{x}\right]\nonumber\\
&=&\frac{14-3\pi}{24}\sigma_0e^2.
\end{eqnarray}
 \end{widetext}
Therefore, we calculated all the terms needed to obtain the Coulomb correction to the conductivity within this
regularization scheme. Using Eqs.\ (\ref{a+b11}), (\ref{b2}), (\ref{b12}), and (\ref{b3}),
we obtain the final result
as found in Ref.\ \onlinecite{HerbutJuricicVafekPRL08}
\begin{equation}\label{cond-PRL}
{\tilde\sigma}^{(c)}={\tilde\sigma}_a+{\tilde\sigma}_b={\tilde\sigma}_a+{\tilde\sigma}_{b1}^{(1)}+{\tilde\sigma}_{b1}^{(2)}+{\tilde\sigma}_{b2}+{\tilde\sigma}_{b3}
=\frac{25-6\pi}{12}\sigma_0e^2,
\end{equation}
which thus yields ${\cal C}=(25-6\pi)/12$, different than one in Eq.\ (\ref{constant-C}) obtained using
dimensional regularization with Pauli matrices in $D=2-\eps$ spatial dimensions.

\section{Direct evaluation of the a. c. conductivity from the Kubo formula in two dimensions}

   In this section we show that the value of the coefficient ${\cal C}$ may also be unambiguously computed directly in two spatial dimensions, provided extra care is taken in evaluations of the integral that defines it. The result is then in agreement with the general dimensional-regularization scheme used in the rest of the paper.

   As suggested by Mishchenko \cite{MishchenkoEPL08}, the expression for the first-order correction to conductivity may also be conveniently written as a sum of three terms
   \begin{equation}
   \sigma' = \sigma_a ' + \sigma_b ' + I
   \end{equation}
   where
   \begin{equation}
   \sigma_a ' = e^2 \omega \int \frac{d^2 \bold{p} d^2 \bold{k}}{(2\pi)^4} V_{ \bold{p}-\bold{k} } \cos\theta_{\bold{p} \bold{k}}
   \frac{\omega^2 - 4 p^2}{p^2 (\omega^2 + 4 p^2)^2 },
   \end{equation}
    \begin{equation}
   \sigma_b ' = e^2 \omega \int \frac{d^2 \bold{p} d^2 \bold{k}}{(2\pi)^4} V_{ \bold{p}-\bold{k} } \cos\theta_{\bold{p} \bold{k}}
   \frac{4- (\omega^2/pk)\cos\theta}{(\omega^2 + 4 k^2) (\omega^2 + 4 p^2)},
   \end{equation}
   \begin{equation}
 I  = - 2 e^2 \omega \int \frac{d^2 \bold{p} d^2 \bold{k}}{(2\pi)^4} V_{ \bold{p}-\bold{k} } \cos\theta_{\bold{p} \bold{k}}
   \frac{k-p \cos\theta}{p^2 k (\omega^2 + 4 p^2)}.
 \end{equation}
 Here we have set the fermi velocity to unity for simplicity, and taken $\omega$ to be the Matsubara frequency.

 Now we show that all three terms are UV convergent, and when summed yield the same result as the dimensional regularization. First,\cite{MishchenkoEPL08}
 \begin{equation}
 \sigma_0 + \sigma_a ' = 2e^2 \omega \int \frac{d^2 \bold{p}}{ (2\pi)^2 } \frac{v_p}{p (1+4 v_p ^2 p^2 )} + O(V^2),
\end{equation}
where
\begin{equation}
v_p = v_F \left(1+ \frac{e^2}{4} \ln\frac{\Lambda}{p}  \right)
\end{equation}
is the renormalized velocity. A simple change of variables then gives
\begin{equation}
\frac{\sigma_a '}{\sigma_0} = \beta_v (e^2),
\end{equation}
where
\begin{equation}
\beta_v (e^2)= -\frac{d v_p}{v_p d  {\rm ln} (p) } = \frac{e^2}{4} +O(e^4)
\end{equation}
is the beta-function for the velocity. The coefficients in the series expansion of $\beta_v (e^2)$ are universal numbers, and  therefore,
\begin{equation}
\frac{\sigma_a '}{\sigma_0}=  \frac{e^2}{4}
\end{equation}
in agreement with ref. \onlinecite{MishchenkoEPL08}. The result also agrees with the brute force numerical integration, in which the integral over the angle is computed first and exactly, to be followed by the UV-convergent integrals over the momenta, which are then computed up to a large cutoff.

  The second integral is completely convergent, and was solved by Mishchenko\cite{MishchenkoEPL08}, with the result
\begin{equation}
\frac{\sigma_b ' }{\sigma_0} = \left( \frac{4}{3} - \frac{\pi}{2} \right) e^2 .
\end{equation}
We have also reproduced this value numerically within a tenth of a percent.

    Finally, the third integral may be written as
\begin{widetext}
\begin{equation}
\frac{I}{\sigma_0} =   - \frac{2 e^2 }{\pi^2} \lim _{\omega\rightarrow 0} \int_0 ^{\Lambda_1 /\omega} \frac{dp}{p (1+ 4 p^2)}  \int_0 ^{\Lambda_2 /\omega} dk  \frac{\partial}{\partial k} \int_0 ^{2\pi} d\theta \cos \theta (p^2 + k^2 - 2 p k \cos \theta)^{1/2}.
\end{equation}
\end{widetext}
where we have carefully retained finite upper cutoffs on the momentum integrals. One finds
\begin{equation}
\frac{I}{\sigma_0} =   - \frac{2 e^2}{ \pi^2} \lim _{\omega\rightarrow 0} \int_0 ^{\Lambda_1 /\omega}  \frac{dp}{p (1+ 4 p^2)} F\left(p, \frac{\Lambda_2}{\omega} \right),
\end{equation}
where
\begin{eqnarray}
F(x,y) &=& \frac{2|x-y|}{3 x y} \left[(x+y)^2 K\left(-\frac{4 x y }{(x-y)^2} \right)\right.   \\ \nonumber
&-&\left.(x^2+y^2) E\left(-\frac{ 4 x y }{(x-y)^2}\right) \right],
\end{eqnarray}
and $K(z)$ and $E(z)$ are the elliptic functions. The single remaining numerical integration quickly converges to a value quite independent of the ratio $\Lambda_1/\Lambda_2$, and we obtain
\begin{equation}
\frac{I}{\sigma_0} = 0.2498 e^2 \rightarrow \frac{e^2}{4},
\end{equation}
where the last equality is the conjectured exact result. All put together gives
\begin{equation}
\frac{\sigma'}{ \sigma_0} = \left[\frac{1}{4} + \left(\frac{4}{3} - \frac{\pi}{2}\right) + \frac{1}{4}\right] e^2 = \left(\frac{11}{6} -\frac{\pi}{2}\right) e^2,
\end{equation}
in agreement with the procedure of dimensional regularization.

It is instructive to see how in this calculation the value ${\cal C}= (25- 6 \pi)/12 $ would arise. If we follow the usually safe practice and take the UV cutoffs to infinity before all the integrals have been performed, in the present case the result turns out to depend on the order of integration. In this way performing exactly the integral over $k$ first we find
 \begin{equation}
\frac{I}{\sigma_0} =   - \frac{2 e^2 }{ \pi^2} \int_0 ^{\infty }  \frac{dp}{p (1+ 4 p^2)}  (-\pi p)= \frac{e^2}{2},
\end{equation}
which ultimately yields $ {\cal C} = (25-6\pi) / 12 $. Computing first numerically the integral over $p$ and then the remaining integral over $k$, on the other hand, leads to $e^2/ 4$ instead. Of course, by the very definition of the integral, the correct way is to take any limits of the integration bounds only after all the integrals have been already performed. It is pleasing to see that this then leads to the same result that is obtained by the general dimensional regularization, which is constructed so to preserve the crucial symmetries of the theory.

\end{document}